\documentclass[12pt]{article}
\usepackage[utf8]{inputenc}

\title{Graphical copula GARCH modeling with dynamic conditional dependence}
  \author{Lupe S. H. Chan \hspace{.2cm}\\
	The Hong Kong University of Science and Technology\\
	and \\
	Amanda M. Y. Chu \\
	The Education University of Hong Kong\\
	and \\
	Mike K. P. So \thanks{Corresponding author}\\
	The Hong Kong University of Science and Technology}

\usepackage[a4paper, total={7in, 9in}]{geometry}
\usepackage{hyperref}
\usepackage{physics}
\usepackage{amsmath,amssymb}
\usepackage{mathtools}
\usepackage[table,xcdraw]{xcolor}
\usepackage[normalem]{ulem}

\usepackage{tikz}
\usepackage{float}

\usepackage{color,soul}
\definecolor{myGreen}{rgb}{0.8,1,0.2}
\definecolor{myRed}{rgb}{0.9,0.65,0.65}

\usepackage{enumerate}
\usetikzlibrary{bayesnet}
\usetikzlibrary{shapes.geometric}%

\usepackage{tikz}
\usepackage{graphicx}
\usepackage{array,multirow,graphicx}
\usepackage{amsmath,siunitx}

\makeatletter

\def\env@cases#1{%
	\let\@ifnextchar\new@ifnextchar
	\left\lbrace\def\arraystretch{1.2}%
	\array{@{}#1@{\quad}l@{}}}
\makeatother

\usepackage[giveninits=true,hyperref=true,
style=apa,
autocite = superscript,
maxbibnames=99,backend=biber,maxcitenames=3,uniquelist=false,
language=auto,    
autolang=other,doi=true,isbn=false,date=year,dashed=false,
]{biblatex} 
\bibliography{GC_GARCH}

\usepackage{tikz}
\usepackage{float}
\usepackage{caption}
\usepackage{subcaption}
\usepackage{bbm}
\usetikzlibrary{calc}
\usepackage{siunitx}
\usetikzlibrary{bayesnet}
\usetikzlibrary{shapes.geometric}%
\usetikzlibrary{shadows,arrows.meta,positioning,backgrounds,fit}
\usetikzlibrary{arrows,decorations.pathreplacing}

\tikzset{%
	materia/.style={draw, fill=blue!10, text width=28em,text centered, minimum height=1.5em},
	cont/.style={draw, fill=yellow!10, text width=28em,text centered, minimum height=1.5em},
	dist/.style={draw, fill=red!10, text width=28em,text centered, minimum height=1.5em},
	mix/.style={draw, fill=green!10, text width=28em,text centered, minimum height=1.5em},
	etape/.style={materia, text width=13em, minimum width=13em, minimum height=3em, rounded corners}, 
	wideEtape/.style={materia, text width=12em, minimum width=12em, minimum height=3em, rounded corners}, 
	contEtape/.style={cont, text width=12em, minimum width=12em, minimum height=3em, rounded corners},
	mixEtape/.style={mix, text width=12em, minimum width=12em, minimum height=3em, rounded corners},
	distEtape/.style={dist, text width=12em, minimum width=12em, minimum height=3em, rounded corners},
	texto/.style={above, text width=6em, text centered},
	linepart/.style={draw, thick, color=black!50, -LaTeX, dashed},
	line/.style={draw, thick, color=black!50, -LaTeX},
	ur/.style={draw, text centered, minimum height=0.01em,draw=white!100,text width=13em},
	result/.style={draw, text centered, minimum height=0.01em,draw=white!100,text width=17em},
	back group/.style={fill=yellow!10,rounded corners, draw=black!50, dashed, inner xsep=15pt, inner ysep=10pt},
	brace_top/.style={
		color=blue,
		decoration={brace},
		decorate
	},
	brace_bottom/.style={
		color=blue,
		decoration={brace, mirror},
		decorate
	},number line/.style={}
}

\usepackage{algpseudocode}

\newcommand\pa{\text{pa}}

\newcommand\diag{\text{diag}}

\usepackage{amsthm}
\theoremstyle{definition}

\newcommand\nbd{\text{nbd}}

\begin{document}
	\maketitle
	\begin{abstract}
		Modeling {returns on large portfolios} is a challenging problem as the number of parameters in the covariance matrix grows as the square of the size of the portfolio. 
		Traditional correlation models, for example, the dynamic conditional correlation (DCC)-GARCH model, often ignore the nonlinear dependencies in the tail of the return distribution. In this paper, we aim to develop a framework to model the nonlinear dependencies dynamically, namely the graphical copula GARCH (GC-GARCH) model. Motivated from the capital asset pricing model, to allow modeling of large portfolios, the number of parameters can be greatly reduced by introducing conditional independence among stocks given some risk factors. The joint distribution of the risk factors is factorized using a directed acyclic graph (DAG) with pair-copula construction (PCC) to enhance the modeling of the tails of the return distribution while offering the flexibility of having complex dependent structures. The DAG induces topological orders to the risk factors, which can be regarded as a list of directions of the flow of information. The conditional distributions among stock returns are also modeled using PCC. Dynamic conditional dependence structures are incorporated to allow the parameters in the copulas to be time-varying. Three-stage estimation is used to estimate parameters in the marginal distributions, the risk factor copulas, and the stock copulas. 
		{The} simulation study shows that the proposed estimation procedure can estimate the parameters and the underlying DAG structure accurately. In the investment experiment of the empirical study, we demonstrate that the GC-GARCH model produces more precise conditional value-at-risk prediction and considerably higher cumulative portfolio returns than the DCC-GARCH model.
	\end{abstract}

	\section{Introduction}
	Portfolio selection for {large portfolios} is often a challenging problem. The mean-variance optimization introduced by \textcite{markowitz} provides a framework to decide the allocation of a set of assets by balancing the risk and return. The problem is also known as the minimum variance (MV) optimization if we do not specify the target return. To take the extreme scenarios into account (for example, catastrophic losses), \textcite{rockafellar2000optimization} proposes an efficient formulation to minimize the Conditional Value-at-Risk (CVaR) of the portfolio. These optimization problems require us to estimate the covariance matrix and the CVaR of the portfolio. However, the number of parameters in the covariance matrix grows quadratically in the size of the portfolio, and the traditional models often omit the nonlinear dependence in the tails due to the assumption of multivariate normality or multivariate $t$, which may give poor CVaR predictions.

	The capital asset pricing model (CAPM) developed by \textcite{sharpe1964capital} and \textcite{lintner1975valuation} has often been adopted in the literature, and uses a smaller number of risk factors to capture the co-movements among a large portfolio of stocks, so as to reduce the number of parameters in the estimation. Several works attempt to apply the CAPM to reduce the dimension in the covariance estimation \parencite{fan2008high,fan2011high,engle2019large}. Nevertheless, these works are limited to multivariate normal distribution or multivariate $t$ distribution. These multivariate distributions fail to capture the nonlinear dependencies in the tails, which are crucial for the portfolio selection when we consider the minimum CVaR problem.
	
	To effectively capture time series properties of high-dimensional returns, we also need to take the dynamic dependence features into account. \textcite{DCC_model} and \textcite{DCC_GARCH} propose the dynamic conditional correlation (DCC)-GARCH models to model the dynamic correlations. However, these models again fail to capture the non-linear dependencies in the tails. The literature provides evidence that assets are more likely to co-move in extreme scenarios \parencite{jang2002asian,jiang2017financial}. It motivates us to develop a model that can reduce the dimension of parameters, and in the meantime, also capture non-linear dependence in the tails dynamically.
	
\textcite{sklar1959fonctions} introduces the framework of the copula, which promotes flexibility to model the marginal distributions of the variables and their multivariate dependencies separately. Several works introduce copula-GARCH models \parencite{jondeau2006copula,huang2009estimating,aloui2013conditional,lee2009copula,mayer2023estimation}, where the marginal time series are assumed to follow GARCH structures, and the multivariate dependencies among variables are modeled using multivariate copulas. Time-varying dependence structures can also be incorporated \parencite{lu2014portfolio,wang2011dynamic,ozun2007portfolio}. A drawback of the multivariate copula is that some copula parameters are applied to all variables to capture the non-linear dependence structure. For example, in the multivariate $t$-copula, the degrees of freedom is shared among all variables to explain the tail dependencies.
	This could still be too restrictive as the tail dependencies may vary across different pairs of variables.
	
	Pair-copula construction (PCC) has been used to further flexibilize the copula-GARCH models, where the joint distribution is decomposed into a product of bivariate copulas based on a hierarchical structure, allowing the copula parameters to vary for different pairs of variables. Vine decomposition proposed by \textcite{joe1997multivariate} is one of the methods to define the hierarchy in the PCC. Vine-copula GARCH models are used in modeling financial returns in the literature \parencite{brechmann2013risk,vine_copula,min2010bayesian,acar2019flexible}. \textcite{min2010bayesian,vine_copula} use the Markov chain Monte Carlo (MCMC) methods for parameter estimation in vine-copula GARCH models. Besides, \textcite{bauer2016pair,pairwise_copula_constructions} attempt to use Bayesian networks (BN) to specify the hierarchical structures among variables in the PCC. Bayesian networks are graphical models to represent dependence structures among variables using directed acyclic graphs (DAG), where the nodes in the DAG represent variables, and the directed edges in the DAG represent the dependence structures. The advantage of using BN is that we can specify topological orders for the variables in the BN, where topological orders give a linear ordering of the variables in the BN. The topological orders of a BN of stock returns can be regarded as the flow of information or risk in the financial market \parencite{chan2023moving,garvey2015analytical}, which provide a natural way to set up a hierarchy in the variables for PCC. {Furthermore, PCCs using BN are more parsimonious than that using vine copula models in general due to its focus on conditional independence} \parencite{bauer2016pair}; {the number of copulas to be estimated in PCCs using BN is smaller than or equal to that using vine decomposition.} 
	
	In this paper, we aim to develop a Graphical Copula GARCH (GC-GARCH) framework to model the nonlinear dependencies in the tails among a large portfolio of stocks (say, 100 stocks). The GC-GARCH model consists of the following four components: 
	\begin{enumerate}[(1)]
		\item The conditional independence of stock returns given the risk factors motivated from the Capital Asset Pricing Model (CAPM).
		\item The use of a directed acyclic graph to define the dependence structures of the risk factors.
		\item To specify the conditional distributions of returns using pair-copula constructions.
		\item To impose time-varying dependence structures through copula parameters.
	\end{enumerate}
	Having these four components, we can greatly reduce the dimensionality in the parameter estimation; in the meantime, we can explain dynamically the nonlinear dependencies among stocks from the market indexes using the copulas, where the market indexes are hierarchized naturally using a DAG. 
	
	The rest of the paper is as follows. In Section \ref{section:model_specification}, we provide details of the methodology of the GC-GARCH model and explain the logic of the above four components. In Section \ref{section:computational_issues}, we discuss the computational issues in the GC-GARCH model. In Section \ref{section:estimation}, we derive the likelihood functions and posterior distributions, and provide algorithms for parameter estimation and DAG estimation. In Section \ref{Section:portfolio_management}, we discuss how we use the GC-GARCH model in portfolio selection. In Section \ref{section:simulation_study}, we provide the results of a simulation study to illustrate that the estimation algorithms in Section \ref{section:estimation} can recover the parameters and the underlying DAG accurately. In Section \ref{section:empirical_study}, we present findings of the empirical study by applying the GC-GARCH model to a portfolio of 92 stocks with 10 market indexes. We also compare the predictive performance between the GC-GARCH model and the DCC-GARCH model. In Section \ref{section:discussion_and_conclusion}, we present our conclusions and a discussion of the paper.


	\section{Methodology}
	\label{section:model_specification}
	We denote the $m$ risk factor returns and $p$ stock returns on day $t$ by $r_{1,t},\ldots,r_{m,t}$, and $r_{m+1,t},\ldots,r_{m+p,t}$, respectively, for $t=1,2,\ldots,T$. To model nonlinear dependence among risk factor returns and stock returns dynamically, we are interested in developing $F^{[t]}(r_{1,t},\ldots,r_{m+p,t})$, the joint distribution function of $r_{1,t},\ldots,r_{m,t},r_{m+1,t},\ldots,r_{m+p,t}$ given $\mathcal{F}_{t-1}$, the information set up to time $t-1$. We further let $D=\{D_1,\ldots,D_T\}$ be the set of data containing all returns from day $1$ to day $T$, where $D_t=\{r_{1,t},\ldots,r_{m+p,t}\}$ contains all returns on day $t$, $t=1,\ldots,T$. This conditional distribution setting is inline with the usual GARCH modeling. In this paper, we develop the graphical copula GARCH (GC-GARCH) model. There are four main features in the GC-GARCH specification: (1) the conditional independence of stock returns, $r_{m+1,t},\ldots,r_{m+p,t}$, given the risk factors, $r_{1,t},\ldots,r_{m,t}$ motivated from the Capital Asset Pricing Model (CAPM) theory; (2) the use of a directed acyclic graph to define the dependence structures of the risk factors; (3) the specification of the conditional distributions using pair-copula construction; and (4) the imposition of time-varying dependence structures in the modeling; more specifically, we allow the correlation parameters in the conditional bivariate copulas to be time varying in a similar way to the dynamic conditional correlation (DCC)-GARCH model \parencite{DCC_model,DCC_GARCH}.
	
	\subsection{Conditional independence of stock returns}
	The first level of construction of the GC-GARCH model is through application of the conditional independence property motivated from the CAPM theory. The joint density function of stock returns and risk factor returns can be factorized into
	
	\begin{equation}
		\begin{aligned}
			f^{[t]}(r_{1,t},\ldots,r_{m+p,t})&=f^{[t]}(r_{m+1,t},\ldots,r_{m+p,t}|r_{1,t},\ldots,r_{m,t})f^{[t]}(r_{1,t},\ldots,r_{m,t})\\
			&= \left[\prod_{j=m+1}^{m+p} f^{[t]}(r_{j,t}|r_{1,t},\ldots,r_{m,t})\right]f^{[t]}(r_{1,t},\ldots,r_{m,t}),
		\end{aligned}
		\label{eqt:full_density_function}
	\end{equation}
	where $f^{[t]}(r_{1,t},\ldots,r_{m,t})$ is the conditional joint density of the risk factor returns given $\mathcal{F}_{t-1}$. It gives a crucial property that the stock return series $r_{i,t}$ is independent of other stock return series $r_{j,t}$ for $i\ne j$ and $i,j>m$, conditional on the risk factor returns $r_{1,t},\ldots,r_{m,t}$. A similar exploration under multivariate $t$ distribution assumption was found in \textcite{risk_factor_mapping}.
	
	\subsection{Directed acyclic graph models for risk factor returns}
\begin{figure}[H]
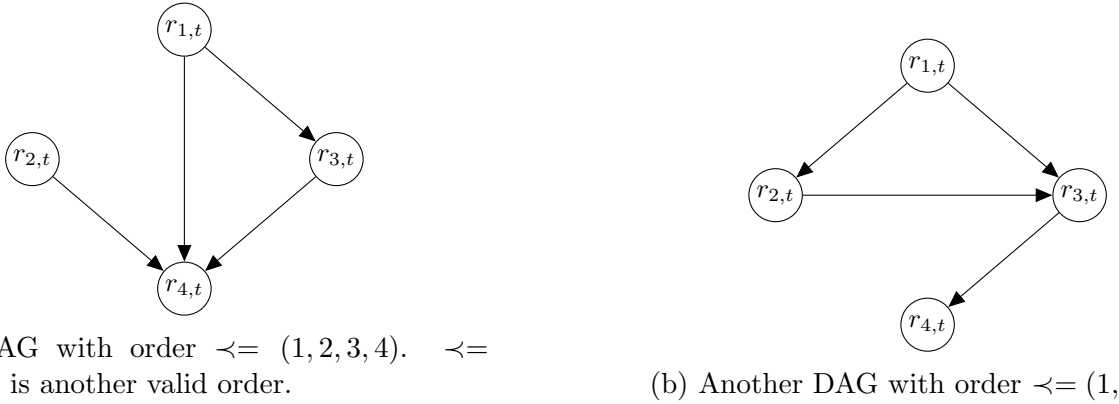

		\centering
		\begin{subfigure}[b]{0.45\textwidth}
			\centering
			\tikz{
				\node[latent] (1) {$r_{1,t}$};%
				\node[latent,below=of 1,xshift=-2cm] (2) {$r_{2,t}$}; %
				\node[latent,below=of 1,xshift=2cm] (3) {$r_{3,t}$}; %
				\node[latent,below=of 2,xshift=2cm] (4) {$r_{4,t}$};
				\edge {1} {3,4} 
				\edge {2} {4}
				\edge {3} {4}
			}
			\caption{A DAG with order $\prec=(1,2,3,4)$. $\prec=(2,1,3,4)$ is another valid order.}
			\label{fig:DAG_example_1a}
		\end{subfigure}
		\hfill
		\begin{subfigure}[b]{0.45\textwidth}
			\centering
			\tikz{
				\node[latent] (1) {$r_{1,t}$};%
				\node[latent,below=of 1,xshift=-2cm] (2) {$r_{2,t}$}; %
				\node[latent,below=of 1,xshift=2cm] (3) {$r_{3,t}$}; %
				\node[latent,below=of 2,xshift=2cm] (4) {$r_{4,t}$};
				\edge {1} {2,3} 
				\edge {2} {3}
				\edge {3} {4}
			}
			\caption{Another DAG with order $\prec=(1,2,3,4)$}
			\label{fig:DAG_example_1b}
		\end{subfigure}
		\caption{Example of two DAGs.}
		\label{fig:Two_dag_example_for_2_orders}
	\end{figure}

	We formulate the conditional joint distribution of the $m$ risk factor returns, i.e., the term 
	$$
	f^{[t]}(r_{1,t},\ldots,r_{m,t})
	$$
	in \eqref{eqt:full_density_function}, using directed acyclic graphs (DAGs) \parencite{koller2009probabilistic}. A principle of DAG is to represent possible causal or conditional independence relationships through a graph. The variables in the DAG are called nodes, and a relationship between any two variables is indicated by a directed edge. \autoref{fig:Two_dag_example_for_2_orders} presents two DAGs with four nodes (i.e., $m=4$) and multiple directed edges. Whenever there is an edge from node $i$ to node $j$, we say that node $i$ is a parent of node $j$, and we notate $i\in\pa(j)$, where $\pa(j)$ is the parent set of node $j$. For example, in \autoref{fig:DAG_example_1a}, $\pa(1)=\pa(2)=\varnothing$ (where $\varnothing$ denotes the empty set), $\pa(3)=\{1\}$, and $\pa(4)=\{1,2,3\}$. Variables in a DAG can be ordered topologically. A topological order, or simply an order, is valid if $i\in \pa(j)$, then node $i$ must be on the left of node $j$ in the order. We notate an order using the symbol $\prec$. For example, an order of the DAG in \autoref{fig:DAG_example_1a} is $\prec=(1,2,3,4)$. Another order $\prec=(2,1,3,4)$ is also valid for the DAG in \autoref{fig:DAG_example_1a}; since nodes $1$ and $2$ are not connected by an edge, we can either order node $1$ first or node $2$ first. With the same order $\prec$, the DAG, however, may not be unique. For example, the order $\prec=(1,2,3,4)$ is valid for both DAGs in \autoref{fig:DAG_example_1a} and \autoref{fig:DAG_example_1b}. 
	
	By using the standard Markov assumptions in the DAG \parencite{pairwise_copula_constructions}, the joint density of $m$ risk factor returns can be decomposed into a product of univarate conditional densities
	
	\begin{equation}
		f^{[t]}(r_{1,t},\ldots,r_{m,t})=\prod_{i=1}^m f^{[t]}(r_{i,t}|\mathbf{r}_{\pa(i),t}),
		\label{eqt:dag_factorization}
	\end{equation}
	where $\mathbf{r}_{\pa(i),t}$ is the set of risk factor returns that are parents of node $i$. For example, we have $f^{[t]}(r_{1,t},\ldots,r_{4,t})=f^{[t]}(r_{1,t})f^{[t]}(r_{2,t})f^{[t]}(r_{3,t}|r_{1,t})f^{[t]}(r_{4,t}|r_{1,t},r_{2,t},r_{3,t})$ for the DAG in \autoref{fig:DAG_example_1a}, and $f^{[t]}(r_{1,t},\ldots,r_{4,t})=f^{[t]}(r_{1,t})f^{[t]}(r_{2,t}|r_{1,t})f^{[t]}(r_{3,t}|r_{1,t},r_{2,t})f^{[t]}(r_{4,t}|r_{3,t})$ for the DAG in \autoref{fig:DAG_example_1b}.
	
	Instead of adopting standard multivariate distribution such as multivariate normal or $t$ distributions, the decomposition in \eqref{eqt:dag_factorization} enables us to flexibly specify the joint distribution of the $m$ risk factors through a graphical representation in the DAG.

	\subsection{Copulas and conditional copulas}
	

	By Sklar's theorem \parencite{sklar1959fonctions}, $F(x_1,\ldots,x_d)$, a $d$-dimensional ($d\in\{1,2,\ldots\}$) cumulative distribution function (CDF) for the variables $(x_1,\ldots,x_d)\in(-\infty,\infty)^d$ can be written as \parencite{copula_book}
	
	$$
	F(x_1,\ldots,x_d)=C(F_1(x_1),\ldots,F_d(x_d)),
	$$
	for some $d$-dimensional copula $C:[0,1]^d\mapsto [0,1]$ and $F_1,\ldots,F_d$ are the marginal CDFs of $x_1,\ldots,x_d$ respectively. The joint density for $x_1,\ldots,x_d$ is
	\begin{equation}
		f(x_1,\ldots,x_d)=c(F_1(x_1),\ldots,F_d(x_d))\prod_{i=1}^d f_i(x_i),
		\label{eqt:d_dim_coupla_density}
	\end{equation}
	where $f_1,\ldots,f_d$ are the marginal density functions of $x_1,\ldots,x_d$ respectively, and the copula density function $c$ is obtained by
	$$
	c(u_1,\ldots,u_d)=\frac{\partial^d C(u_1,\ldots,u_d)}{\partial u_1 \ldots \partial u_d},
	$$
	where $u_i=F_i(x_i)$ for $i=1,\ldots,d$.
	
	The conditional distribution in \eqref{eqt:dag_factorization} are often modeled using multivariate Gaussian distributions \parencite{GBN}, which may performs poorly in extreme scenarios in financial applications. Instead, we factorize the conditional distributions in \eqref{eqt:dag_factorization} using a similar logic of vine decomposition based on pair-copula construction (PCC) \parencite{vine_copula}. Using \eqref{eqt:d_dim_coupla_density}, we can express the conditional density of variables $x$ and $y$ given a random vector $z$ by $f(x,y|z)=c_{x,y|z}(F(x|z),F(y|z))f(x|z)f(y|z)$, where $c_{x,y|z}$ is a conditional copula density function given $z$. Therefore, we can express $f(x|y,z)$ as
	\begin{equation}
		f(x|y,z)=\frac{f(x,y|z)}{f(y|z)}=\frac{c_{x,y|z}(F(x|z),F(y|z))f(x|z)f(y|z)}{f(y|z)}=c_{x,y|z}(F(x|z),F(y|z))f(x|z).
		\label{eqt:copula_expression_of_conditional_distribution}
	\end{equation}
	Using \eqref{eqt:copula_expression_of_conditional_distribution} and taking $x$ as $r_{j,t}$, $y$ as $r_{m,t}$, and $z$ as $r_{1,t},\ldots,r_{m-1,t}$ for $j=m+1,\ldots,m+p$, the conditional density of return on the $j$-th stock can be expressed as
	$$
	\begin{aligned}
		& f^{[t]}(r_{j,t}|r_{1,t},\ldots,r_{m,t})\\ =& c_{j,m|1,\ldots,m-1}^{[t]}(F^{[t]}(r_{j,t}|r_{1,t}\ldots,r_{m-1,t}),F^{[t]}(r_{m,t}|r_{1,t},\ldots,r_{m-1,t}))f^{[t]}(r_{j,t}|r_{1,t},\ldots,r_{m-1,t}),
	\end{aligned}
	$$
	and recursively, for $i>1$, we have
	\begin{equation}
		f^{[t]}(r_{j,t}|r_{1,t},\ldots,r_{m,t})=\left[\prod_{i=1}^m c_{j,i|1,\ldots,i-1}^{[t]}(F^{[t]}(r_{j,t}|r_{1,t},\ldots,r_{i-1,t}),F^{[t]}(r_{i,t}|r_{1,t},\ldots,r_{i-1,t}))\right]\cdot f_j^{[t]}(r_{j,t}),
		\label{eqt:stock_return_pdf_factorization}
	\end{equation}
	where $c_{j,i|1,\ldots,i-1}^{[t]}(\cdot,\cdot)$ is the conditional bivariate copula density between the $j$-th stock return $r_{j,t}$ and the $i$-th risk factor return $r_{i,t}$ given the information on all the first $i-1$ risk factor returns $r_{1,t},\ldots,r_{i-1,t}$ and $\mathcal{F}_{t-1}$, and $f_j^{[t]}(r_{j,t})$ is the conditional marginal distribution of $r_{j,t}$ given $\mathcal{F}_{t-1}$, {for $i=2,\ldots,m$ and $j=m+1,\ldots,m+p$. For $i=1$, we have}
	$$
	c_{j,i|1,\ldots,i-1}^{[t]}(F^{[t]}(r_{j,t}|r_{1,t},\ldots,r_{i-1,t}),F^{[t]}(r_{i,t}|r_{1,t},\ldots,r_{i-1,t})) \equiv c_{j,1}(F^{[t]}(r_{j,t}),F^{[t]}(r_{1,t})).
	$$
	
	
	In the same way, to express the conditional distributions in the DAG for the risk factor returns in \eqref{eqt:dag_factorization} into a product of conditional bivariate copula density, we label the $m$ risk factor returns to give the order $\prec=(1,2,\ldots,m)$. Then, denote the parent set of $r_{i,t}$ as $\pa(i)=\{i[1],\ldots,i[n(i)]\}$, where $i[1]<\ldots<i[n(i)]$, and $n(i)$ is the number of parents of $r_{i,t}$. With the above setting, we can express the conditional density of the return of the $i$-th risk factor to
	\begin{equation}
		f^{[t]}(r_{i,t}|\mathbf{r}_{\pa(i),t})=\left[\prod_{k=1}^{n(i)} c_{i,i[k]|i[1],\ldots,i[k-1]}^{[t]}(F^{[t]}(r_{i,t}|r_{i[1],t},\ldots,r_{i[k-1],t}),F^{[t]}(r_{i[k],t}|r_{i[1],t},\ldots,r_{i[k-1],t}))\right]\cdot f_i^{[t]}(r_{i,t}).
		\label{eqt:risk_factor_return_pdf_factorization}
	\end{equation}
	
	For example, for the DAG in \autoref{fig:DAG_example_1a}, we have the parent set $\pa(4)=\{1,2,3\}$, the number of parents $n(4)=3$, $4[1]=1$, $4[2]=2$ and $4[3]=3$. Therefore, \eqref{eqt:risk_factor_return_pdf_factorization} implies 
	$$
	\begin{aligned}
		f^{[t]}(r_{4,t}|\mathbf{r}_{\pa(4),t})
		&=c_{4,3|1,2}^{[t]}(F^{[t]}(r_{4,t}|r_{1,t},r_{2,t}),F^{[t]}(r_{3,t}|r_{1,t},r_{2,t}))
		c_{4,2|1}^{[t]}(F^{[t]}(r_{4,t}|r_{1,t}),F^{[t]}(r_{2,t}|r_{1,t}))\\&c_{4,1}^{[t]}(F^{[t]}(r_{4,t}),F^{[t]}(r_{1,t}))\cdot f^{[t]}_4(r_{4,t}).
	\end{aligned}
	$$
	Similarly, we have $\pa(3)=\{1\}$, $n(3)=1$ and $3[1]=1$. Applying \eqref{eqt:risk_factor_return_pdf_factorization} again implies
	$$
	f^{[t]}(r_{3,t}|\mathbf{r}_{\pa(3),t})= c_{3,1}^{[t]}(F^{[t]}(r_{3,t}),F^{[t]}(r_{1,t})) \cdot f^{[t]}_3(r_{3,t}).
	$$
	Note that $\pa(r_{1,t})=\pa(r_{2,t})=\varnothing$ and thus $f^{[t]}(r_{2,t}|\mathbf{r}_{\pa(2),t})=f^{[t]}_2(r_{2,t})$ and $f^{[t]}(r_{1,t}|\mathbf{r}_{\pa(1),t})=f^{[t]}_1(r_{1,t})$. Combining these factorization results and \eqref{eqt:risk_factor_return_pdf_factorization}, we can express the joint density of $m$ risk factor returns as
	$$
	\begin{aligned}
		f^{[t]}(r_{1,t},r_{2,t},r_{3,t},r_{4,t})&=\Bigg[ c_{4,3|1,2}^{[t]}(F^{[t]}(r_{4,t}|r_{1,t},r_{2,t}),F^{[t]}(r_{3,t}|r_{1,t},r_{2,t}))
		c_{4,2|1}^{[t]}(F^{[t]}(r_{4,t}|r_{1,t}),F^{[t]}(r_{2,t}|r_{1,t}))\\&c_{4,1}^{[t]}(F^{[t]}(r_{4,t}),F^{[t]}(r_{1,t}))\Bigg] \cdot c_{3,1}^{[t]}(F^{[t]}(r_{3,t}),F^{[t]}(r_{1,t}))\cdot \prod_{i=1}^4 f^{[t]}_i(r_{i,t}).
	\end{aligned}
	$$

	In general, combining the DAG decomposition of the $m$ risk factors in  \eqref{eqt:dag_factorization}, the copula factorization of the stock returns in \eqref{eqt:stock_return_pdf_factorization}, and the copula factorization for the risk factor returns in \eqref{eqt:risk_factor_return_pdf_factorization}, the joint density of all returns $r_{1,t}\ldots r_{m+p,t}$ in \eqref{eqt:full_density_function} can be decomposed into a product of conditional bivariate copulas and marginal densities:
	\begin{equation}
		\begin{aligned}
			f^{[t]}(r_{1,t},\ldots,r_{m+p,t})&=\left[\prod_{j=m+1}^{m+p}\prod_{i=1}^m c_{j,i|1,\ldots,i-1}^{[t]}(F^{[t]}(r_{j,t}|r_{1,t},\ldots,r_{i-1,t}),F^{[t]}(r_{i,t}|r_{1,t},\ldots,r_{i-1,t}))\right]\cdot\\
			&\left[\prod_{i=1}^m \prod_{k=1}^{n(i)}c_{i,i[k]|i[1],\ldots,i[k-1]}^{[t]}(F^{[t]}(r_{i,t}|r_{i[1],t},\ldots,r_{i[k-1],t}),F^{[t]}(r_{i[k],t}|r_{i[1],t},\ldots,r_{i[k-1],t})) \right]\cdot\\
			&\prod_{j=1}^{m+p}f_{j}^{[t]}(r_{j,t}).
		\end{aligned}
		\label{eqt:full_factorization}
	\end{equation}
	
	In financial econometrics studies, the number of stocks, $p$, is much larger than the number of risk factors, $m$. Therefore, from the joint density in \eqref{eqt:full_factorization}, the number of bivariate copulas is at most $mp+m(m-1)/2$, equality holds if all pairs of nodes in the DAG are connected, in which case the number of edges in the DAG is $\binom{m}{2}=m(m-1)/2$. The number of parameters is of $O(p)$ instead of $O(p^2)$ in usual dynamic covariance modeling of $p$ stocks.
	
	\subsection{Dynamic conditional dependence and tail dependence}
	The decomposition in \eqref{eqt:full_factorization} also provides flexibility in the choice of conditional
	dependence, including linear correlation and the tail dependence parameters
	through conditional bivariate copulas. To incorporate time-varying dependence features in the GC-GARCH model, we follow the idea in \textcite{vine_copula}, which inspired by the DCC-GARCH models of \textcite{DCC_model} and \textcite{DCC_GARCH}, to allow the correlation parameters in the conditional copulas $c^{[t]}_{j,i|1,\ldots,i-1}$ and $c^{[t]}_{i,i[k]|i[1],\ldots,i[k-1]}$ to be time-varying. In this paper, we focus on the use of bivariate $t$-copulas with correlation parameters $\varphi_{j,i|1,\ldots,i-1}^{[t]}$ and $\varphi^{[t]}_{i,i[k]|i[1],\ldots,i[k-1]}$. In general, we denote the conditional copula and copula density at time $t$ for two arbitrary variables $r_{x,t}$ and $r_{y,t}$ conditional to the set $z$ as $C_{xy|z}^{[t]}$ and $c_{xy|z}^{[t]}$ respectively.  In the simulation study and empirical study, we use the $t$-copula to illustrate the GC-GARCH model. The conditional $t$-copula at time $t$ with correlation parameter $\varphi=\varphi_{xy|z}^{[t]}$ and degrees of freedom $v=v_{xy|z}$ is given by \parencite{demarta2005t}
	$$
	\begin{aligned}
		C^{[t]}_{xy|z}(u_{x|z,t},u_{y|z,t})&= \int_{-\infty}^{t_{v}^{-1}(u_{x|z,t})}\int_{-\infty}^{t_{v}^{-1}(u_{y|z,t})} f_{2;v,\varphi}(x,y) dy dx,
	\end{aligned}
	$$
	where $u_{x|z,t}=F^{[t]}(r_{x,t}|r_{z,t})$, $u_{y|z,t}=F^{[t]}(r_{y,t}|r_{z,t})$, $t_{v}^{-1}(\cdot)$ is the inverse CDF of the univariate $t$ distribution with degrees of freedom $v$, $f_{2;v,\varphi}(\cdot,\cdot)$ is the density function of the bivariate $t$ distribution with degrees of freedom $v$ and correlation parameter $-1< \varphi < 1$, defined as
	$$
	f_{2;v,\varphi}(x,y)=\frac{1}{2\pi\sqrt{1-{\varphi}^2}}\left(
	1+\frac{ x^2+y^2-2\varphi xy }{v(1-\varphi^2)}
	\right)^{-\frac{v+2}{2}}.
	$$
	We impose the finite covariance condition that $v>2$ \parencite{demarta2005t}. Using \eqref{eqt:d_dim_coupla_density}, the conditional $t$-copula density at time $t$ with correlation parameter $\varphi=\varphi_{xy|z}^{[t]}$ and degrees of freedom $v=v_{xy|z}$ is given by
	$$
	\begin{aligned}
		c_{xy|z}^{[t]}(u_{x|z,t},u_{y|z,t}) &= \frac{f_{2;v,\varphi}(t^{-1}_v(u_{x|z,t}),t^{-1}_v(u_{y|z,t}))}{f_{1;v}(t^{-1}_v(u_{x|z,t}))f_{1;v}(t^{-1}_v(u_{y|z,t}))},
	\end{aligned}
	$$
	where $f_{1;v}(\cdot)$ is the density function of the univariate $t$-distribution with degrees of freedom $v$, where
	$$
	f_{1;v}(x)=\frac{\Gamma\left(\frac{v+1}{2}\right)}{\sqrt{v\pi}\Gamma\left(\frac{v}{2}\right)}\left(1+\frac{x^2}{v}\right)^{-\frac{v+1}{2}}.
	$$

	We assume that the conditional correlations change dynamically similar to the dynamics in \textcite{vine_copula}, inspired by the DCC-GARCH models of \textcite{DCC_GARCH} and \textcite{DCC_model}. The dynamic conditional correlation $\varphi^{[t]}_{xy|z}$ between returns $r_{x,t}$ and $r_{y,t}$ given $r_{z,t}$ and $\mathcal{F}_{t-1}$ is given by
	\begin{equation}
		\varphi^{[t]}_{xy|z} = (1-a_{xy|z}-b_{xy|z})\bar \varphi_{xy|z}+a_{xy|z}\xi_{xy|z,t-1}+b_{xy|z}\varphi^{[t-1]}_{xy|z},
		\label{eqt:copula_parameter_dynamic}
	\end{equation}
	where $\bar \varphi_{xy|z}$ is the long-run correlation and $\xi_{xy|z,t-1}$ is the sample correlation at time $t-1$. The stationary conditions are $0\leq a_{xy|z},b_{xy|z} < 1$, $a_{xy|z}+b_{xy|z}< 1$ and $-1< \bar\varphi_{xy|z} < 1$. The sample correlation is obtained given the past $m_{sc}$-period of data and $r_{z,t}$, and is defined as
	$$
	\xi_{xy|z,t-1}=\frac{\sum_{i=1}^{m_{sc}} \tilde{r}_{x|z,t-i}\tilde{r}_{y|z,t-i}}{\sqrt{\sum_{i=1}^{m_{sc}}\tilde{r}_{x|z,t-i}^2 \sum_{j=1}^{m_{sc}} \tilde{r}_{y|z,t-j}^2}},
	$$
	where $\tilde{r}_{x|z,t}=t_{v_{xy|z}}^{-1}(F^{[t]}(r_{x,t}|r_{z,t}))$ and $\tilde{r}_{y|z,t}=t_{v_{xy|z}}^{-1}(F^{[t]}(r_{y,t}|r_{z,t}))$. We pick ${m_{sc}}=2$ in this paper for the simulation study and empirical study.
	
	In financial application, we often want to estimate the unconditional correlations. In this case, computations are needed to convert the conditional correlation $\varphi_{xy|z}^{[t]}$ in the GC-GARCH model into unconditional correlation as in \textcite{vine_copula}. The unconditional correlation can be obtained iteratively using the formula in \textcite{rummel1976understanding},
	$$
	\varphi_{xy|z_{-j}}^{[t]}=\varphi_{x,y|z}^{[t]}\sqrt{(1-{\varphi_{xz_j|z_{-j}}^{[t]}}^2)(1-{\varphi_{yz_j|z_{-j}}^{[t]}}^2)}+\varphi_{xz_j|z_{-j}}^{[t]}\varphi_{yz_j|z_{-j}}^{[t]},
	$$
	where $z_j$ is the $j$th component of the vector $z$, and $z_{-j}$ is the vector obtained by removing the $j$th component in $z$.

	{The $t$-copula has often been used in financial return data modeling due to its ability to capture the phenomenon of dependent extreme values }\parencite{demarta2005t,t_copula_rm_eg1,vine_copula}. {The tail dependence coefficient of the conditional $t$-copula $C_{xy|z}^{[t]}$ with correlation parameter $\varphi_{xy|z}^{[t]}$ and degrees of freedom $v_{xy|z}$ is defined as}
	$$
	\lambda_{xy|z}^{[t]}=2t_{v+1}\left(-\sqrt{\frac{(v_{xy|z}+1)(1-\varphi_{xy|z}^{[t]})}{(1+\varphi_{xy|z}^{[t]})}}\right),
	$$
	{where $t_{d}(\cdot)$ is the CDF of the standard $t$-distribution with degrees of freedom $d>0$.
		The tail dependence depends on two parameters, the correlation and the degrees of freedom. Note that, even when the correlation is zero, the $t$-copula still gives dependence in the tails. The $t$-copula captures the dependence of extreme values which is often observed in financial modeling. We report the estimate of the degrees of freedom $v_{xy|z}$ for each copula as a measure of tail dependence.}

	\section{Computational issues}
	\label{section:computational_issues}
	\subsection{Marginal distributions}
	Similar to the approach in \textcite{vine_copula}, we assume that the each of the returns $r_{1,t},\ldots,r_{m+p,t}$ follows a GARCH(1,1)-$t$ model, i.e., a GARCH(1,1) model with $t$ distributed innovations,
	\begin{equation}
		\begin{aligned}
			r_{i,t}&=\sigma_{i,t}\varepsilon_{i,t},\\
			\sigma_{i,t}^2&=\omega_i+\alpha_ir_{i,t-1}^2+\beta_i\sigma_{i,t-1}^2,
		\end{aligned}
		\label{eqt:marginal_dynamics}
	\end{equation}
	where $\varepsilon_{i,t}$'s are independently and identically distributed standardized $t$ innovations with degrees of freedom $v_i$, $i=1,\ldots,m+p$. The constraints for positive variance and covariance stationary are $\omega_i>0$, $\alpha_i\geq 0$, $\beta_i\geq 0$, and $\alpha_i+\beta_i<1$ for $i=1,\ldots,m+p$. The conditional density function of $r_{i,t}$ under GARCH(1,1)-$t$ in \eqref{eqt:marginal_dynamics} is given by
	
	\begin{equation}
		f_i^{[t]}(r_{i,t})=\frac{\Gamma\left(\frac{v_i+1}{2}\right)}{\sigma_{i,t}\sqrt{(v_i-2)\pi}\Gamma\left(\frac{v_i}{2}\right)}\left(1+\frac{r_{i,t}^2}{\sigma_{i,t}^2(v_i-2)}\right)^{-\frac{v_i+1}{2}},
		\label{eqt:marginal_pdf}
	\end{equation}
	for $t=1,\ldots,T$. To ensure the variance is finite and \eqref{eqt:marginal_pdf} is well-defined, we require $v_i>2$. \eqref{eqt:marginal_pdf} is derived using the transformation
	$$
	r_{i,t}=\frac{\sigma_{i,t}}{\sqrt{\frac{v_i}{v_i-2}}}T_{v_i},
	$$
	where $T_{v_i}$ follows a (unstandardized) $t$-distribution with degrees of freedom $v_i$.
	
	\subsection{Conditional distributions}
	Similar to \textcite{vine_copula}, the conditional distribution functions of the form $F^{[t]}(x_t|v_t)$ in \eqref{eqt:full_factorization} have to be computed recursively using the $h$-function associated with the conditional copulas $C_{x,v_j|v_{-j}}^{[t]}$ \parencite{Joe1996FamiliesO}:
	\begin{equation}
		F^{[t]}(x_t|v_t)=\frac{\partial C_{x,v_j|v_{-j}}^{[t]}(F^{[t]}(x_t|v_{-j,t}),F^{[t]}(v_{j,t}|v_{-j,t}))}{\partial F^{[t]}(v_{j,t}|v_{-j,t})}:=h_{x,v_j|v_{-j}}^{[t]}(F^{[t]}(x_t|v_{-j,t}),F^{[t]}(v_{j,t}|v_{-j,t})),
		\label{eqt:h_function}
	\end{equation}
	where $v_t$ is a random vector at time $t$, $v_{j,t}$ is the $j$th element of $v_t$, and $v_{-j,t}$ is the vector obtained by removing $v_{j,t}$ from $v_t$. Conversely, the inverse of the $h$-function at time $t$ in \eqref{eqt:h_function} is also useful if we want to reduce the dimension of the conditioning set of $F^{[t]}(x_t|v_t)$. If both $F^{[t]}(x_t|v_t)$ and $F^{[t]}(v_{j,t}|v_{-j,t})$ are available, then $F^{[t]}(x_t|v_{-j,t})$ can be evaluated using the inverse $h$-function
	\begin{equation}
		F^{[t]}(x_t|v_{-j,t})={h^{[t]}}^{-1}_{x,v_j|v_{-j}}(F^{[t]}(x_t|v_t),F^{[t]}(v_{j,t}|v_{-j,t})).
		\label{eqt:h_inv_function}
	\end{equation}
	In the simulation study and empirical study, we use the $t$-copulas in the GC-GARCH model. The $h$-function and inverse $h$-function of the $t$-copula with correlation parameter $\varphi=\varphi^{[t]}_{xy|z}$ and degrees of freedom $v=v_{xy|z}$ are respectively given by
	$$
	h_{xy|z}^{[t]}(u_{xt},u_{yt})=t_{v+1}\left(
	\frac{t_{v}^{-1}(u_{xt})-\varphi t_{v}^{-1}(u_{yt})}
	{\sqrt{\frac{({v}+ [t_{v}^{-1}(u_{yt})]^2)(1-\varphi^2)}{{v}+1}}}
	\right)
	$$
	and
	$$
	h^{{[t]}^{-1}}_{xy|z}(u_{xt},u_{yt})=t_{v}\left(
	t_{{v}+1}^{-1}(u_{xt})\sqrt{\frac{({v}+ [t_{v}^{-1}(u_{yt} )]^2)(1-\varphi^2)}{{v}+1}}+\varphi t_{v}^{-1}(u_{yt})
	\right),
	$$
	where $u_{xt}=F_{x}^{[t]}(r_{xt})$ and $u_{yt}=F_{y}^{[t]}(r_{yt})$ are the cumulative distributions of two return variables $r_{xt}$ and $r_{yt}$ at time $t$.

	\subsection{Computation of the likelihood function}
	\label{section:computation_of_likelihood_function}
	The joint density in \eqref{eqt:full_factorization} contains three parts, which have to be computed sequentially: the marginal distributions, $f_i^{[t]}$, the conditional copulas among the risk factors, $c_{i,i[k]|i[1],...,i[k-1]}^{[t]}$, and the conditional copulas between the risk factors and stocks, $c_{j,i|1,...,i-1}^{[t]}$. We define the following parameter sets: 
	\begin{enumerate}
		\item $\Theta_{1i}=\{\omega_i,\alpha_i,\beta_i,v_i\}$ be the set of parameters in the marginal distribution of the $i$th stock, where $i=1,\ldots,m+p$. We also denote $\Theta_1=\{\Theta_{1i}\}_{i=1}^{m+p}$.
		\item $\Theta_2=\{ \bar \varphi_{i,i[k]|z_{ik}},a_{i,i[k]|z_{ik}},b_{i,i[k]|z_{ik}},v_{i,i[k]|z_{ik}}: k=1,\ldots,n(i)\text{ and } i=2,\ldots,m\}$ be the set of parameters in the DAG copulas, where $z_{ik}=\{i[1],\ldots,i[k-1]\}$, and
		\item $\Theta_{3j}=\{\bar\varphi_{j,i|1,\ldots,i-1},a_{j,i|1,\ldots,i-1},b_{j,i|1,\ldots,i-1},v_{j,i|1,\ldots,i-1}:i=1,\ldots,m\}$ be the set of parameters in the copulas of the stock $j$, where $j=m+1,\ldots,m+p$. We also denote $\Theta_3=\{\Theta_{3j}\}_{j=m+1}^{m+p}$.
	\end{enumerate}
	The steps to compute of the joint density are depicted below. Starting at $t=1$,
	\begin{enumerate}
		\item (Marginal distribution) We first compute the cumulative distribution of the marginal distributions $F^{[t]}(r_{1,t}),\ldots,F^{[t]}(r_{m+p,t})$ and the log density functions $\log f^{[t]}_1(r_{1,t})$, $\ldots,\log f^{[t]}_{m+p}(r_{m+p,t})$.
		\item (Conditional copulas of the DAG) For each $i=2,\ldots,m$ and if $r_{i,t}$ has at least one parent (i.e., $n(i)\geq 1$), then for $k=1,\ldots,n(i)$,
		\begin{enumerate}[(i)]
			\item if $t=1$, initialize $\varphi_{i,i[k]|i[1],\ldots,i[k-1]}^{[1]}$ as the long-run correlation $\bar\varphi_{i,i[k]|i[1],\ldots,i[k-1]}$. Otherwise, $\varphi_{i,i[k]|i[1],\ldots,i[k-1]}^{[t]}$ is computed according to the dynamics in \eqref{eqt:copula_parameter_dynamic}.
			\item Compute and store the log copula density with correlation parameter $\varphi_{i,i[k]|i[1],\ldots,i[k-1]}^{[t]}$
			and degrees of freedom $v_{i,i[k]|i[1],\ldots,i[k-1]}$:
			$$
			\log c_{i,i[k]|i[1],\ldots,i[k-1]}^{[t]}(F^{[t]}(r_{i,t}|r_{i[1],t},\ldots,r_{i[k-1],t}),F^{[t]}(r_{i[k],t}|r_{i[1],t},\ldots,r_{i[k-1],t})).
			$$
			\item If $k<n(i)$, compute and store $F^{[t]}(r_{i,t}|r_{i[1],t},\ldots,r_{i[k],t})$ and $F^{[t]}(r_{i[k+1],t}|r_{i[1],t},\ldots,r_{i[k],t})$ using the $h$-function in \eqref{eqt:h_function}.
		\end{enumerate}
		\item (Conditional copulas between the stocks and the risk factors) For each stock $j=m+1,\ldots,m+p$, we compute as follows: For $i=1,\ldots,m$,
		\begin{enumerate}[(i)]
			\item if $t=1$, initialize $\varphi_{j,i|1,\ldots,i-1}^{[1]}$ as the long-run correlation $\bar\varphi_{j,i|1,\ldots,i-1}$. Otherwise, $\varphi_{j,i|1,\ldots,i-1}^{[t]}$ is computed according to the dynamics in \eqref{eqt:copula_parameter_dynamic}.
			\item Compute and store the log copula density with correlation parameter $\varphi_{j,i|1,\ldots,i-1}^{[t]}$
			and degrees of freedom $v_{j,i|1,\ldots,i-1}$:
			$$
			\log c_{j,i|1,\ldots,i-1}^{[t]}(F^{[t]}(r_{j,t}|r_{1,t},\ldots,r_{i-1,t}),F^{[t]}(r_{i,t}|r_{1,t},\ldots,r_{i-1,t})).
			$$
			\item If $i<m$, compute and store $F^{[t]}(r_{j,t}|r_{1,t},\ldots,r_{i,t})$ and $F^{[t]}(r_{i+1,t}|r_{1,t},\ldots,r_{i,t})$ using the $h$-function in \eqref{eqt:h_function}. 
		\end{enumerate}
		\item Set $t$ {to} $t+1$ and go back to step 1, until $t=T$.
	\end{enumerate}
	Finally, we can calculate the log likelihood function
	$$
	\begin{aligned}
		& \ell(\Theta|D) = \sum_{t=1}^T \log f^{[t]}(r_{1,t},\ldots,r_{m+p,t})\\  
		&=\left(\sum_{i=1}^{m+p}\ell_{1i}(\Theta_{1i}|D)\right)+\ell_2(\Theta_1,\Theta_2|D)+\left(\sum_{j=m+1}^{m+p}\ell_{3j}(\Theta_1,\Theta_2,\Theta_{3j}|D)\right),
	\end{aligned}
	$$
	with
	\begin{align}
		\ell_{1i}(\Theta_{1i}|D) &=\sum_{t=1}^T \log f_{i}^{[t]}(r_{i,t}), \text{for }i=1,\ldots,m+p, \label{eqt:marginal_likelihood} \\
		\ell_2(\Theta_1,\Theta_2|D) &=\sum_{i=2}^m \sum_{k=1}^{n(i)} \sum_{t=1}^T\log c_{i,i[k]|i[1],\ldots,i[k-1]}^{[t]}(F^{[t]}(r_{i,t}|r_{i[1],t},\ldots,r_{i[k-1],t}),F^{[t]}(r_{i[k],t}|r_{i[1],t},\ldots,r_{i[k-1],t})),  \label{eqt:ell_2} \\
		\ell_{3j}(\Theta_1,\Theta_2,\Theta_{3j}|D) &=\sum_{i=1}^m \sum_{t=1}^T\log c_{j,i|1,\ldots,i-1}^{[t]}(F^{[t]}(r_{j,t}|r_{1,t},\ldots,r_{i-1,t}),F^{[t]}(r_{i,t}|r_{1,t},\ldots,r_{i-1,t})), \label{eqt:ell_3j}\\
		&\text{for }j=m+1,\ldots,m+p, \nonumber
	\end{align}
	and $\Theta=\{\Theta_1,\Theta_2,\Theta_3\}$. In the estimation of parameters in DAG copulas, we need to compute starting from $i=2$ to $m$ sequentially since the conditional distributions depend on the parents. However, the parameters in each of the marginal distribution can be estimated separately. For the stock copulas, $r_{j,t}$ only depend on the risk factor returns $r_{1,t},\ldots,r_{m,t}$ but not other stock returns $r_{j',t}$, for all $j'\in \{m+1,\ldots,m+p\}\setminus \{j\}$. This implies that we can estimate the parameters in the stock copulas separately for each stock. Thus, we express the log likelihood function as a sum of marginal likelihood functions and a sum of stock copula likelihood functions because the parameters can be estimated separately for each term in the two summations. We discuss the detailed procedures for the estimation in Section \ref{section:estimation}.
	
	\subsection{The reduced DAG space}
	\label{section:reduced_DAG_space}
\begin{figure}[H]
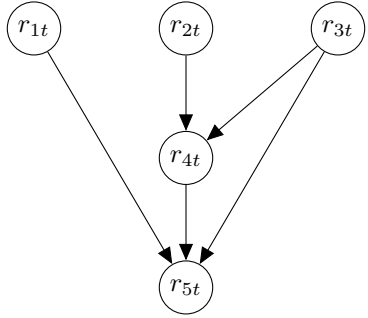

		\centering
		\tikz{
			\node[latent] (1) {$r_{1t}$};%
			\node[latent,xshift=2cm] (2) {$r_{2t}$}; %
			\node[latent,xshift=4cm] (3) {$r_{3t}$}; %
			\node[latent,below=of 2] (4) {$r_{4t}$};
			\node[latent,below=of 4] (5) {$r_{5t}$};
			\edge {1} {5} 
			\edge {2} {4}
			\edge {3} {4,5}
			\edge {4} {5}
		}
		\caption{An example to illustrate the difficulty of calculating the conditional distributions.}    \label{fig:example_difficult_conditional_probability}
	\end{figure}

	The conditional density of the $i$th risk factor return, $i=1,\ldots,m$, in \eqref{eqt:risk_factor_return_pdf_factorization} is dependent on the conditional distribution functions $F^{[t]}(r_{i,t}|r_{i[1],t},\ldots,r_{i[k-1],t})$ and $F^{[t]}(r_{i[k],t}|r_{i[1],t},\ldots,r_{i[k-1],t})$, which are computed recursively using the $h$-function in \eqref{eqt:h_function}. However, the computation of the second conditional distribution function $F^{[t]}(r_{i[k],t}|r_{i[1],t},\ldots,r_{i[k-1],t})$ may be complicated for some DAGs. To illustrate this, \autoref{fig:example_difficult_conditional_probability} contains a DAG with five variables as an example. Using the DAG factorization in \eqref{eqt:dag_factorization}, the joint density function of $r_{1,t},r_{2,t},r_{3,t},r_{4,t},r_{5,t}$ can be expressed as
	\begin{equation}
		f^{[t]}(r_{1,t},r_{2,t},r_{3,t},r_{4,t},r_{5,t})=f^{[t]}(r_{1,t})f^{[t]}(r_{2,t})f^{[t]}(r_{3,t})f^{[t]}(r_{4,t}|r_{2,t},r_{3,t})f^{[t]}(r_{5,t}|r_{1,t},r_{3,t},r_{4,t}).
		\label{eqt:Full_DAG_reduce_space_example}
	\end{equation}
	Choosing the topological order $\prec=(1,2,3,4,5)$ (as there are multiple orders that are compatible to the DAG in \autoref{fig:example_difficult_conditional_probability}), we further consider the pair-copula construction for the term $f^{[t]}(r_{5,t}|r_{1,t},r_{3,t},r_{4,t})$ in \eqref{eqt:Full_DAG_reduce_space_example} using \eqref{eqt:risk_factor_return_pdf_factorization} :
	\begin{equation}
		\begin{aligned}
			f^{[t]}(r_{5,t}|r_{1,t},r_{3,t},r_{4,t})=&c_{5,1}^{[t]}(F^{[t]}(r_{5,t}),F^{[t]}(r_{1,t}))
			c_{5,3|1}^{[t]}(F^{[t]}(r_{5,t}|r_{1,t}),F^{[t]}(r_{3,t}|r_{1,t}))\cdot \\
			&c_{5,4|3,1}^{[t]}(F^{[t]}(r_{5,t}|r_{3,t},r_{1,t}),F^{[t]}(r_{4,t}|r_{3,t},r_{1,t})).
		\end{aligned}
		\label{eqt:DAG_factorize_example_reduced_space}
	\end{equation}
	On the other hand, the conditional density for $r_{4,t}$ in \eqref{eqt:Full_DAG_reduce_space_example} is factorized as 
	\begin{equation}
		f^{[t]}(r_{4t}|r_{2t},r_{3t})=c_{4,2}^{[t]}(F^{[t]}(r_{4t}),F^{[t]}(r_{2t}))c_{4,3|2}^{[t]}(F^{[t]}(r_{4t}|r_{2t}),F^{[t]}(r_{3t}|r_{2t})),
		\label{eqt:r4_factorization}
	\end{equation}
	which depends only on two copulas $c_{4,2}(\cdot,\cdot)$ and $c_{4,3|2}(\cdot,\cdot)$. When we compute the conditional distribution function $F^{[t]}(r_{4t}|r_{3t},r_{1t})$ in the last term in \eqref{eqt:DAG_factorize_example_reduced_space}, we apply the $h$ function in \eqref{eqt:h_function}:
	\begin{equation}
		\begin{aligned}
			F^{[t]}(r_{4t}|r_{3t},r_{1t}) =
			F^{[t]}(r_{4t}|r_{3t}) 
			= h_{4,3}^{[t]}(F^{[t]}(r_{4t}),F^{[t]}(r_{3t})),
		\end{aligned}
		\label{eqt:reduced_space_example}
	\end{equation}
	where the first equality in \eqref{eqt:reduced_space_example} is due to the fact that $r_{4t}$ and $r_{1t}$ are independent in the Bayesian network in \autoref{fig:example_difficult_conditional_probability} ($r_{4t}$ and $r_{1t}$ are independent when $r_{5t}$ is not given). 
	The second equality in \eqref{eqt:reduced_space_example} requires the $h$ function $h_{4,3}^{[t]}(u,v)=\partial C_{4,3}^{[t]}(u,v)/\partial v$, where $u=F^{[t]}(r_{4t})$ and $v=F^{[t]}(r_{3t})$, but the copula $C_{4,3}^{[t]}(\cdot,\cdot)$ is not defined in \eqref{eqt:r4_factorization} under the DAG in \autoref{fig:example_difficult_conditional_probability}. Therefore, we cannot use the $h$ function to calculate $F^{[t]}(r_{4t}|r_{3t})$. Alternatively, we can compute this conditional distribution function using the formula
	\begin{equation}
		\begin{aligned}
			F^{[t]}(r_{4t}|r_{3t})&=\int_{-\infty}^{\infty}F^{[t]}(r_{4t}|r_{3t},r_{2t})f^{[t]}(r_{2t})dr_{2t}\\
			& \equiv  \int_{-\infty}^{\infty}h_{4,3|2}^{[t]}(F^{[t]}(r_{4t}|r_{2t}),F^{[t]}(r_{3t}|r_{2t}))f^{[t]}(r_{2t})dr_{2t},
		\end{aligned}
		\label{eqt:difficult_integration}
	\end{equation}
	where the $h$ function $h_{4,3|2}^{[t]}(\cdot,\cdot)$ is directly computable from the copula $C_{4,3|2}^{[t]}(\cdot,\cdot)$. However, the integration in \eqref{eqt:difficult_integration} may not be easily computable and  the use of numerical integration is required. To conclude, \eqref{eqt:reduced_space_example} is not easily computable because the condition in the conditional probability in \eqref{eqt:reduced_space_example} does not include the first parent of $r_{4t}$, i.e., $r_{2t}$. We need to marginalize $r_{2t}$ from the copula as in \eqref{eqt:difficult_integration} to compute the conditional distribution $F^{[t]}(r_{4t}|r_{3t})$. 
	
	To reduce computation burden, we consider the DAGs in the reduced space, such that each of the conditional probabilities in the equation 
	\begin{equation}
		\begin{aligned}
			&c_{i,i[k]|i[1],\ldots,i[k-1]}^{[t]}(F^{[t]}(r_{i,t}|r_{i[1],t},\ldots,r_{i[k-1],t}),F^{[t]}(r_{i[k],t}|r_{i[1],t},\ldots,r_{i[k-1],t}))
		\end{aligned}
		\label{eqt:target_copula_for_check}
	\end{equation}
	contains a ``cummulative" parent set.  Appendix \ref{section:subset_DAG} {contains an algorithm to subset the DAG space to reduce computation burden.}

	\subsection{Generating returns from the GC-GARCH model}
	\label{section:generate_return}

	In the simulation study, we simulate a hypothetical data set to evaluate the performance of the estimation procedures. To simulate $(m+p)$-dimensional time series from the joint density function in \eqref{eqt:full_factorization}, we use a method similar to the one in \textcite{vine_copula}. For each variable in the DAG, we generate a value from $U(0,1)$, the uniform distribution over the interval $(0,1)$, to represent a realization of the conditional CDF of $r_{i,t}$ given its parent set, i.e., $u_{i|\mathbf{r}_{\pa(i)},t}=F^{[t]}(r_{i,t}|\mathbf{r}_{\pa(i)})$, for $i=1,\ldots,m$. 
	For each stock variable, we similarly generate a value from $U(0,1)$ to represent a realization of the conditional CDF of $r_{j,t}$ given the risk factor returns $r_{1,t}\ldots,r_{i-1,t}$, notated by $u_{j|1,\ldots,i-1,t}=F^{[t]}(r_{j,t}|r_{1,t},\ldots,r_{i-1,t})$. We then obtain $F^{[t]}(r_{1,t}),\ldots,F^{[t]}(r_{m+p,t})$ by applying the inverse $h$-function in \eqref{eqt:h_inv_function} to reduce the conditioning set by 1 each time until it is empty. 
	
	For example, suppose that we want to simulate returns from a GC-GARCH model with $m=4$ risk factors with the DAG structure in \autoref{fig:DAG_example_1a} and $p$ stocks.  For $t=1,\ldots,T$, we first generate $u_{1,t},u_{2,t},u_{3|1,t},u_{4|123,t}$ independently from $\text{Uniform}[0,1]$ for the risk factor returns. We then determine the unconditional cumulative distribution functions of the risk factor returns $F^{[t]}(r_{1,t}),\ldots,F^{[t]}(r_{4,t})$:
	$$
	\begin{aligned}
		F^{[t]}(r_{1,t})&=u_{1,t},\\
		F^{[t]}(r_{2,t}) &= u_{2,t},\\
		F^{[t]}(r_{3,t}) &= h_{3,1}^{{[t]}^{-1}}(u_{3|1,t},u_{1,t}),\\
		F^{[t]}(r_{4,t}) & =  h_{4,1}^{{[t]}^{-1}}(h_{4,2|1}^{{[t]}^{-1}}(h_{4,3|1,2}^{{[t]}^{-1}}(u_{4|1,2,3,t},u_{3|1,t}),u_{2,t}),u_{1,t}).
	\end{aligned}
	$$
	For $j=m+1,\ldots,p$, we generate $u_{j|1,2,3,4}$ from $\text{Uniform}[0,1]$, which represents a realization of $F^{[t]}(r_{j,t}|r_{1,t},\ldots,r_{4,t})$, the conditional distribution of the stock returns $r_{j,t}$ given the risk factor returns. Then, we apply the inverse $h$-function in \eqref{eqt:h_inv_function} recursively to obtain the unconditional cumulative distribution functions of the stock return $r_{j,t}$:
	
	$$
	F^{[t]}(r_{j,t})=h_{j,1}^{{[t]}^{-1}} h_{j,2}^{{[t]}^{-1}}(h_{j,3|1,2}^{{[t]}^{-1}}(h_{j,4|1,2,3}^{{[t]}^{-1}}(u_{j|1,2,3,4},u_{4|1,2,3}),u_{3|1,t}),u_{2,t}),u_{1,t}).
	$$
	Finally, as we assume the returns follow GARCH(1,1) models with $t$ distributed innovations as described in \eqref{eqt:marginal_dynamics}, by applying the inverse of the $t$ distribution for each $F^{[t]}(r_{1,t}),\ldots,F^{[t]}(r_{m+p,t})$, we obtain a vector of simulated returns $r_{1t},\ldots,r_{m+p,t}$.

	\section{Estimation}
	\label{section:estimation}
	\subsection{Parameter estimation}
	\label{section:parameter_estimation}
	Recall that we divide the parameters in the GC-GARCH model into three groups in Section \ref{section:computation_of_likelihood_function}.
	In the literature, the estimation of the pair copulas is conducted individually for each copula \parencite{brechmann2013risk}, referred to as sequential estimation. The sequential estimates can be obtained quickly due to the reduction in dimension.  The sequential estimates are shown to be consistent and often perform well \parencite{HOBAEKHAFF201291,Haff2013ParameterEF}. 
	
	For $\Theta_{1i}$'s, the parameters in the marginal distributions are first estimated using a numerical solver in \texttt{R}. This is the method of inference functions for margins (IFM) introduced by \textcite{joe1997multivariate,joe2005asymptotic}. For $\Theta_2$, the parameters in the DAG copulas, we use the sequential estimates of the parameters as starting values for a full estimation procedure using Markov chain Monte Carlo (MCMC) sampling \parencite{Haff2013ParameterEF}. For $\Theta_{3j}$'s, the parameters in the stock copulas, however, we do not conduct MCMC sampling as the number of stocks in the portfolio, $p$, is large in the simulation study and empirical study, therefore conducting MCMC sampling for each of the stocks is not feasible in practice. 
	
	Similar to the IFM, we first estimate $\Theta_{2}$, the parameters in the DAG copulas. Then, we insert the estimates of $\Theta_{2}$ back into the likelihood function and estimate $\Theta_{3j}$'s, the parameters of the stock copulas. The rationale behind this is that the risk factors are used to explain the co-movements of the stock returns. Thus, the parameters in the DAG copulas should be estimated alone without including the stocks. For example, the parameters of the copulas in the DAG should be the same no matter whether we include $p=50$ or $p=100$ stocks in the GC-GARCH model. 
	
	We have already derived the likelihood functions in Section \ref{section:computation_of_likelihood_function}. We further derive the posterior distributions of the DAG copulas for MCMC sampling. The procedures of the parameter estimation in the GC-GARCH model are as follows. 
	
	\begin{enumerate}
		\item Each return series follows a GARCH(1,1) models with $t$ distributed innovations in \eqref{eqt:marginal_dynamics}. For $i=1,\ldots,m+p$, we obtain the maximum likelihood estimates $\hat\Theta_{1i}$ from the log likelihood function of the $i$-th return series in \eqref{eqt:marginal_likelihood} using a numerical optimizer in \texttt{R}, with positive variances and stationary covariance constraints in \eqref{eqt:marginal_dynamics}.
		
		\item Using the sequential estimates of $\Theta_2$ as starting values, we conduct MCMC sampling for the parameters $\Theta_2$, whose posterior distribution is
		\begin{equation}
			\begin{aligned}
				& \log  f(\Theta_2|\hat\Theta_1,D) \\
				=& \ell_2(\hat\Theta_1,\Theta_2|D) + \log \pi(\Theta_2)+\text{constant},
			\end{aligned}
			\label{eqt:posterior_DAG_copula}
		\end{equation}
		where $\ell_2(\hat\Theta_1,\Theta_2|D)$ is given by \eqref{eqt:ell_2} in regard to the DAG copulas, and $\pi(\Theta_2)$ is the prior distribution for $\Theta_2$. We choose a uniform prior for $\Theta_2$, which is proportional to 1 if the parameters satisfy the stationary conditions and the finite covariance condition, i.e., 
		\begin{equation}
			\begin{aligned}
				&\pi(\Theta_2)\\
				\propto& \prod_{i=2}^m \prod_{k=1}^{n(i)} \mathbf{1}(0\leq a_{i,i[k]|z_{ik}},b_{i,i[k]|z_{ik}} < 1, a_{i,i[k]|z_{ik}}+b_{i,i[k]|z_{ik}} < 1, -1< \bar\varphi_{i,i[k]|z_{ik}} < 1,v_{i,i[k]|z_{ik}}>2),
			\end{aligned}
			\label{eqt:DAG_parameter_prior}
		\end{equation}
		where $z_{ik}=\{i[1],\ldots,i[k-1]\}$ and $\mathbf{1}(\cdot)$ is an indicator function that takes 1 if the condition inside the indicator function is satisfied, and 0 otherwise. Note that we use the IFM method so that we insert $\hat\Theta_1$, the maximum likelihood estimates of the parameters in the marginal distributions into $\log  f(\Theta_2|\hat\Theta_1,D)$. We use $\overline \Theta_2$, the medians of the samples drawn from the posterior distribution as the Bayes estimates of $\Theta_2$.
		
		\item For each stock $j=m+1,\ldots,m+p$, we conduct sequential estimation for $\Theta_{3j}$ by optimizing the function $\ell_{3j}(\hat\Theta_1,\overline\Theta_2,\Theta_{3j})$ in \eqref{eqt:ell_3j}, with stationary condition constraints in \eqref{eqt:copula_parameter_dynamic}. Note that we substitute $\hat\Theta_1$ into \eqref{eqt:ell_3j} due to the method of IFM, and $\overline\Theta_2$ into \eqref{eqt:ell_3j} as we estimate the DAG parameters separately.
	\end{enumerate}

	\subsection{MCMC sampling for parameters in DAG copulas}
	\label{section:DAG_copula_MCMC_estimation}
	MCMC sampling is used for the estimation of $\Theta_2$. We first obtain the sequential estimates of the parameters as the initial values for the MCMC sampling, denoted by $\Theta^{(0)}_2$. We adopt the Robust adaptive Metropolis (RAM) algorithm \parencite{RAM} for the MCMC sampling. Starting from $n=1$, we do the following steps:
	
	\begin{enumerate}
		\item We set a proposal
		$$
		\Theta_2^* = \Theta_2^{(n-1)}+S_{n-1}U_n,
		$$
		where $U_n$ is a $d$-dimension vector drawn from the proposal distribution $q(\cdot)$, $d=\abs{\Theta_2}$ is the number of parameters in $\Theta_2$, and $S_{n-1}$ is a matrix that captures the correlation between the parameters. We use the multivariate normal distribution with mean equals a $d$-dimensional vector of zeros and covariance matrix equals the $d$-dimensional identity matrix as the proposal distribution. However, changing all parameters at once may lead to poor mixing in the MCMC sampling. We instead change a subset of parameters in $\Theta_2$ in each step. This can be done by setting the entries in $U_n$ corresponding to the variables that are not in the subset to 0.
		\item With a probability of $\alpha_n=\min\{1,f(\Theta_2^*|\hat\Theta_1,D)/f(\Theta_2^{(n-1)}|\hat\Theta_1,D)\}$, the proposal is accepted and we set $\Theta_2^{(n)}=\Theta_2^*$. Otherwise, we set $\Theta_2^{(n)}=\Theta_2^{(n-1)}$.
		\item Update the lower-diagonal matrix with positive diagonal elements $S_n$ satisfying the equation
		$$
		S_{n}S_{n}^T = S_{n-1}\left(I+\eta_n(\alpha_n-\alpha_*)\frac{U_nU_n^T}{\norm{U_n}^2}\right)S_{n-1}^T,
		$$
		where $I$ is the $d$ by $d$ identity matrix, $\alpha_*\in(0,1)$ is the target mean acceptance probability of the algorithm, and $\{\eta_n\}_{n\geq 1}$ is a step size sequence decaying to zero, where $\eta_n\in (0,1]$. It can be done using Cholesky decomposition.
		\item Set $n\leftarrow n+1$ and go back to step 1. 
	\end{enumerate}
	The advantage of the RAM algorithm is that the matrix $S_n$ captures the correlation between parameters and allows the algorithm to attain a given mean acceptance rate $\alpha_*$, which helps improve the convergence compared to the traditional MCMC sampling algorithms. $S_0$ is initialized with the $d$ by $d$ diagonal matrix. $\alpha_*$ is set to 23.4\%, which is a typical choice in literature \parencite{10.2307/2245134}. The step size sequence $\eta_n$ is often defined as $\eta_n=n^{-\gamma}$, where $\gamma\in(1/2,1]$ \parencite{RAM}. In the simulation study and empirical study, we choose $\gamma=2/3$. The sequence $\eta_n$ plays an important role to ensure the ergodicity of the MCMC algorithm.

	
	

	\subsection{Structural learning of the DAG}
	\label{section:structural_learning}
	
	In practice, the underlying DAG of the risk factor returns is often unknown, and we need to estimate the DAG from the data. Score-based learning using MCMC sampling is adopted, which searches for DAGs that are around the posterior modes. In practice, there could be multiple DAGs that have similar scores. Specifically, to construct a score function, consider the term $\ell_2(\hat\Theta_1,\Theta_2|D)$ in \eqref{eqt:ell_2} with regard to the DAG parameters, in which we insert $\hat\Theta_1$, the maximum likelihood estimates of the parameters in the marginal distribution due to the IFM method. Note that we do not include the likelihood function for stock copulas because we do not estimate the structure of the risk factors using the stock returns. To emphasize that the likelihood function is a function of a network $G$, we rewrite the likelihood function as $\exp(\ell_2(\hat\Theta_1,\Theta_2,G|D))\equiv P(D|\hat\Theta_1,\Theta_2,G)$. In DAG structural learning, we need to calculate the marginal likelihood that marginalizes out the parameters:
	\begin{equation}
		P(D|\hat\Theta_1,G)=\int P(D|\hat\Theta_1,\Theta_2,G)\pi(\Theta_2|G) d\Theta_2,
		\label{eqt:marginalization_of_DAG_likelihood}
	\end{equation}
	where $\pi(\Theta_2|G)$ is the prior of $\Theta_2$ given $G$. Note that the prior distribution of $\Theta_2$ in \eqref{eqt:DAG_parameter_prior} is dependent on the DAG $G$ as $G$ can be varying. Under this case, we rewrite the prior for $\Theta_2$ in \eqref{eqt:DAG_parameter_prior} as $\pi(\Theta_2|G)$ in \eqref{eqt:marginalization_of_DAG_likelihood}.
	
	The integration in the marginal likelihood in \eqref{eqt:marginalization_of_DAG_likelihood} is, however, difficult to calculate in practice since the integrand involves a product of $t$-copulas. An approximation is to use the Bayesian information criterion (BIC) derived by \textcite{BIC}:
	$$
	\log P(D|\hat\Theta_1,G)\approx \log P(D|\hat\Theta_1,\hat\Theta_2,G)-\frac{\abs{\Theta_2}}{2}\log T,
	$$
	where $\abs{\Theta_2}$ is the number of parameters in $\Theta_2$, $\hat\Theta_2$ is the maximum likelihood estimates of $\Theta_2$, and $T$ is the number of days of observations in the data set $D$. However, calculating the maximum likelihood of $\Theta_2$ of the function $P(D|\hat\Theta_1,\Theta_2,G)$ is in general time consuming, since the dimension of $\Theta_2$ is moderately large. It is not feasible since it often involves at least thousands of iterations in the MCMC sampling. Thus, we can further approximate the BIC using the sequential estimates, which perform well in the literature \parencite{HOBAEKHAFF201291,Haff2013ParameterEF}. Denote $\tilde \Theta_2$ to be the sequential estimates obtained by optimizing each copula in $P(D|\hat\Theta_1,\Theta_2,G)$ separately. The score function used in the DAG structural learning is
	\begin{equation}
		\log P(D|\hat\Theta_1,G)\approx \log P(D|\hat\Theta_1,\tilde\Theta_2,G)-\frac{\abs{\Theta_2}}{2}\log T.
		\label{eqt:approximate_BIC}
	\end{equation}

	Now we discuss the MCMC scheme for the DAG structural learning. We use the Structure MCMC proposed by \textcite{structure_mcmc}. It samples graphs by conducting local edge movements. Let $\nbd(G)$ be the set of neighborhood of the graph $G$, defined as the set containing the graphs that are equal to $G$ except with a one-edge difference, and in the reduced DAG space introduced in Section \ref{section:reduced_DAG_space}. A neighbor graph $G'$ is sampled randomly, and thus the transition kernel is given by
	$$
	q(G'|G)=\frac{1}{\abs{\nbd(G)}},
	$$
	where $\abs{\nbd(G)}$ is the number of graphs in $\nbd(G)$. Under the Bayesian framework, we also need to derive the posterior distribution $P(G|\hat\Theta_1,D)$, which is given by
	$$
	P(G|\hat\Theta_1,D)\propto P(D|\hat\Theta_1,G)\pi(G) \approx \exp(\log P(D|\hat\Theta_1,\tilde\Theta_2,G)-\frac{\abs{\Theta_2}}{2}\log T)\pi(G),
	$$
	where we approximate the likelihood using \eqref{eqt:approximate_BIC} and $\pi(G)$ is the prior distribution of the graph $G$. We set a flat prior $\pi(G)\propto 1$ if $G$ is a DAG and inside the reduced DAG space since we do not have any prior information with regard to the structure of the risk factor returns. We then apply the Metropolis-Hasting algorithm to sample graphs from the posterior distribution $P(G|\hat\Theta_1,D)$. We sample a graph $G'$ from $\nbd(G)$. The graph $G'$ is accepted with the probability
	$$
	\alpha(G,G')=\min\left\{1,\frac{P(G'|\hat\Theta_1,D)q(G|G')}{P(G|\hat\Theta_1,D)q(G'|G)}\right\}.
	$$

	\section{Portfolio selection}
	\label{Section:portfolio_management}
	\subsection{Minimum variance (MV) portfolio and minimum conditional value-at-risk (MCVaR) portfolio}
	\label{section:MVP_MCVaRP}
	
	The GC-GARCH model can be used for portfolio selection. Suppose that we take $G^*$, the maximum a posteriori (MAP) network in the structural learning as the underlying network in the GC-GARCH model. Consider a portfolio of $p$ assets and $m$ risk factors.
	Suppose that we invest in a portfolio assigning weight $w_{j,t}$ to stock $j$ at the end of day $t$, where $j=m+1,\ldots,m+p$ and $\sum_{j=m+1}^{m+p}w_{j,t}=1$. The portfolio return on day $t+1$ (tomorrow) is given by
	$$
	r_{PF,t+1}=\sum_{j=m+1}^{m+p} w_{j,t} r_{j,t+1} \equiv \mathbf{w}_t^T \mathbf{r}_{t+1},
	$$
	where $\mathbf{w}_t=(w_{m+1,t},\ldots,w_{m+p,t})^T$ is a vector of portfolio weights and $\mathbf{r}_{t+1}=(r_{m+1,t+1},\ldots,r_{m+p,t+1})^T$ is a vector of stock returns on date $t+1$. Note that the returns are indexed by $t+1$ but the portfolio weights are indexed by $t$, because the portfolio weight is determined before we observe $\mathbf{r}_{t+1}$. We assume that the portfolio weights are determined 1 day before. The variance of the portfolio return on date $t+1$ is
	$$
	\sigma_{PF,t}^2=\sum_{i=m+1}^{m+p}\sum_{j=m+1}^{m+p}w_{i,t}w_{j,t}\sigma_{ij,t+1}\equiv \mathbf{w}_t^T \Sigma_{t+1} \mathbf{w}_t,
	$$
	where $\sigma_{ij,t+1}$ is the conditional covariance between stocks $i$ and $j$ on date $t+1$ given $\mathcal{F}_t$, and $\Sigma_{t+1}=[\sigma_{ij,t+1}]_{i,j=m+1}^{m+p}$ is the covariance matrix of the stock returns. We estimate $\Sigma_{t+1}$ by simulating returns from the fitted GC-GARCH model.
	
	Using the minimum variance optimization proposed by \textcite{markowitz}, we determine the optimal portfolio weights $\mathbf{w_t}$ that minimize the 1-day ahead predicted variance of the portfolio return given $\mathcal{F}_t$. Mathematically, we determine $\mathbf{w}_t$ for the problem
	
	\begin{equation}
		\min_{\mathbf{w_t}} \mathbf{w}_t^T\hat\Sigma_{t+1}\mathbf{w}_t~~\text{ 
			such that  }\sum_{j=m+1}^{m+p} w_{j,t} = 1,
		\label{eqt:markovitz}
	\end{equation}
	where $\hat\Sigma_{t+1}$ is the 1-day ahead prediction of the conditional covariance of $\mathbf{r}_{t+1}$ matrix given $\mathcal{F}_t$ from the GC-GARCH model.
	
	To estimate $\hat\Sigma_{t+1}$ for the problem \eqref{eqt:markovitz}, we simulate $K=20,000$ vectors of stock returns on day $t+1$ from the GC-GARCH model, $\mathbf{r}_{1,t+1},\ldots,\mathbf{r}_{K,t+1}$, using the methodology in Section \ref{section:generate_return}. We first estimated the 1-day ahead prediction of the conditional correlation matrix of $\mathbf{r}_{t+1}$:
	$$
	\hat \Upsilon_{t+1} = \diag(\tilde\Sigma_{t+1})^{-1/2} \tilde\Sigma_{t+1} \diag(\tilde\Sigma_{t+1})^{-1/2},
	$$
	where
	\begin{equation}
		\tilde \Sigma_{t+1} = \frac{1}{K}{\sum_{k=1}^{K} \mathbf{r}_{k,t+1} \mathbf{r}_{k,t+1}^T},
		\label{eqt:Sigma_not_accurate}
	\end{equation}
	and $\diag(\tilde\Sigma_{t+1})$ is the diagonal matrix containing the diagonal elements in $\tilde\Sigma_{t+1}$. Then, we estimate the 1-day ahead conditional covariance matrix of $\mathbf{r}_{t+1}$ by
	\begin{equation}
		\hat \Sigma_{t+1} = \Lambda_t^{1/2} \hat \Upsilon_{t+1} \Lambda_t^{1/2},
		\label{eqt:Sigma_more_accurate}
	\end{equation}
	where {$\Lambda_t$} is the diagonal matrix, whose diagonals are the exact 1-day ahead predictions of the variances obtained from the marginal GARCH(1,1) models. Note that both \eqref{eqt:Sigma_not_accurate} and \eqref{eqt:Sigma_more_accurate} are estimators of the 1-day ahead covariance matrix, whereas \eqref{eqt:Sigma_more_accurate} matches the variances to the exact 1-day ahead predictions of the variances. We prefer to use \eqref{eqt:Sigma_more_accurate}.

	As discussed in \textcite{brechmann2013risk}, since only the second moment is taken into account in the minimum variance portfolio problem, the information regarding to the tails is not considered. To also include the information in the tails, we  solve the minimum conditional value-at-risk (MCVaR) portfolio problem:
	
	\begin{equation}
		\min_{\mathbf{w_t}} CVaR_{t+1}^\alpha(\mathbf{w}_t^T\mathbf{r}_{t+1})~~\text{ 
			such that  }\sum_{j=m+1}^{m+p} w_{j,t} = 1,
		\label{eqt:min_CVaR}
	\end{equation}
	where
	$$
	CVaR_{t+1}^\alpha(\mathbf{w}_t^T\mathbf{r}_{t+1})=-E(\mathbf{w}_t^T\mathbf{r}_{t+1}|\mathbf{w}_t^T\mathbf{r}_{t+1}<q_{t+1}^\alpha(\mathbf{w}_t^T\mathbf{r}_{t+1}),\mathcal{F}_t)
	$$
	and $q_{t+1}^\alpha(\mathbf{w}_t^T\mathbf{r}_{t+1})$ is the $\alpha$th quantile of the 1-day ahead portfolio return $\mathbf{w}_t^T\mathbf{r}_{t+1}$. \textcite{rockafellar2000optimization} and \textcite{VaR_optim_comp} discuss an alternative method to approximate the MCVaR portfolio problem by generating returns from the model. We approximate the solution in \eqref{eqt:min_CVaR} by sampling $K=20,000$ vectors of 1-day ahead stock returns from the GC-GARCH model, and solve
	\begin{equation}
		\min_{\mathbf{w_t},\mathbf{u},a} \left( -a + \frac{1}{K\alpha}\sum_{k=1}^K u_k \right)
		\label{eqt:min_CVaR_alternative}
	\end{equation}
	subject to the linear constraints
	$$
	\begin{aligned}
		&\sum_{j=m+1}^{m+p} w_{j,t} = 1,\\
		&u_k\geq 0, \text{ for }k=1,\ldots,K, \text{and}\\
		&\mathbf{w}_t^T \mathbf{r}_{k,t+1}-a+u_k\geq 0, \text{ for }k=1,\ldots,K,
	\end{aligned}
	$$
	where $\mathbf{u}=(u_1,\ldots,u_K)$ are auxiliary variables, $a$ is the VaR level, and $\mathbf{r}_{k,t+1}$ is the $k$th sampled vector of 1-day ahead stock returns, $k=1,\ldots,K$.
	
	If we estimate the portfolio weights by solving the MCVaR portfolio problem \eqref{eqt:min_CVaR_alternative}, the minimized objective function $\left( -a + {(K\alpha)^{-1}}\sum_{k=1}^K u_k \right)$ is exactly the 1-day ahead prediction of CVaR. If we want to predict the 1-day ahead CVaR from the portfolio weights obtained in the MV portfolio problem in \eqref{eqt:markovitz}, this can be done by generating $K=20,000$ portfolio returns and estimate the 1-day ahead CVaR by
	\begin{equation}
		CVaR_{t+1}^\alpha = -\frac{1}{g_t} \sum_{k=1}^{K} \mathbf{w}_t^T \mathbf{r}_{k,t+1} \mathbf{1}(\mathbf{w}_t^T \mathbf{r}_{k,t+1}<\hat q_{t+1}^\alpha(\mathbf{w}_t^T\mathbf{r}_{t+1})),
		\label{eqt:CVaR_MVP}
	\end{equation}
	where $\hat q_{t+1}^\alpha(\mathbf{w}_t^T\mathbf{r}_{t+1})$ is the $\alpha$th sample quantile of the generated $K=20,000$ portfolio returns and $g_t=\sum_{k=1}^{K} \mathbf{1}(\mathbf{w}_t^T \mathbf{r}_{k,t+1}< \hat q_{t+1}^\alpha(\mathbf{w}_t^T\mathbf{r}_{t+1}))$ is the number of generated portfolio returns that are smaller than $\hat q_{t+1}^\alpha(\mathbf{w}_t^T\mathbf{r}_{t+1})$.

	\subsection{MV and MCVaR portfolios with model averaging}
	\label{section:model_averaging}
	In Section \ref{section:MVP_MCVaRP}, we solve the MV portfolio problem in \eqref{eqt:markovitz} and the MCVaR portfolio problem in \eqref{eqt:min_CVaR_alternative} using the MAP graph $G^*$. However, in the structural learning, there could be multiple graphs that have similar network scores as $G^*$. Then, we can conduct model averaging over the highest scoring $N_g$ networks for the portfolio selection instead of using only the MAP network. Model averaging methods are shown to outperform the model selection method in the literature \parencite{10.5555/3023549.3023619,dash2004model}. We denote $G^{(j)}$ to be the $j$th highest scoring network (then, $G^{(1)}=G^*$). To estimate the 1-day ahead conditional covariance matrix by modeling averaging, we simulate $K=20,000$ vectors of stock returns from each estimated GC-GARCH model with network $G^{(j)}$ (we denote such estimated model as $\mathcal{M}^{(j)}$), $j=1,\ldots,N_g$. We denote $r_{k,t+1}^{(j)}$ to be the $k$th vector of stock returns generated from $\mathcal{M}^{(j)}$. The 1-day ahead conditional covariance matrix under $\mathcal{M}^{(j)}$, $\Sigma_{t+1}^{(j)}$, is estimated similarly with the formula \eqref{eqt:Sigma_more_accurate} using the vectors $r_{k,t+1}^{(j)}$, $k=1,\ldots,K$. The model-averaged 1-day ahead conditional covariance matrix is given by
	$$
	\begin{aligned}
		\overline\Sigma_{t+1}&=\sum_{j=1}^{N_g} \Sigma_{t+1}^{(j)}P(\mathcal{M}^{(j)})
	\end{aligned}
	$$
	where
	$$
	\begin{aligned}
		P(\mathcal{M}^{(j)}) & = \frac{P(G^{(j)}|D,\hat\Theta_{1})}{\sum_{j=1}^{N_g}P(G^{(j)}|D,\hat\Theta_{1})}\\
		\approx & \frac{\exp(BIC^{(j)})}{\sum_{j=1}^{N_g}\exp(BIC^{(j)})}.
	\end{aligned}
	$$
	Note that $\hat\Theta_{1}$ is the set of estimated parameters in the marginal distributions that we estimate separately due to the IFM method. As in the structural learning, the marginalized posterior probabilities are difficult to calculate, and we use the BICs in \eqref{eqt:approximate_BIC} to approximate them. $BIC^{(j)}$ is the score of $G^{(j)}$ in the structural learning in \eqref{eqt:approximate_BIC}. The probability $P(\mathcal{M}^{(j)})$ is the marginalized posterior probability of $G^{(j)}$ normalized by the sum of the marginalized posterior probabilities of all candidates $G^{(1)},\ldots,G^{(N_g)}$ (to ensure that $\sum_{j=1}^{N_g}P(\mathcal{M}^{(j)})=1$). We also assume a uniform prior for $G^{(j)}$ so that
	$$
	P(G^{(j)}|D,\hat\Theta_{1})\propto P(D|\hat\Theta_{1},G^{(j)})\approx \exp(BIC^{(j)}).
	$$
	Then, we solve the MV problem in \eqref{eqt:markovitz} using the covariance matrix $\overline\Sigma_{t+1}$.
	
	To solve the MCVaR problem, we draw $K=20,000$ vectors of returns from the GC-GARCH models $\mathcal{M}^{(1)},\ldots,\mathcal{M}^{(N_g)}$, where the probability of sampling from the model $\mathcal{M}^{(j)}$ is $P(\mathcal{M}^{(j)})$. Then, we use the sampled $K$ vectors to solve the MCVaR portfolio problem in \eqref{eqt:min_CVaR_alternative}. The estimation methods of CVaR of the portfolio returns are discussed in Section \ref{section:MVP_MCVaRP}. The sampled $K$ vectors can be viewed as the realizations from the mixture distribution of $\mathcal{M}^{(1)},\ldots,\mathcal{M}^{(N_g)}$ with weights $P(\mathcal{M}^{(1)}),\ldots,P(\mathcal{M}^{(N_g)})$.

	\section{Simulation study}
	\label{section:simulation_study}
	\subsection{Simulation settings}
	In this simulation study, we aim to test our estimation method in Section \ref{section:estimation}. We conduct the following:
	
	\begin{enumerate}[1.]
		\item When the underlying DAG is known, we show that the parameters in the copulas can be accurately estimated using the method in Section \ref{section:parameter_estimation}. 
		\item We show that the structural learning algorithm in Section \ref{section:structural_learning} can search for structures that are close to the underlying DAG.
	\end{enumerate}
	
	We conduct the following two sets of simulation studies:
	\begin{enumerate}[(S1)]
		\item $m=8$ risk factors, $p=100$ stocks and $T=1000$ days of observations. We pick a relative simple network.
		\item $m=10$ risk factors, $p=100$ stocks and $T=1000$ days of observations. We pick a more complicated graph in this case.
	\end{enumerate}
	
	We repeat each experiment for 200 replications using the same DAG and the same set of parameters. We first simulated the common set of parameters:
	\begin{enumerate}[(i)]
		\item The parameters in the marginal distributions in \eqref{eqt:marginal_dynamics} are generated from the following distributions: $w_i\sim U(0.01,0.2)$, $b_i\sim U(0.8,0.96)$, $a_i\sim (1-b_i)\times U(0.3,0.7)$, and $ v_i\sim U(5,10)$. $U(k_1,k_2)$ denotes a uniform distribution over the interval $(k_1,k_2)$ for $k_1<k_2$. The method ensures that $\omega_i>0$, $a_i,b_i\geq 0$ and $a_i+b_i<1$ for all $i=1,\ldots,m+p$.
		\item The parameters in the DAG copulas and the stock copulas in \eqref{eqt:copula_parameter_dynamic} are generated from the following distributions: $\bar\varphi_{xy|z}\sim U(-1,1)$, $ b_{xy|z}\sim U(0.8,0.96)$, $a_{xy|z}\sim (1-b_{xy|z})\times U(0.3,0.7)$, and $v_{xy|z}\sim U(5,10)$. The method ensures that $0\leq a_{xy|z},b_{xy|z} < 1$, $a_{xy|z}+b_{xy|z} < 1$, and $-1 < \bar\varphi_{xy|z} < 1$.
	\end{enumerate}
	In each replication, the data set is generated using the method in Section \ref{section:generate_return}. We use a simpler setting under (S1) to illustrate the ideas in the proposed GC-GARCH model, whereas (S2) is a more complicated case that is closer to the cases in the empirical study. We present the simulation results for (S1) and (S2) below. We first present the performances of the structural learning. Then, we present the performances of the estimation of the parameters in the DAG copulas and stock copulas.
	
	\subsection{Structural learning evaluation}
	MCMC sampling is used to sample graphs from the posterior modes using the methods in Section \ref{section:structural_learning}. For each replication, we conduct the MCMC sampling for $200$ iterations. Let $G_1,\ldots,G_{200}$ be the sampled graphs. We discard the first $B$ sampled graphs in the burn-in step, where $B$ is selected using the Geweke's statistic (to be introduced later in this section). Using the graphs after burn-in, $G_{B+1},\ldots,G_{200}$, we estimate the probability of the existence of an edge from $r_{i,t}$ to $r_{j,t}$ by
	\begin{equation}
		P(a_{ij}=1|D)=\frac{1}{200-B}\sum_{k=B+1}^{200} \mathbf{1}(a_{ij}(G_k)=1),
		\label{eqt:edge_feature}
	\end{equation}
	where $a_{ij}$ denotes the $(i,j)$th entry of the adjacency matrix of the true graph (which is assumed to be random in Bayesian analysis), and $\mathbf{1}(a_{ij}(G_k)=1)$ equals 1 if the edge from $r_{i,t}$ to $r_{j,t}$ exists in the graph $G_k$, and equals 0 otherwise. \eqref{eqt:edge_feature} counts the proportion of graphs that contain the edge from $r_{i,t}$ to $r_{j,t}$, and is called an edge feature in the literature \parencite{order_mcmc}.

	The edges features are used to predict the existences of edges. We predict the edge from $r_{it}$ to $r_{jt}$ to exist if $P(a_{ij}=1|D)>c$ for some threshold $c\in [0,1]$. By altering the values of $c$, we trace a receiver operating characteristic (ROC) curve. We use the completed partially directed acyclic graph (CPDAG) \parencite{CPDAG} of the underlying network to evaluate the accuracy of the prediction, as the CPDAG represents an equivalence class of a set of graphs, where the graphs in the equivalence class have the same set of conditional independence. The CPDAG contains two types of edges. The first type is a directed edge, indicating that all graphs in the class agree with the direction. The second type is a bi-directed edge, indicating that some of the graphs disagree with the direction of an edge. The area under the ROC curve (AUROC) is used to measure the performance of the graph prediction.
	
	On top of the AUROC, we also use the following metrics. Let $TP(c)$, $TN(c)$, $FP(c)$ and $FN(c)$ be respectively the numbers of true positive, true negative, false positive and false negative predictions when we use the threshold $c$. We have the following five metrics
	
	\begin{enumerate}[(1)]
		\item The accuracy
		$$
		ACC(c) = \frac{TP(c)+TN(c)}{TP(c)+FN(c)+FP(c)+TN(c)},
		$$
		\item the false discovery rate
		$$
		FDR(c) = \frac{FP(c)}{TP(c)+FP(c)},
		$$
		\item the false omission rate
		$$
		FOR(c) = \frac{FN(c)}{FN(c)+TN(c)},
		$$
		\item the sensitivity
		$$
		SEN(c) = \frac{TP(c)}{TP(c)+FN(c)},\text{ and}
		$$
		\item the specificity
		$$
		SPE(c) = \frac{TN(c)}{TN(c)+FP(c)}.
		$$
	\end{enumerate}
	The accuracy (ACC) measures the proportion of correct prediction, which gives an overall performance of the prediction. The false discovery rate (FDR) measures the proportion of incorrectly predicted edges out of all predicted edges. The false omission rate (FOR) measures the proportion of incorrectly omitted edges out of all omitted edges. The sensitivity (SEN) measures the accuracy of detecting edges that exist in the underlying network, and the specificity (SPE) measures the accuracy of omitting edges that do not exist in the underlying network. Yet the SEN and SPE have been used to calculate the AUROC by changing the threshold $c$, and we also include the numerical results for reference.

	To evaluate the convergence of $\{G_k\}_{k=B+1}^{200}$, we calculate the Geweke's statistic \parencite{geweke} of a characteristic of the graph. We need to transform the network into a univariate time series $\{d(G_k)\}_{k=B+1}^{200}$, for some function $d:\{0,1\}^{m\times m}\mapsto \mathbb{R}$, to calculate the Geweke's statistic. We pick
	$$
	d(G_k)=\sum_{i=1}^m\sum_{j=1}^m \mathbf{1}(a_{ij}(G_k)\ne a_{ij}),
	$$
	where $a_{ij}(G_k)$ is the $(i,j)$th entry of the adjacency matrix of $G_k$, and $a_{ij}$ is the $(i,j)$th entry of the adjacency matrix of the true network, which is known in the simulation study. $\mathbf{1}(a_{ij}(G_k)\ne a_{ij})$ equals 1 if the $(i,j)$th entries in the adjacency matrices of $G_k$ and the true network, respectively, disagree with each other, and 0 zero otherwise. $d(G)$ is to keep track if the differences between the sampled graphs and the true graph are stabilized. The Geweke's statistic tests if the average in the first 10\% of $\{d(G_k)\}_{k=B+1}^{200}$ is significantly different from the average in the last 50\% of $\{d(G_k)\}_{k=B+1}^{200}$. If the p-value is greater than 0.01, this indicates that the difference of the two averages is not significant, and thus we can conclude that the samples are converged.

\begin{figure}[H]
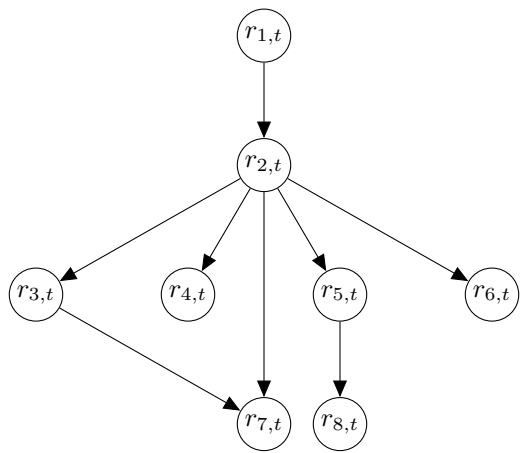

		\centering
		\tikz{
			\node[latent] (1) {$r_{1,t}$};%
			\node[latent,below=of 1] (2) {$r_{2,t}$}; %
			\node[latent,below=of 2,xshift=-3cm] (3) {$r_{3,t}$}; %
			\node[latent,below=of 2,xshift=-1cm] (4) {$r_{4,t}$}; %
			\node[latent,below=of 2,xshift=1cm] (5) {$r_{5,t}$}; %
			\node[latent,below=of 2,xshift=3cm] (6) {$r_{6,t}$}; %
			\node[latent,below=of 4,xshift=1cm] (7) {$r_{7,t}$}; %
			\node[latent,below=of 5] (8) {$r_{8,t}$}; %
			\edge {1} {2} 
			\edge {2} {3,4,5,6}
			\edge {2,3}{7}
			\edge {5} {8}
		}
		\caption{The underlying network in the simulation (S1).}
		\label{fig:true_graph_S1}
	\end{figure}

	The underlying network in simulation (S1) is shown in \autoref{fig:true_graph_S1}. We conduct 200 replications, where we generate different data sets from the same DAG and the same set of parameters. The burn-in $B$ in each replication is selected to minimize the Geweke's statistic. The average AUROC out of 200 replications is 75.4\%, showing that the structural learning performs well. The ACC, FDR, FOR, SEN and SPE in simulation (S1) using different thresholds ($c=0.01,0.10,0.50,0.90,0.99$) is shown in \autoref{tab:3_rates_m_8}. The ACC, SEN and SPE are moderatly high and the FDR and FOR are small, together with a good AUROC, indicating that many of the existing edges in the underlying graph are detected. The p-values of the Geweke's statistics in 187 out of 200 replications are greater than 0.01, supporting that the burn-in sample is converged.
	
	We do the same for the simulation (S2). The underlying network in simulation (S2) is shown in \autoref{fig:true_graph_S2}. It contains more nodes and edges than the graph in simulation (S1), and we expect that the structural learning in simulation (S2) will be more challenging. The average AUROC out of 200 replications is 70.1\%. The ACC, FDR, FOR, SEN and SPE in simulation (S1) is shown in \autoref{tab:3_rates_m_10}. We observe that the false discovery rate becomes higher than the case in simulation (S1). This indicates that the structural learning algorithms tend to include more edges, especially in higher dimensions. Nevertheless, the algorithm has a sensitivity up to 0.642. The p-values of the Geweke's statistics in 184 out of 200 replications are greater than 0.01. We still get a moderately good AUROC and convergence for a more challenging case.
	
	However, we do not know the underlying network of a real data set. These two simulation results give some clues of the performance of the structural learning algorithm. The algorithm is able to cover a moderately high portion of existing edges in the underlying network. Then, in the empirical study, we also conduct 200 iterations for the structural learning.

	\begin{table}[]
		\centering
		\begin{tabular}{llllll}
			Threshold $c$ & ACC & FDR & FOR & SEN & SPE \\ \hline
			0.01 & 0.810 & 0.347 & 0.111 & 0.723 & 0.845 \\
			0.10 & 0.818 & 0.315 & 0.125 & 0.676 & 0.874 \\
			0.50 & 0.819 & 0.238 & 0.163 & 0.537 & 0.932 \\
			0.90 & 0.781 & 0.204 & 0.218 & 0.323 & 0.965 \\
			0.99 & 0.766 & 0.202 & 0.234 & 0.254 & 0.971
		\end{tabular}
		\caption{The average ACC, FDR, FOR, SEN and SPE of the structural learning in simulation (S1) out of 200 replications at different thresholds.}
		\label{tab:3_rates_m_8}
	\end{table}

	\begin{figure}[H]
		\centering
		\tikz{
			\node[latent] (1) {$r_{1,t}$};%
			\node[latent,below= of 1,xshift=-2cm] (2) {$r_{2,t}$}; %
			\node[latent,below=of 2,xshift=-4 cm] (3) {$r_{3,t}$}; %
			\node[latent,below=of 2] (4) {$r_{4,t}$}; %
			\node[latent,below=of 4,xshift=1cm] (5) {$r_{5,t}$}; %
			\node[latent,below=of 5] (6) {$r_{6,t}$}; %
			\node[latent,below=of 5,xshift=2cm] (7) {$r_{7,t}$}; %
			\node[latent,below=of 3,xshift=-2cm] (8) {$r_{8,t}$}; 
			\node[latent,below=of 6,xshift=1cm] (9) {$r_{9,t}$};
			\node[latent,below=of 9,xshift=-3cm] (10) {$r_{10,t}$};
			%
			\edge {1} {2,3,4,5,7,9} 
			\edge {2} {3,4,7}
			\edge {3}{7,8,10}
			\edge {4} {5}
			\edge {5} {6}
			\edge {7} {9}
			\edge {9} {10}
		}
		\caption{The underlying network in the simulation (S2).}
		\label{fig:true_graph_S2}
	\end{figure}
	
	\begin{table}[]
		\centering
		\begin{tabular}{llllll}
			Threshold $c$ & Accuracy & FDR & FOR & SEN & SPE \\ \hline
			0.01 & 0.744 & 0.447 & 0.153 & 0.642 & 0.786 \\
			0.10 & 0.748 & 0.440 & 0.163 & 0.609 & 0.804 \\
			0.50 & 0.750 & 0.420 & 0.192 & 0.500 & 0.852 \\
			0.90 & 0.755 & 0.397 & 0.204 & 0.440 & 0.883 \\
			0.99 & 0.752 & 0.397 & 0.211 & 0.409 & 0.891
		\end{tabular}
		\caption{The average ACC, FDR, FOR, SEN and SPE of the structural learning in simulation (S2) out of 200 replications at different thresholds.}
		\label{tab:3_rates_m_10}
	\end{table}
	
	\subsection{Parameter estimation evaluation}
	The parameter estimations are separated into three phases: We first estimate $\Theta_1$, the parameters in the marginal distributions. We then estimate $\Theta_2$, the parameters in the DAG copulas using MCMC sampling. Finally, for each stock $j$, we estimate $\Theta_{3j}$, the parameters in the stock copulas $c_{j,i|1,\ldots,i-1}^{[t]}$ using sequential estimation starting from $i=1$ to $i=m$. We do not report the estimation results for the parameters in the marginal distributions as our interest is the copula parameters.
	
	\subsubsection{Parameters in the DAG copulas}
	
\begin{figure}[H]
		\begin{subfigure}{0.5\textwidth}
			\centering
			\includegraphics[width=7cm]{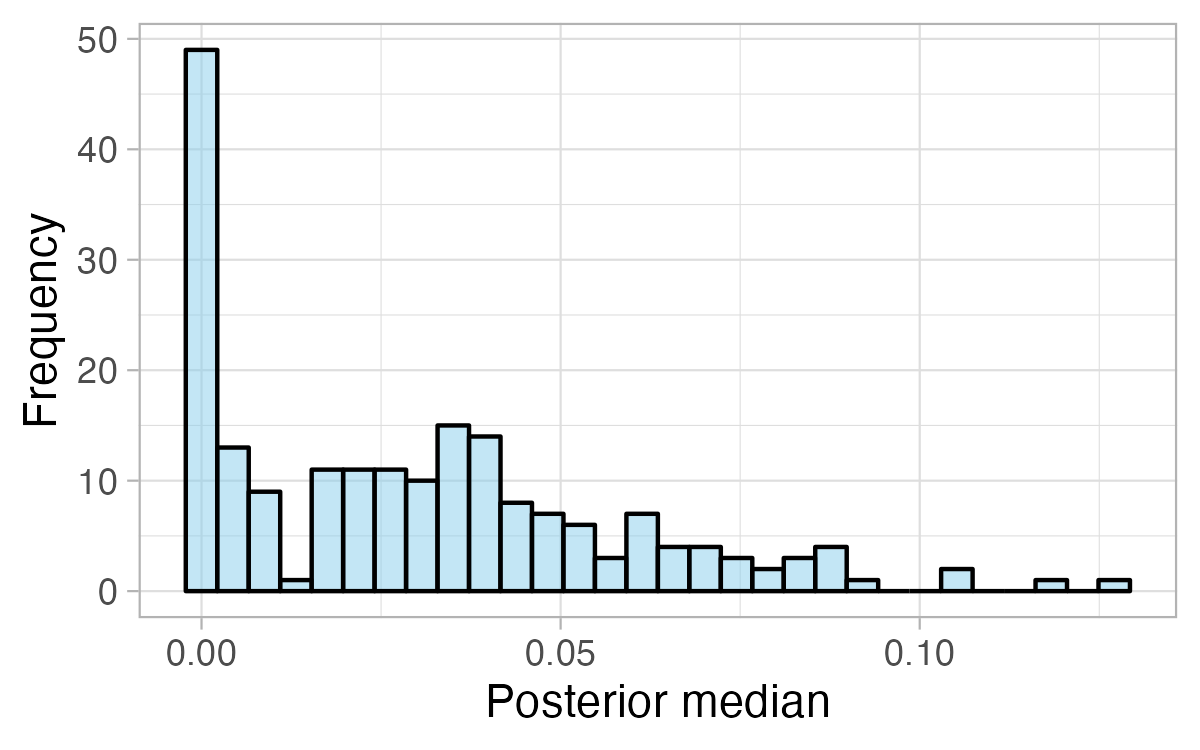}
			\caption{}
			\label{fig:a21_example}
		\end{subfigure}
		\begin{subfigure}{0.5\textwidth}
			\centering
			\includegraphics[width=7cm]{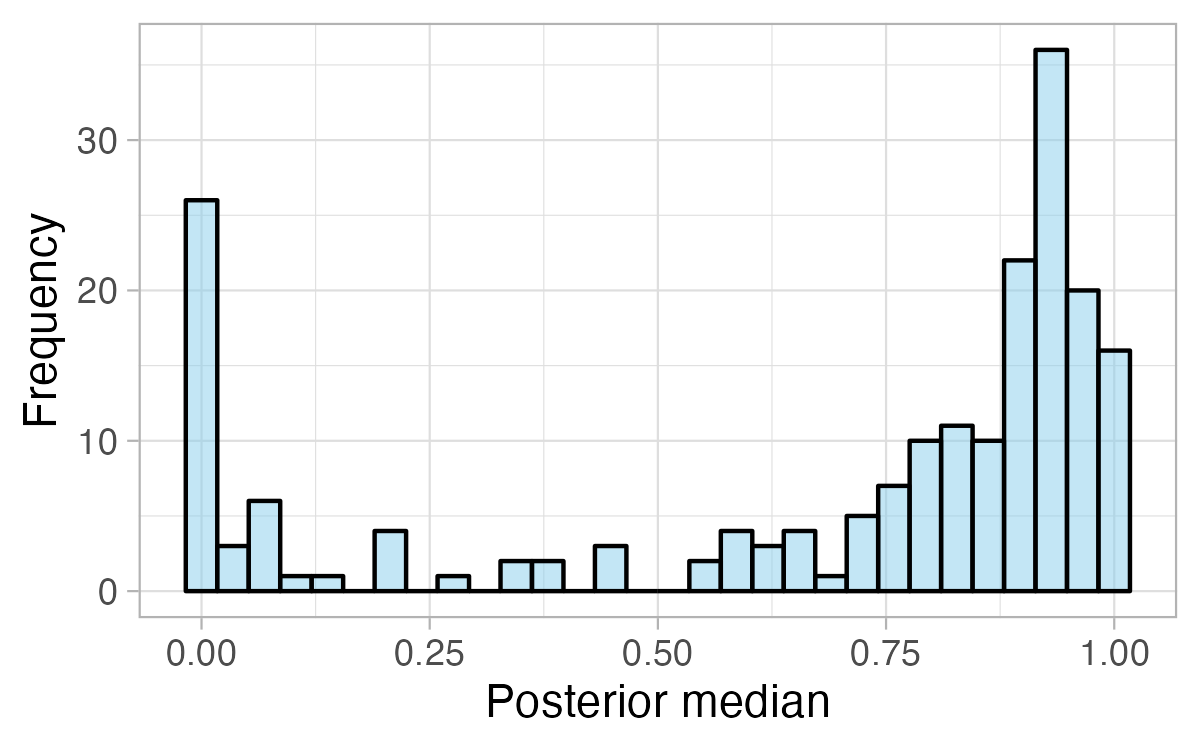}
			\caption{}
			\label{fig:b21_example}
		\end{subfigure}
		\mbox{}\\
		\begin{subfigure}{0.5\textwidth}
			\centering
			\includegraphics[width=7cm]{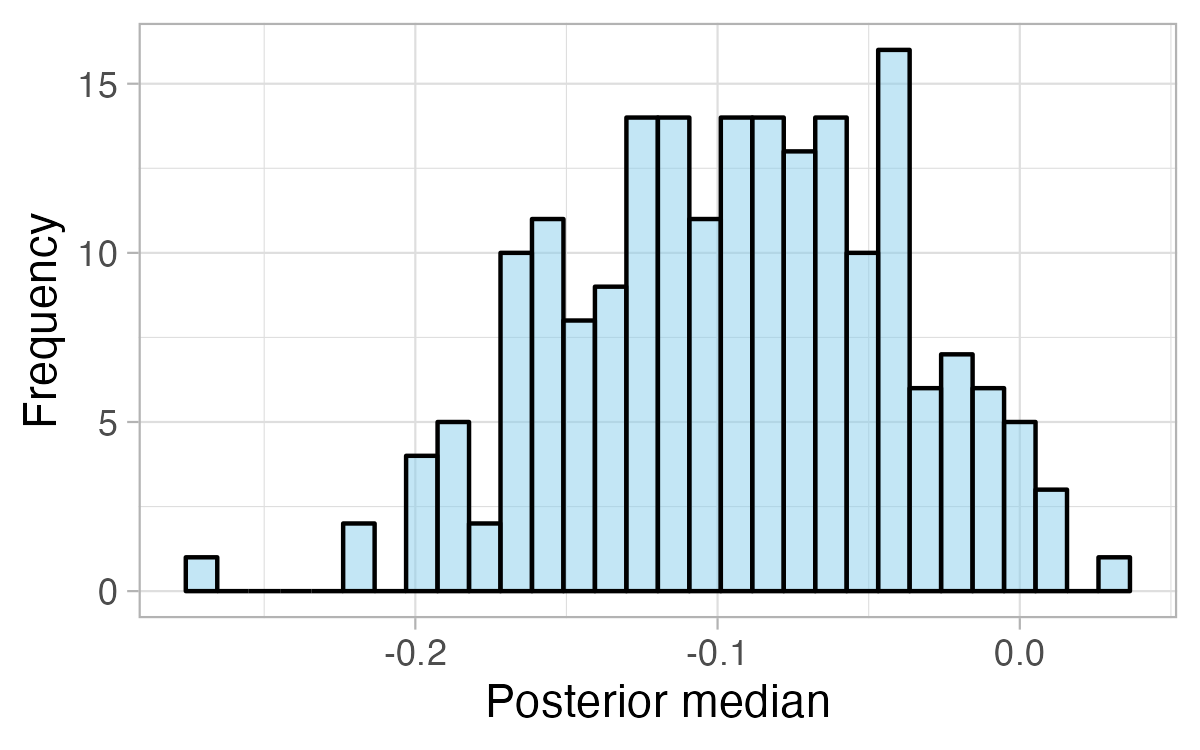}
			\caption{}
			\label{fig:phibar21_example}
		\end{subfigure}
		\begin{subfigure}{0.5\textwidth}
			\centering
			\includegraphics[width=7cm]{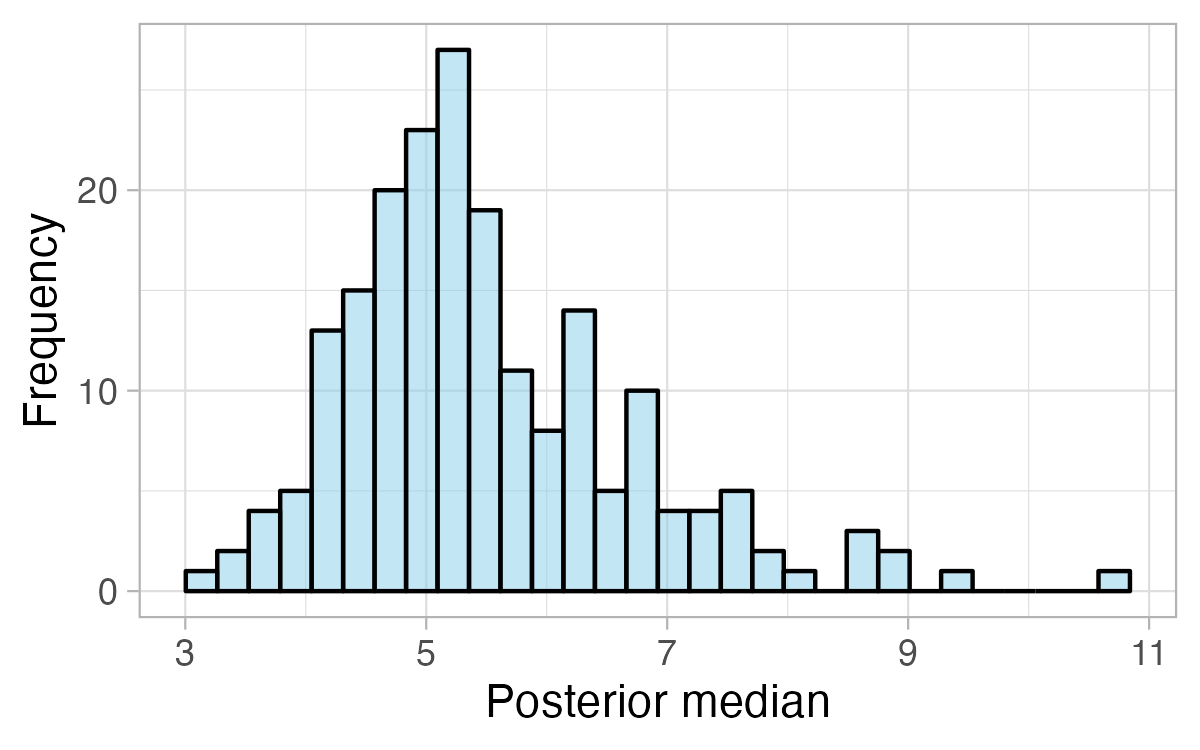}
			\caption{}
			\label{fig:v2121_example}
		\end{subfigure}
		\caption{The distributions of the posterior median estimates of (a) $a_{2,1}$, (b) $b_{2,1}$, (c) $\bar\varphi_{2,1}$ and (d) $v_{2,1}$, across 200 replications, for the copula $c_{2,1}^{[t]}$ in simulation (S1).}
		\label{fig:C21_example}
	\end{figure}

	We conduct 20,000 iterations for learning the parameters in the MCMC sampling for the DAG copulas in each replication. The burn-in is chosen automatically by minimizing the average Geweke's statistics of the chains of the parameters. We first present the estimation results for simulation (S1). We present the true values, the mean, the mean absolute error (MAE) of the mean, the median, the MAE of the median, the 5th quantile ($q_{0.05}$), and the 95th quantile ($q_{0.95}$) of the posterior median estimates out of the 200 replications in \autoref{tab:m_8_dag_median}. A true value marked with an asterisk indicates that $(q_{0.05},q_{0.95})$, the 90\% credible interval contains the true value. All the true values are contained in the 95\% credible intervals except for the parameter $\bar\varphi_{8,5}$. The MAEs of the mean and median of $\bar\varphi_{xy|z}$ are very small, indicating that the estimation is precise. The MAEs for $a_{xy|z}$ $b_{xy|z}$ are, however, larger relative to the magnitudes of the parameters. This can be explained as follows. \autoref{fig:C21_example} shows the distributions of the posterior median estimates of (a) $a_{2,1}$, (b) $b_{2,1}$, (c) $\bar\varphi_{2,1}$ and (d) $v_{2,1}$, across 200 replications, for the copula $c_{2,1}^{[t]}$ in simulation (S1) as an example. We observe that some of the estimates of $a_{xy|z}$ and $b_{xy|z}$ are close to zero.  This is because the data sets contain noises, and the algorithm may not be able to detect the dynamic dependencies of some copulas from the data sets in a few replications. Nevertheless, the median of the estimates of $a_{xy|z}$ and $b_{xy|z}$ are close to the true values as seen in \autoref{tab:m_8_dag_median}. The estimation results for simulation (S2) is in Appendix \ref{section:parameters_in_the_DAG_copulas} as the tables are too long to be included in the paper. The results are similar to those of simulation (S1), supporting that the MCMC estimation method in Section \ref{section:DAG_copula_MCMC_estimation} is reliable.


	\begin{table}[H]
		\centering
		\scriptsize	
		\begin{tabular}{llllllll}
			Parameter & True value & Mean of estimates & Mean MAE & Median of estimates & Median MAE & $q_{0.05}$ & $q_{0.95}$ \\ \hline
			$a_{2,1}$ & $\mathbf{0.04^*}$ & 0.0298 & 0.0277 & 0.0269 & 0.0247 & 0 & 0.0829 \\
			$a_{3,2}$ & $\mathbf{0.04^*}$ & 0.0331 & 0.025 & 0.0323 & 0.0198 & 0 & 0.0708 \\
			$a_{4,2}$ & $\mathbf{0.04^*}$ & 0.0395 & 0.0214 & 0.0379 & 0.0168 & 0.0015 & 0.0745 \\
			$a_{5,2}$ & $\mathbf{0.05^*}$ & 0.0466 & 0.0294 & 0.0458 & 0.0228 & 0.0004 & 0.0942 \\
			$a_{6,2}$ & $\mathbf{0.02^*}$ & 0.017 & 0.0191 & 0.0111 & 0.0159 & 0 & 0.0581 \\
			$a_{7,2}$ & $\mathbf{0.06^*}$ & 0.051 & 0.0361 & 0.0515 & 0.0301 & 0 & 0.1129 \\
			$a_{7,3|2}$ & $\mathbf{0.09^*}$ & 0.0898 & 0.0221 & 0.0896 & 0.0171 & 0.0577 & 0.1246 \\
			$a_{8,5}$ & $\mathbf{0.12^*}$ & 0.0977 & 0.0307 & 0.096 & 0.0315 & 0.0497 & 0.1478 \\ \hline
			$b_{2,1}$ & $\mathbf{0.89^*}$ & 0.6826 & 0.3549 & 0.8675 & 0.2568 & 0.0003 & 0.9947 \\
			$b_{3,2}$ & $\mathbf{0.91^*}$ & 0.7973 & 0.2811 & 0.9066 & 0.1516 & 0.0165 & 0.9895 \\
			$b_{4,2}$ & $\mathbf{0.88^*}$ & 0.7956 & 0.222 & 0.8664 & 0.1169 & 0.1002 & 0.9475 \\
			$b_{5,2}$ & $\mathbf{0.84^*}$ & 0.6854 & 0.3141 & 0.8219 & 0.2102 & 0.0044 & 0.967 \\
			$b_{6,2}$ & $\mathbf{0.96^*}$ & 0.6985 & 0.3853 & 0.9211 & 0.2747 & 0.0001 & 0.9962 \\
			$b_{7,2}$ & $\mathbf{0.81^*}$ & 0.6308 & 0.3239 & 0.7719 & 0.2463 & 0.0002 & 0.9878 \\
			$b_{7,3|2}$ & $\mathbf{0.86^*}$ & 0.8412 & 0.0756 & 0.8495 & 0.0423 & 0.7593 & 0.9124 \\
			$b_{8,5}$ & $\mathbf{0.8^*}$ & 0.7747 & 0.0851 & 0.7897 & 0.0661 & 0.6134 & 0.8878 \\ \hline
			$\bar\varphi_{2,1}$ & $\mathbf{-0.1^*}$ & $-$0.0945 & 0.0545 & -0.0936 & 0.045 & -0.1897 & -0.009 \\
			$\bar\varphi_{3,2}$ & $\mathbf{0.4^*}$ & 0.3774 & 0.0609 & 0.3758 & 0.0508 & 0.2867 & 0.481 \\
			$\bar\varphi_{4,2}$ & $\mathbf{0.76^*}$ & 0.7346 & 0.0331 & 0.7365 & 0.0331 & 0.6806 & 0.7868 \\
			$\bar\varphi_{5,2}$ & $\mathbf{0.59^*}$ & 0.5651 & 0.0462 & 0.562 & 0.0422 & 0.4874 & 0.6411 \\
			$\bar\varphi_{6,2}$ & $\mathbf{0.34^*}$ & 0.3073 & 0.0657 & 0.3045 & 0.0572 & 0.2089 & 0.4085 \\
			$\bar\varphi_{7,2}$ & $\mathbf{0.27^*}$ & 0.2496 & 0.0471 & 0.2497 & 0.041 & 0.171 & 0.3346 \\
			$\bar\varphi_{7,3|2}$ & $\mathbf{0.41^*}$ & 0.4011 & 0.0853 & 0.4042 & 0.0653 & 0.2541 & 0.5323 \\
			$\bar\varphi_{8,5}$ & $-0.49$ & $-$0.3225 & 0.0624 & -0.3229 & 0.1678 & -0.428 & -0.2258 \\ \hline
			$v_{2,1}$ & $\mathbf{5.29^*}$ & 5.5084 & 1.212 & 5.2921 & 0.8857 & 3.8757 & 7.7017 \\
			$v_{3,2}$ & $\mathbf{5.22^*}$ & 5.765 & 1.1881 & 5.615 & 0.9539 & 4.1212 & 8.0312 \\
			$v_{4,2}$ & $\mathbf{7.26^*}$ & 8.3123 & 2.359 & 7.6156 & 1.8309 & 5.568 & 12.7916 \\
			$v_{5,2}$ & $\mathbf{9.71^*}$ & 10.3178 & 3.1879 & 9.7887 & 2.4571 & 6.3559 & 16.1122 \\
			$v_{6,2}$ & $\mathbf{5.18^*}$ & 5.7413 & 1.5622 & 5.4904 & 1.1309 & 3.8794 & 8.7478 \\
			$v_{7,2}$ & $\mathbf{7.07^*}$ & 7.9327 & 2.4533 & 7.3082 & 1.6834 & 5.3139 & 12.8052 \\
			$v_{7,3|2}$ & $\mathbf{6.59^*}$ & 7.342 & 2.3234 & 6.8102 & 1.7867 & 4.5986 & 11.9898 \\
			$v_{8,5}$ & $\mathbf{7.27^*}$ & 10.6446 & 3.1184 & 10.0816 & 3.613 & 6.1601 & 16.0641
		\end{tabular}
		\caption{The true values, the mean, mean absolute error (MAE) of the mean, the median, the MAE of the median, the 5th quantile ($q_{0.05}$), and the 95th quantile ($q_{0.95}$) of the posterior median estimates out of the 200 replications for simulation (S1).}
		\label{tab:m_8_dag_median}
	\end{table}

	\subsubsection{Parameters in the stock copulas}
	\label{section:parameters_in_the_stock_copulas}
\begin{figure}[H]
		\centering
		\includegraphics[width=13cm]{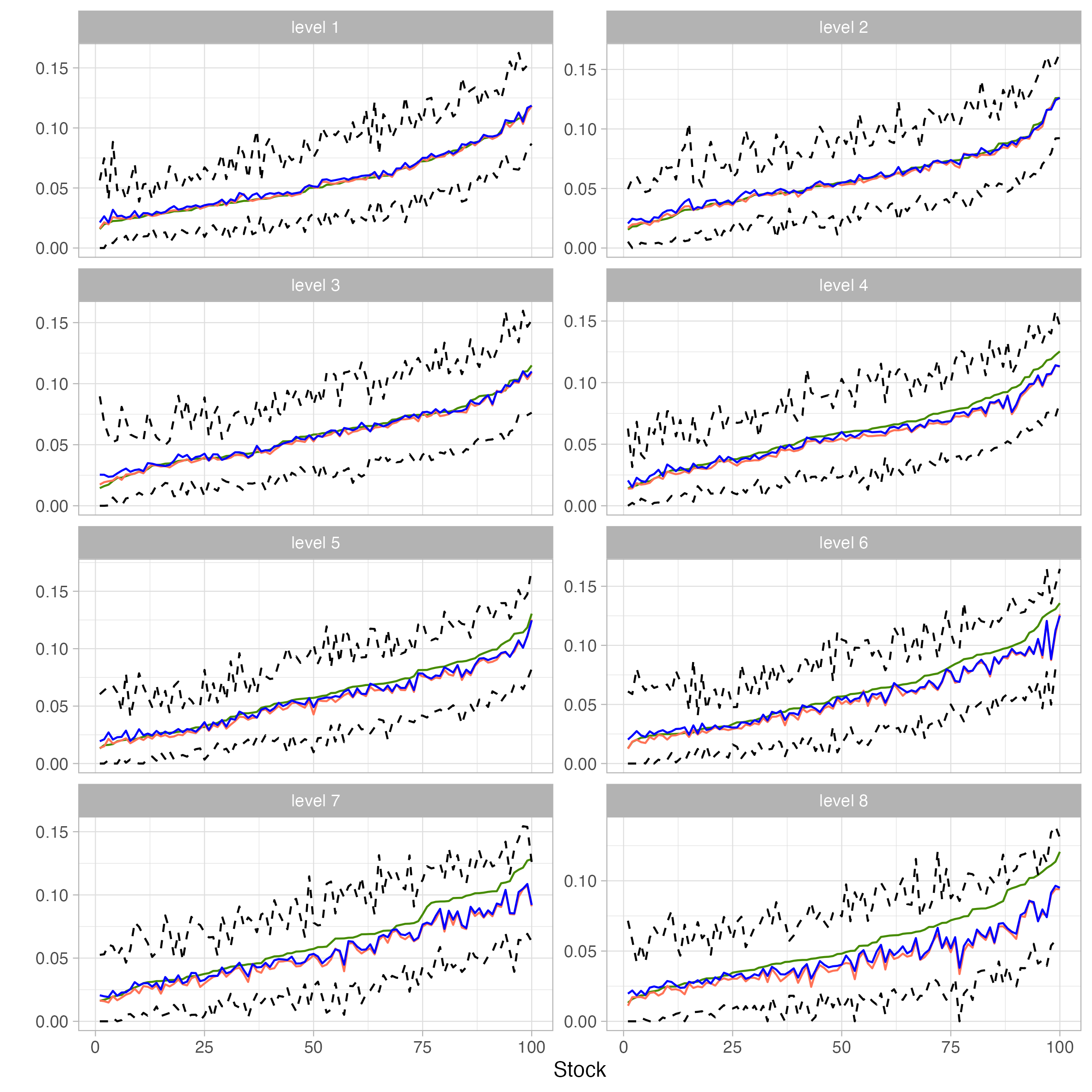}
		\caption{A graph summarizing the parameter estimates for $a_{xy|z}$ by level in simulation (S1). The copulas are in ascending order based on the true values of $a_{xy|z}$ in each level for a better visualization. The green curves indicate true values, the blue curves indicate mean estimates, the orange curves indicate the median estimates, and the pair of dashed curves represent the 90\% credible intervals of the sequential estimates out of the 200 replications.}
		\label{fig:axy_stock_m_8}
	\end{figure}

\begin{figure}[H]
		\centering
		\includegraphics[width=13cm]{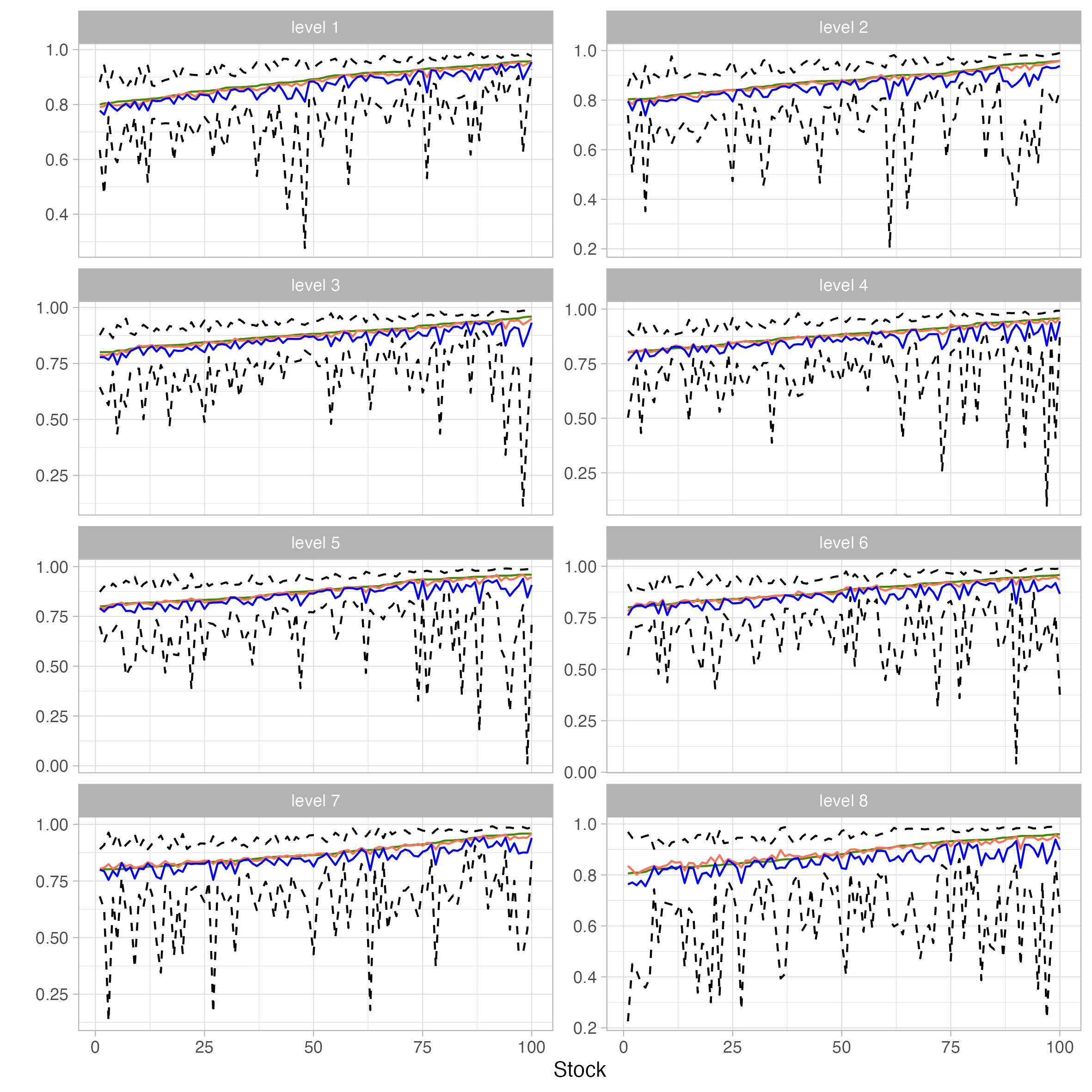}
		\caption{A graph summarizing the parameter estimates for $b_{xy|z}$ by level in simulation (S1). The copulas in ascending order based on the true values of $b_{xy|z}$ in each level for a better visualization. The green curves indicate true values, the blue curves indicate mean estimates, the orange curves indicate the median estimates, and the pair of dashed curves represent the 90\% credible intervals of the sequential estimates out of the 200 replications.}
		\label{fig:bxy_stock_m_8}
	\end{figure}

\begin{figure}[H]
		\centering
		\includegraphics[width=13cm]{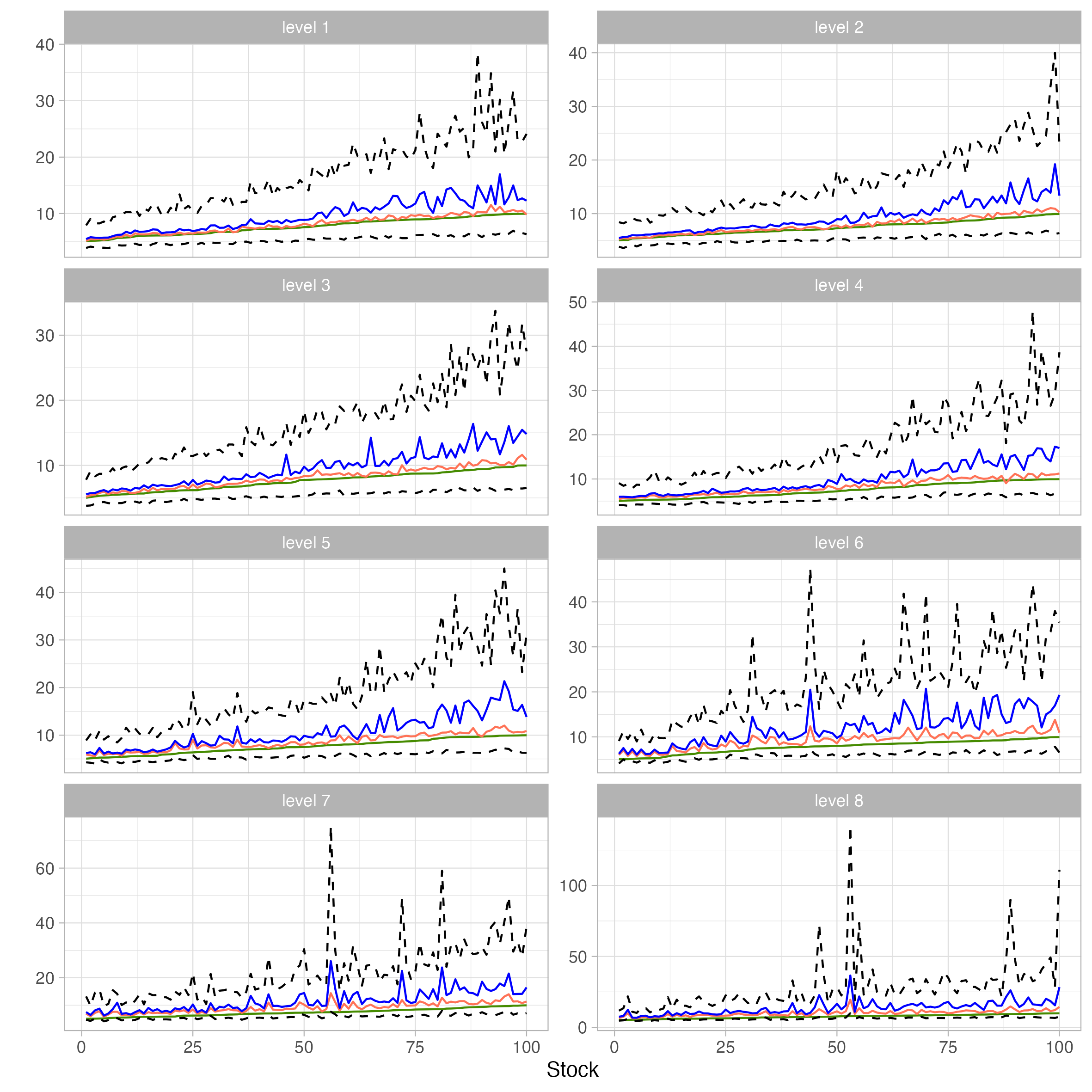}
		\caption{A graph summarizing the parameter estimates for $v_{xy|z}$ by level in simulation (S1). The copulas are in ascending order based on the true values of $v_{xy|z}$ in each level for a better visualization. The green curves indicate true values, the blue curves indicate mean estimates, the orange curves indicate the median estimates, and the pair of dashed curves represent the 90\% credible intervals of the sequential estimates out of the 200 replications.}
		\label{fig:vxy_stock_m_8}
	\end{figure}

\begin{figure}[H]
		\centering
		\includegraphics[width=13cm]{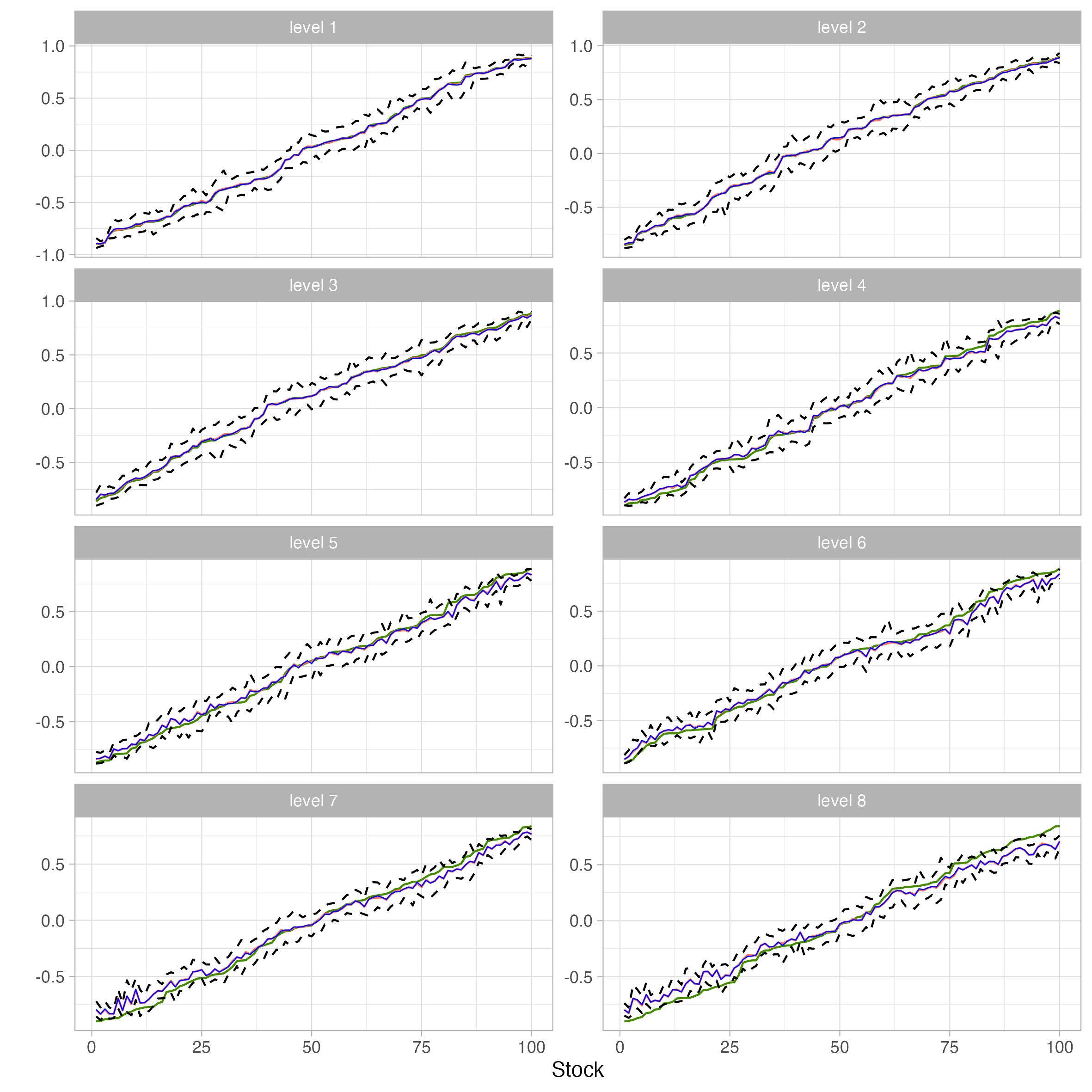}
		\caption{A graph summarizing the parameter estimates for $\bar\varphi_{xy|z}$ by level in simulation (S1). The copulas are in ascending order based on the true values of $\bar\varphi_{xy|z}$ in each level for a better visualization. The green curves indicate true values, the blue curves indicate mean estimates, the orange curves indicate the median estimates, and the pair of dashed curves represent the 90\% credible intervals of the sequential estimates out of the 200 replications.}
		\label{fig:phibar_stock_m_8}
	\end{figure}

	The parameters in the stock copulas are estimated using sequential estimation as discussed in Section \ref{section:parameter_estimation}. \ref{section:comparing_sequential_estimates} contains a comparison with the sequential estimates and the MCMC estimates of the parameters in the stock copulas, and the results suggest that the sequential estimates work as well as the MCMC estimates. Since there are $m\times p$ estimates for, respectively, $a_{j,i|1,\ldots,i-1}$, $b_{j,i|1,\ldots,i-1}$, $\bar\varphi_{j,i|1,\ldots,i-1}$, and $v_{j,i|1,\ldots,i-1}$, it is more efficient to present the results using graphs. We also partition the parameters into $m$ levels, where the first level contains the parameters $a_{j,1}$, $b_{j,1}$ $\bar\varphi_{j,1}$ and $v_{j,1}$, and the $i$th level contains $a_{j,i|1,\ldots,i-1}$, $b_{j,i|1,\ldots,i-1}$ $\bar\varphi_{j,i|1,\ldots,i-1}$ and $v_{j,i|1,\ldots,i-1}$, for $i=2,\ldots,m$. We first present the sequential estimation results for the stock parameters in Simulation (S1). \autoref{fig:axy_stock_m_8}, \autoref{fig:bxy_stock_m_8}, \autoref{fig:vxy_stock_m_8}, and \autoref{fig:phibar_stock_m_8} show the graphs summarizing the parameter estimates for respectively $a_{xy|z}$, $b_{xy|z}$, $v_{xy|z}$ and $\bar\varphi_{xy|z}$ by level. The green curves indicate true values, the blue and orange curves indicate respectively the means and medians of the sequential estimates out of 200 replications, and the pair of dashed curves represent the 90\% credible interval out of 200 replications. The x-axes are in ascending order according to the true values of the parameters in each level for a better visualization.
	
	Both the means and medians of the sequential estimates of $a_{xy|z}$ are close to the true values for the first three levels as observed from \autoref{fig:axy_stock_m_8}. The 90\% credible intervals are also narrow, indicating that the estimates are precise. The estimates tend to be slightly smaller than the true values in deeper levels. This happens because the estimates in the deeper levels rely on the estimates in previous levels, which may lead to larger errors. We expect the errors in deeper levels would have mild effects on the model. This is because, as we will see later in the empirical study, usually the long-run correlations are small (and thus not useful to capture the linear dependence among stocks) and the degrees of freedom are large (and thus not useful to capture the tail dependence) in deeper levels. Then, as long as the estimates of the first several levels are accurate, the model can already capture the linear and non-linear dependencies among stocks accurately.
	
	Both the means and medians of the estimates of $b_{xy|z}$ are close to the true values, even in the deeper levels as observed from \autoref{fig:bxy_stock_m_8}. However, the 90\% credible intervals are wild, especially in deeper levels. This could be due to the ``information'' stored in the data set. Different data sets may contain different noises so that the estimation algorithm may not detect the temporal dependence for the correlation, and thus the algorithm may give a small value of $b_{xy|z}$. Nevertheless, the algorithm can recover the true values of $b_{xy|z}$ on average as the means and medians of the sequential estimates out of 200 replications are close to their true values.
	
	The means and medians of the estimates of $v_{xy|z}$ are close to the true values in all levels as observed in \autoref{fig:vxy_stock_m_8}. The 90\% credible intervals are also narrow whenever the true degrees of freedom are small. This shows that the algorithm is able to detect strong tail dependence precisely, which is crucial in practice. However, the 90\% credible intervals are wilder when the degrees of freedom are larger and in deeper levels. This is again due to the ``information'' stored in the data set. For larger degrees of freedom (and thus weaker tail dependence), the data may not contain much information about the co-moving tail cases, and thus the estimates will be less precise. But it is not an issue in practice, since we are able to capture the strong dependence from the data set precisely, which is more important in application.
	
	The means and medians of the sequential estimates of $\bar\varphi_{xy|z}$ are close to their true values and the 90\% credible intervals are very narrow as observed in \autoref{fig:phibar_stock_m_8}. This shows that the sequential estimation performs very well in estimating long-run correlations.
	
	The sequential estimation results for the stock copulas in Simulation (S2) are contained in Appendix \ref{section:appendix_simulation_S2_stock_parameters}. The estimation results are similar to those in this section, and the comments in this section are also applicable to the results in Appendix \ref{section:appendix_simulation_S2_stock_parameters}. This shows that the algorithm also works well for a more complicated network with $m=10$ variables.

	\section{Empirical study}
	\label{section:empirical_study}
	\subsection{Data}
	In this section, we consider a portfolio of $p=92$ stocks in \autoref{tab:list_of_stocks} in Appendix \ref{section:List_of_stocks} and $m=10$ market indexes in \autoref{tab:risk_factor_description}. We conduct a full GC-GARCH estimation and also conduct an investment experiment using a moving-window approach, in which we determine the portfolio weights for rebalancing by refitting the GC-GARCH model weekly.
	
	The daily closing prices of $m+p=102$ stocks/market indexes are available from 4 January 2017 to 5 May 2023 ($T+1=1,561$ trading days of closing prices and $T=1,560$ returns can be calculated from the closing prices). Let $S_{i,t}$ be the closing price of the $i$th stocks/market indexes on day $t$ for $i=1,\ldots,m+p$, $t=0,\ldots,T$. $t=0$ corresponds to 4 Jan 2017. The log returns on $T$ trading days of the stocks/market indexes, { $r_{i,t}= 100\times \left[ \log(S_{i,t})-\log(S_{i,t-1})\right]$ for $i=1,\ldots,m+p$ and $t=1,\ldots,T$}, are used to fit the GC-GARCH model. Similar to the simulation study, we say that a copula is in level $k$ if the copula has $k$ conditioning variables.

	\begin{figure}[H]
		\centering
		\includegraphics[width=8cm]{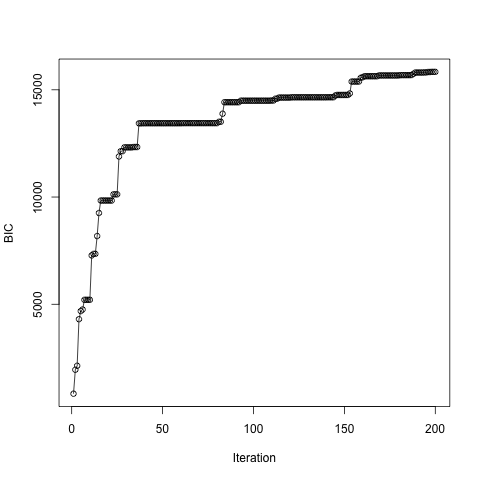}
		\caption{The time series of the BIC in the structural learning of the full estimation in the empirical study.}
		\label{fig:BIC_full_estimation}
	\end{figure}

	\begin{table}[]
		\scriptsize
		\begin{tabular}{lllll}
			Name & Symbol & Order & Number of constituents & Portfolio included \\ \hline
			Hang   Seng Composite LargeCap Index & HSLI & 1 & 122 & 67 \\
			Hang   Seng China Enterprises Index & HSCEI & 2 & 50 & 33 \\
			HK-Listed   Biotech Index & HSHKBIO & 3 & 73 & 3 \\
			Hang   Seng Index & HSI & 4 & 76 & 58 \\
			Hang   Seng ESG 50 Index & HSESG50 & 5 & 50 & 33 \\
			China   H-Financials Index & H.FIN & 6 & 30 & 20 \\
			Hang   Seng Climate Change 1.5$\deg$C Target Index & HSC15TI & 7 & 212 & 21 \\
			Hang   Seng TECH Index & HSTECH & 8 & 30 & 4 \\
			Hang   Seng Composite MidCap Index & HSMI & 9 & 52 & 4 \\
			Growth   Enterprise Market Index & HKSPGEM & 10 & 49 & 0
		\end{tabular}
		\caption{The name, symbol, topological order, number of constituent and the number of stocks included in the portfolio of the $p=92$ stocks for each risk factor. The topological orders of the stocks are based on the MAP network in \autoref{fig:empirical_all_data_MAP_network}.}
		\label{tab:risk_factor_description}
	\end{table}

	\subsection{Full estimation}
	
\begin{figure}[H]
		\centering
		\includegraphics[width=10cm]{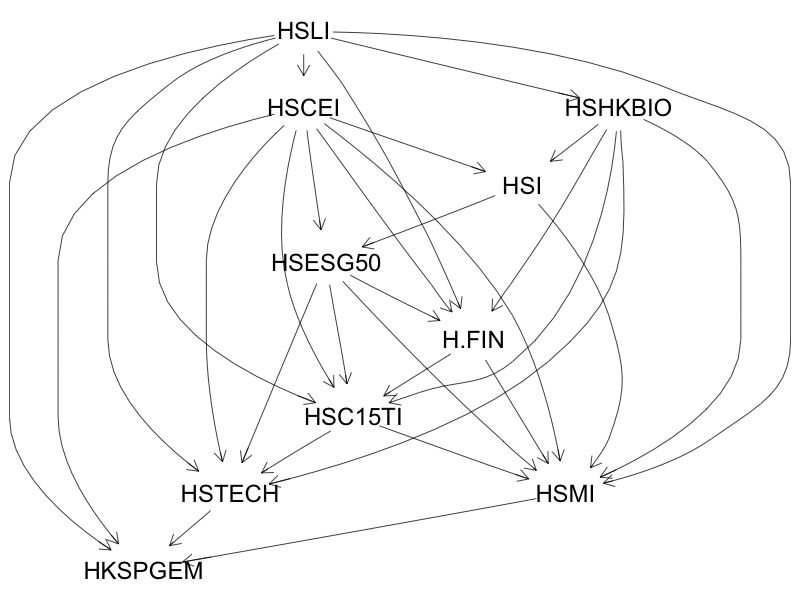}
		\caption{The MAP network.}
		\label{fig:empirical_all_data_MAP_network}
	\end{figure}

	We first estimate the GC-GARCH model using all $T=1560$ trading days of returns. We conduct a structural learning with 200 iterations using the method in Section \ref{section:structural_learning}. \autoref{fig:BIC_full_estimation} shows the time series of the BIC in the structural learning of the full estimation. The burn-in is chosen automatically by minimizing the Geweke's statistic of the BIC. The structure with the highest score is shown in \autoref{fig:empirical_all_data_MAP_network}. The topological orders of the risk factors are shown in \autoref{tab:risk_factor_description}. The parameter estimates of the DAG copulas are shown in \autoref{tab:empirical_study_full_parameter_estimates}. 
	
	The Hang Seng Composite LargeCap index (HSLI) is at the top of the order in the network in \autoref{fig:empirical_all_data_MAP_network}, and connected to seven risk factors, showing that HSLI can be regarded as the ``source'' of the market movements. The HSLI also contains 67 stocks in our portfolio as shown in \autoref{tab:risk_factor_description}. We expect that the HSLI can explain a high portion of the co-movements of the stocks. The HSCEI, HSI and HSHKBIO are also ranked in top orders in the network. Moreover, the risk factors in the network are highly interconnected, indicating that these risk factors are closely dependent.

	The parameter estimates in the DAG copulas are shown in \autoref{tab:empirical_study_full_parameter_estimates}. Most of $a_{j,i|1,\ldots,i-1}+b_{j,i|1,\ldots,i-1}$ are close to 1 (as most of $b_{j,i|1,\ldots,i-1}$ themselves are close to 1), indicating that the correlations are highly persistent. The long-run correlations tend to be smaller in magnitude in deeper levels; for example, the long-run correlations of (HSMI,HSLI), (HSMI,HSCEI$|$HSLI), (HSMI,HSHKBIO$|$HSLI,HSCEI), and (HSMI,HSI$|$HSLI,HSCEI,HSHKBIO) are respectively 0.9350, 0.0554, 0.4562, and $-$0.1924. This means that high portions of the correlations among risk factors are explained by the first level. 
	
	Some degrees of freedom of the risk factor pairs are small, indicating that the dependence in the tails for some pairs of risk factor are strong, captured by the GC-GARCH model. For example, the degrees of freedom of (HSCEI,HSLI), (HSI,HSCEI), (HSESG50,HSCEI) and (HSESG50,HSI$|$HSCEI) are respectively 5.59, 3.81, 4.51 and 5.53. It indicates that there are high levels of tail dependence among the risk factors.

\begin{figure}[H]
		\centering
		\includegraphics[width=13cm]{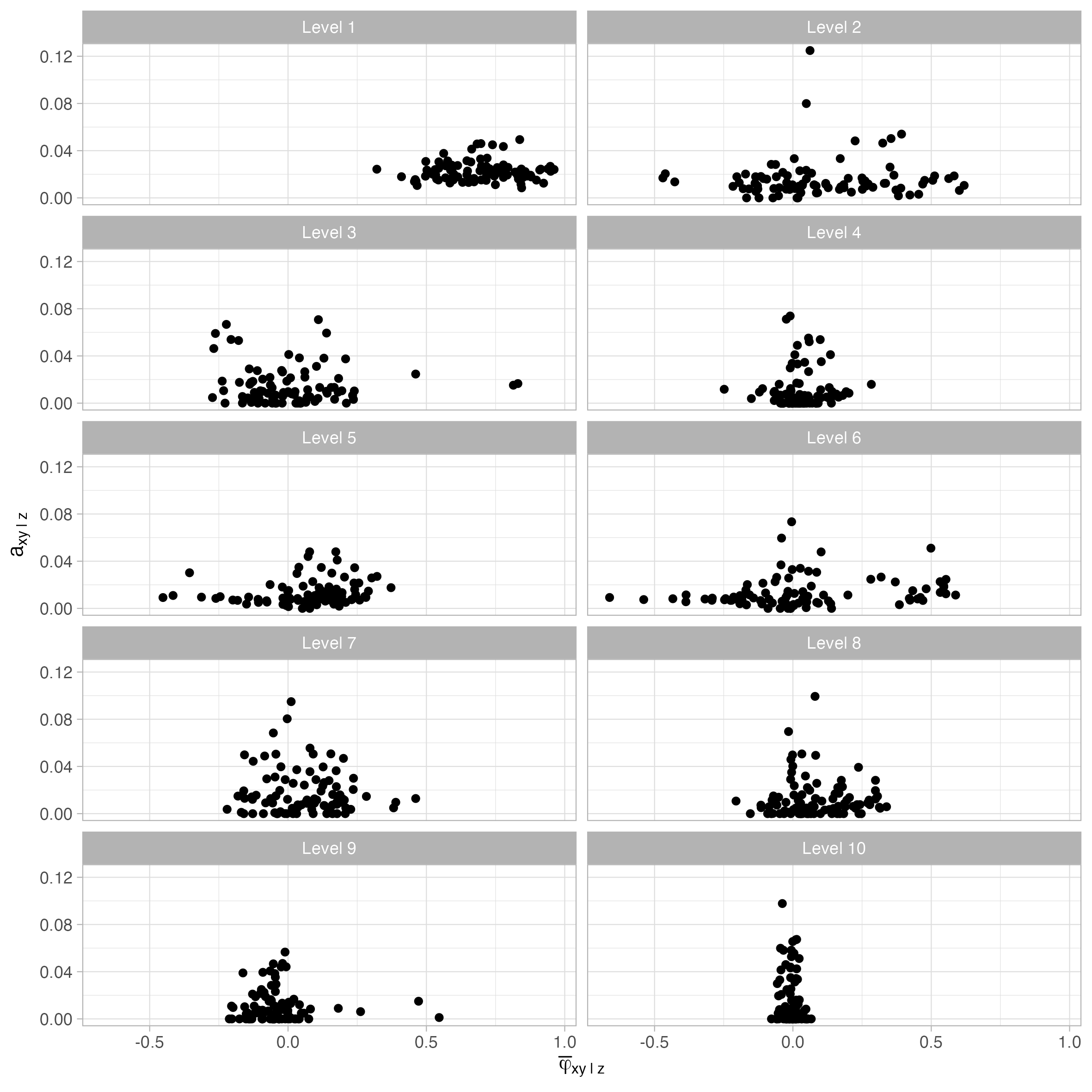}
		\caption{The scatterplots of $a_{xy|z}$ versus $\bar\varphi_{xy|z}$ by level. A point corresponds to a stock copulas $c_{j,i|1,\ldots,i-1}^{[t]}$. The $k$-th level contains the copulas $c_{j,k|1,\ldots,k-1}^{[t]}$ for $j=m+1,\ldots m+p$.}
		\label{fig:axy_vs_phibar}
	\end{figure}

\begin{figure}[H]
		\centering
		\includegraphics[width=13cm]{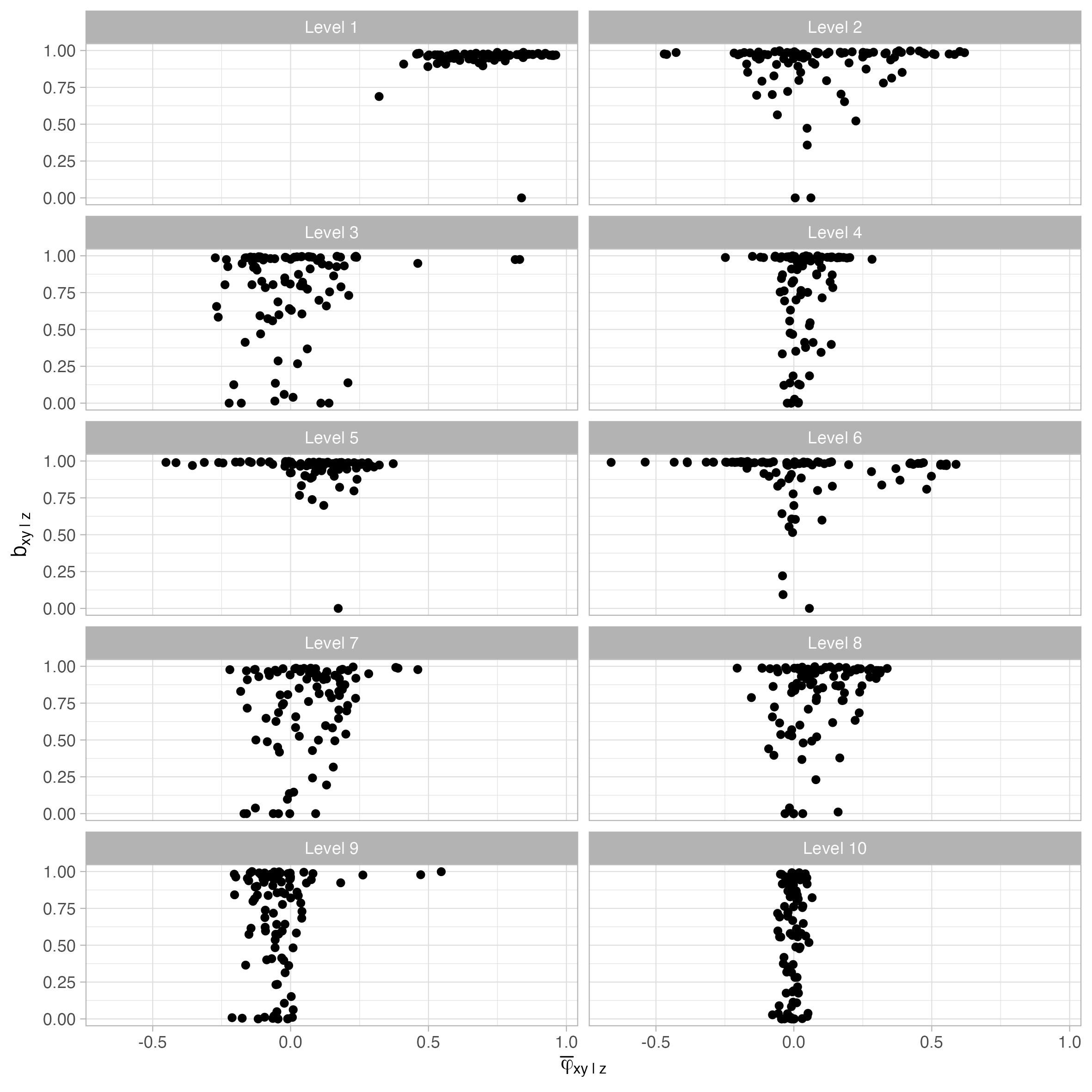}
		\caption{The scatterplots of $b_{xy|z}$ versus $\bar\varphi_{xy|z}$ by level. A point corresponds to a stock copulas $c_{j,i|1,\ldots,i-1}^{[t]}$. The $k$-th level contains the copulas $c_{j,k|1,\ldots,k-1}^{[t]}$ for $j=m+1,\ldots m+p$.}
		\label{fig:bxy_vs_phibar}
	\end{figure}

\begin{figure}[H]
		\centering
		\includegraphics[width=13cm]{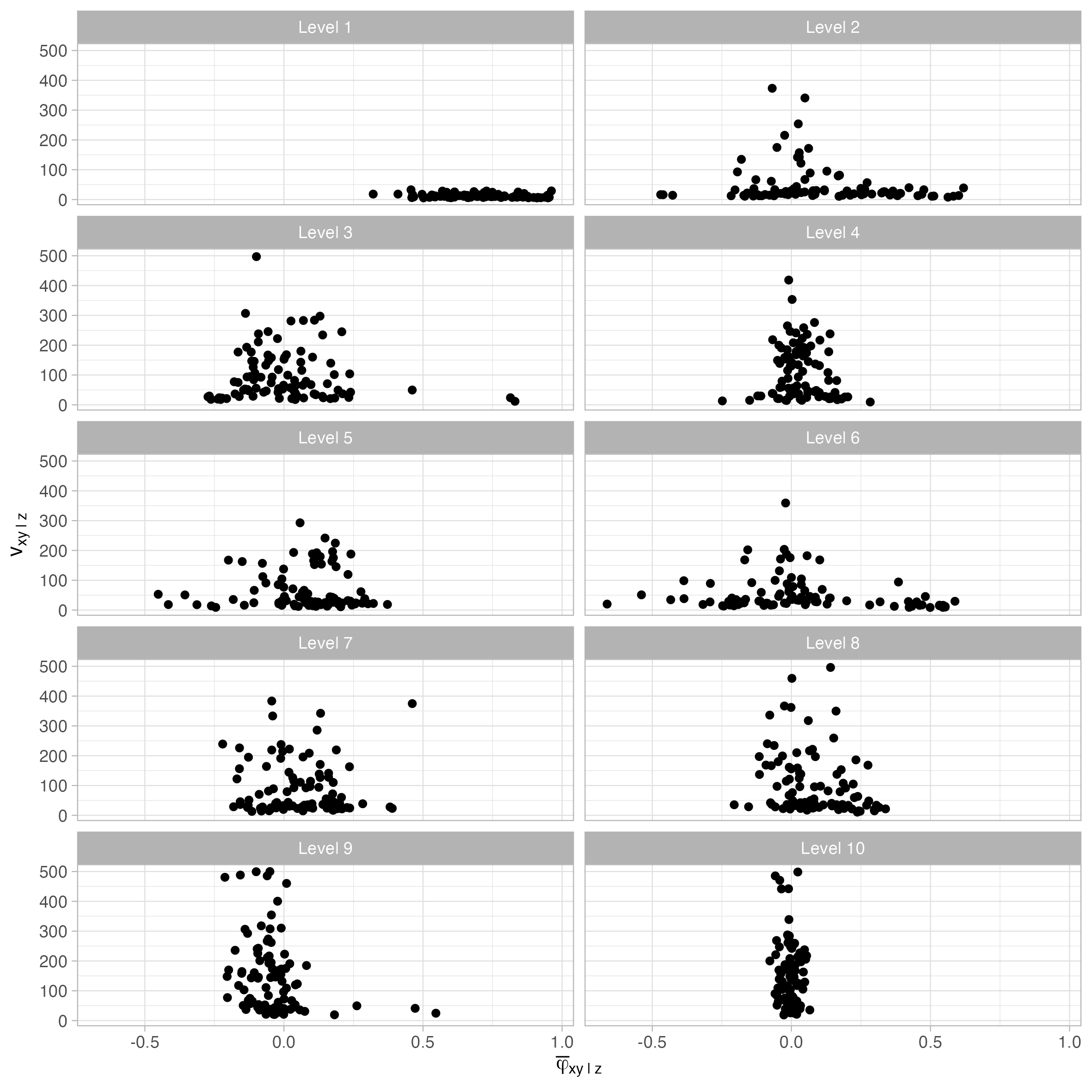}
		\caption{The scatterplots of $v_{xy|z}$ versus $\bar\varphi_{xy|z}$ by level. A point corresponds to a stock copulas $c_{j,i|1,\ldots,i-1}^{[t]}$. The $k$-th level contains the copulas $c_{j,k|1,\ldots,k-1}^{[t]}$ for $j=m+1,\ldots m+p$.}
		\label{fig:vxy_vs_phibar}
	\end{figure}

	As there are $m\times p=920$ parameters for each $a_{j,i|1,\ldots,i-1}$, $b_{j,i|1,\ldots,i-1}$, $v_{j,i|1,\ldots,i-1}$ and $\bar\varphi_{j,i|1,\ldots,i-1}$ in the stock copulas, we use scatter plots in \autoref{fig:axy_vs_phibar}, \autoref{fig:bxy_vs_phibar} and \autoref{fig:vxy_vs_phibar} to report the parameter estimates of $a_{j,i|1,\ldots,i-1}$, $b_{j,i|1,\ldots,i-1}$ and $v_{j,i|1,\ldots,i-1}$. $\bar\varphi_{j,i|1,\ldots,i-1}$'s are used as the x-axes. We report the estimates by level, where level $i$ contains the parameters in the copulas of the form $C^{[t]}_{j,i|1,\ldots,i-1}$. By reporting by different levels, we can observe the effects of the risk factors to the stocks clearly. The y-axes correspond to the estimates of the above three parameters, and the x-axes correspond to the estimates of $\bar\varphi_{j,i|1,\ldots,i-1}$. The reason for putting $\bar\varphi_{j,i|1,\ldots,i-1}$ in the x-axes is that we observe that the estimates of $a_{j,i|1,\ldots,i-1}$, $b_{j,i|1,\ldots,i-1}$ and $v_{j,i|1,\ldots,i-1}$ are associated with the values of the long-run correlations $\bar\varphi_{j,i|1,\ldots,i-1}$.  $a_{j,i|1,\ldots,i-1}$ and $b_{j,i|1,\ldots,i-1}$ tend to be smaller, and $v_{j,i|1,\ldots,i-1}$ tends to be large whenever the long-run correlation $\bar\varphi_{j,i|1,\ldots,i-1}$ is close to zero. This could be due to the ``information'' contained in the data set; when the long-run correlation $\bar\varphi_{j,i|1,\ldots,i-1}$ is close to zero, the ``information'' stored in the data set would be relatively weak, and thus the algorithm may not be able to detect the temporal dependence, captured by $a_{j,i|1,\ldots,i-1}$ and $b_{j,i|1,\ldots,i-1}$, and the tail dependence, captured by $v_{j,i|1,\ldots,i-1}$ (as a larger degrees of freedom indicates weaker tail dependence). From \autoref{fig:bxy_vs_phibar}, we observe that most of the estimates of $b_{j1}$ in level 1 (correspond to the risk factor HSLI) are close to 1, and most of the long-run correlations $\bar\varphi_{j1}$ are larger than 0.5, showing that the correlations of these stocks between HSLI are dynamic and most of these stocks are strongly correlated with HSLI. The estimates of $b_{ji|1,\ldots,i-1}$ tend to be smaller in deeper levels, and the long-run correlations in deeper levels also tend to move towards 0, showing that high portions of the correlations are explained by the risk factors in the first several levels. \autoref{fig:vxy_vs_phibar} shows that the degrees of freedom in the first level are small, and tend to be larger in deeper levels, indicating that the tail dependence between the stocks and respectively the first several risk factors are stronger. High portions of tail dependence are explained by the first several levels. Recall we have mentioned, in Section \ref{section:parameters_in_the_stock_copulas}, that the estimates could become slightly imprecise (i.e., the credible intervals are wider) in deeper levels. We claim that the effects would be mild as the correlations are getting smaller and the degrees of freedom are getting larger in deeper levels as observed in \autoref{fig:vxy_vs_phibar}. Nevertheless, we have shown that the estimates are accurate on average even in deeper levels in the simulation study.

	\begin{table}[]
		\centering
		\scriptsize
		\begin{tabular}{lllll}
			Copula & $\bar\varphi_{xy|z}$ & $a_{xy|z}$ & $b_{xy|z}$ & $v_{xy|z}$ \\ \hline
			HSCEI,HSLI & 0.9911 & 0.0102 & 0.9776 & 5.59 \\
			HSHKBIO,HSLI & 0.7601 & 0.0275 & 0.9347 & 12.65 \\
			HSI,HSCEI & 0.9910 & 0.0122 & 0.9734 & 3.81 \\
			HSI,HSHKBIO$|$HSCEI & 0.0560 & 0.0000 & 0.9996 & 24.34 \\
			HSESG50,HSCEI & 0.9783 & 0.0132 & 0.9731 & 4.51 \\
			HSESG50,HSI$|$HSCEI & 0.9361 & 0.0143 & 0.9732 & 5.53 \\
			H.FIN,HSLI & 0.9666 & 0.0135 & 0.9847 & 9.90 \\
			H.FIN,HSCEI$|$HSLI & 0.9114 & 0.0020 & 0.9975 & 17.72 \\
			H.FIN,HSHKBIO$|$HSLI,HSCEI & -0.0724 & 0.0022 & 0.9936 & 19.53 \\
			H.FIN,HSESG50$|$HSLI,HSCEI,HSHKBIO & 0.5499 & 0.0247 & 0.9643 & 18.36 \\
			HSC15TI,HSLI & 0.9929 & 0.0062 & 0.9758 & 6.16 \\
			HSC15TI,HSCEI$|$HSLI & 0.4995 & 0.0108 & 0.9679 & 9.57 \\
			HSC15TI,HSHKBIO$|$HSLI,HSCEI & 0.5294 & 0.0149 & 0.9568 & 13.22 \\
			HSC15TI,HSESG50$|$HSLI,HSCEI,HSHKBIO & 0.4742 & 0.0191 & 0.9646 & 27.08 \\
			HSC15TI,H.FIN$|$HSLI,HSCEI,HSHKBIO,HSESG50 & 0.0680 & 0.0128 & 0.9624 & 18.11 \\
			HSMI,HSLI & 0.9350 & 0.0153 & 0.9537 & 13.18 \\
			HSMI,HSCEI$|$HSLI & 0.0554 & 0.0055 & 0.9818 & 19.91 \\
			HSMI,HSHKBIO$|$HSLI,HSCEI & 0.4562 & 0.0243 & 0.8494 & 15.61 \\
			HSMI,HSI$|$HSLI,HSCEI,HSHKBIO & -0.1924 & 0.0264 & 0.9003 & 20.49 \\
			HSMI,HSESG50$|$HSLI,HSCEI,HSHKBIO,HSI & 0.0285 & 0.0112 & 0.9880 & 18.61 \\
			HSMI,H.FIN$|$HSLI,HSCEI,HSHKBIO,HSI,HSESG50 & -0.2130 & 0.0087 & 0.9910 & 15.87 \\
			HSMI,HSC15TI$|$HSLI,HSCEI,HSHKBIO,HSI,HSESG50,H.FIN & 0.7572 & 0.0016 & 0.9979 & 18.08 \\
			HSTECH,HSLI & 0.9680 & 0.0226 & 0.9605 & 12.91 \\
			HSTECH,HSCEI$|$HSLI & -0.3211 & 0.0106 & 0.9783 & 19.93 \\
			HSTECH,HSHKBIO$|$HSLI,HSCEI & 0.3041 & 0.0029 & 0.9960 & 23.84 \\
			HSTECH,HSESG50$|$HSLI,HSCEI,HSHKBIO & -0.4858 & 0.0246 & 0.9692 & 15.94 \\
			HSTECH,HSC15TI$|$HSLI,HSCEI,HSHKBIO,HSESG50 & 0.3677 & 0.0078 & 0.9905 & 19.65 \\
			HKSPGEM,HSLI & 0.5129 & 0.0810 & 0.5853 & 9.56 \\
			HKSPGEM,HSCEI$|$HSLI & -0.0213 & 0.0000 & 0.5636 & 21.31 \\
			HKSPGEM,HSMI$|$HSLI,HSCEI & 0.1800 & 0.0667 & 0.0000 & 31.26 \\
			HKSPGEM,HSTECH$|$HSLI,HSCEI,HSMI & 0.0563 & 0.0023 & 0.9968 & 26.78
		\end{tabular}
		\caption{The parameter estimates of the DAG copulas in the MAP network in \autoref{fig:empirical_all_data_MAP_network}.}
		\label{tab:empirical_study_full_parameter_estimates}
	\end{table}

	\subsection{Moving-window investment experiment}
	
	We conduct an investment experiment where we rebalance the portfolio on the first trading day of each week using the methodology in Section \ref{Section:portfolio_management}. We aim to compare the investment performances between the models (1) GC-GARCH model using the MAP network, (2) GC-GARCH model with model averaging using the $N_g=3$ highest scoring networks, and (3) DCC-GARCH model proposed by \textcite{DCC_model} and \textcite{DCC_GARCH}.
	
	The experiment is conducted with a moving-window size of $w=750$ days. We rebalance the portfolio weights on the first trading day of a week. To determine the portfolio weights, we use the data on the previous $w=750$ trading days (up to the last trading days of the week, which is typically a Friday) to fit the model, and solve the MV and MCVaR problems in \eqref{eqt:markovitz} and \eqref{eqt:min_CVaR_alternative}.

	Suppose that $t_1,t_2,\ldots,t_L\in\{1,\ldots,T\}$ are the last trading days of all weeks in the data set, where $L=331$ is the number of weeks in the data set. The first model can be fitted using the first 749 trading days (in the week 159). The reason for using 749 and not 750 for the first window is that, the 749th trading day in the data set is a Friday, which fits the setting that we fit a model at the end of a week. We further reserve the first 32 weeks of model fits for the sake of an investment strategy, to be introduced later. Then, the first investment can be done on 31 August 2020, which is the first day in week 192, where the portfolio weights is determined at the end of week $159+32=191$. Let $R_{PF,t}=\exp(r_{PF,t})-1$ be the simple rate of return of the portfolio on day $t$. Note that the portfolio weights are determined on last days of each of week $t_{191},t_{192},\ldots,t_{330}$, and we make investment on the first trading days of each week $t_{191}+1,t_{192}+1,\ldots,t_{330}+1$. The simple rate of return of the portfolio in week $i$ ($i=192,\ldots,331$) is given by {$R_{PF,\text{week }i} = \prod_{\tau=t_{i-1}+1}^{t_i}(1+R_{PF,\tau})$}. Suppose that we invest $10,000$ dollars on day $t_{191}+1$ in week 192 (the first week we invest), the cumulative value of the portfolio as of week $i$ ($i=192,\ldots,331$) is given by
	\begin{equation}
		P_i = 10,000 \prod_{\tau=t_{191}+1}^{t_i}(1+R_{PF,\tau}).
		\label{eqt:proceeds}
	\end{equation}
	We also evaluate the quality of prediction, following \textcite{ES_estimation}, we compute the cost function with coverage level $\alpha$,
	\begin{equation}
		C(\alpha) = \frac{1}{g(\alpha)}\sum_{i=192}^{331}\abs{-R_{PF,t_{i-1}+1} -CVaR^{\alpha}_{t_{i-1}+1}} \mathbf{1}(-R_{PF,t_{i-1}+1}\geq CVaR^{\alpha}_{t_{i-1}+1}),
		\label{eqt:cost_function}
	\end{equation}
	for $\alpha=0.0005,0.001,0.005,0.01,0.05$, where $CVaR^{\alpha}_{t_{i-1}+1}$ is the predicted conditional value-at-risk with coverage level $\alpha$ on day $t_{i-1}+1$, estimated using the methods of the portfolio returns introduced in Section \ref{section:MVP_MCVaRP}, and $g(\alpha)=\sum_{i=192}^{331}\mathbf{1}(-R_{PF,t_{i-1}+1}\geq CVaR^{\alpha}_{t_{i-1}+1})$ is the number of days that losses exceed the predicted CVaRs from weeks $192$ to $331$.
	
	There are two investment strategies:
	
	\begin{itemize}
		\item (Strategy 1) We rebalance the portfolio on the first day of every week. We determine the portfolio weights by solving (i) the MV portfolio problem in \eqref{eqt:markovitz}, and (ii) the MCVaR portfolio problems with $\alpha=0.0005,0.001,0.005,0.01,0.05$.
		\item (Strategy 2) We avoid investing in the stock market whenever we predict a high level of 1-day ahead portfolio CVaR. Mathematically, at the end of week $i$, we first calculate the average 1-day ahead predicted CVaR with coverage level $\alpha$ in the past $w_S$ weeks (excluding week $i$):
		\begin{equation}
			\overline{CVaR}^{\alpha}_{(i-w_S):(i-1)}=\frac{1}{w_S} \sum_{i'=i-w_S}^{i-1} CVaR_{t_{i'}+1}^\alpha.
			\label{eq:average_CVaR}
		\end{equation}
		We then predict $CVaR_{t_i+1}^\alpha$. We decide to invest in the stock market only when {$CVaR_{t_i+1}^\alpha$ $\leq$ $\overline{CVaR}^{\alpha}_{(i-w_S):(i-1)}$}, i.e., the potential loss is predicted to be lower than or equal to the $w_S$-week average. We take $w_S=8,16,32$ to test if the strategy performs uniformly well with different choice of $w_S$. We only solve the MCVaR portfolio problems for this strategy as this strategy is designed using CVaR.
	\end{itemize}
	To ensure that we are comparing the performance of strategies 1 and 2 on the same trading days, we reserve the first 32 weeks of model fits as we need $32$ weeks to calculate \eqref{eq:average_CVaR} when $w_S=32$. The first investment is done on 31 August 2020, which is the first day in week 192, for both strategies 1 and 2 with $w_S=8,16,32$ and $\alpha=0.0005,0.001,0.005,0.01,0.05$.

	We compare the investment performances of between the models (1) GC-GARCH model using the MAP network, (2) GC-GARCH model with model averaging using the $N_g=3$ highest scoring networks (we call it MA for simplicity), and (3) DCC-GARCH model. We determine the portfolio weights using (i) MCVaR portfolio with different $\alpha$ values and (ii) MV portfolio for strategy 1. We also estimate the CVaR from the MV portfolios using \eqref{eqt:CVaR_MVP} because we want to compare the performance of the MCVaR portfolio to the traditional MV portfolio. We expect the MV portfolio to work poorly as the MV portfolio problem did not consider the tail behavior. We only consider the MCVaR portfolio with different $\alpha$ values for strategy 2. We first compare the cost function and the number of days that losses exceed the CVaR. We then compare the investment performances by plotting the time series of the cumulative values of the portfolios in \eqref{eqt:proceeds}. 
	
	\autoref{tab:back_testing} shows the cost function in \eqref{eqt:cost_function} and the number of days that the losses exceed the CVaR (or the number of exceedance) for models (1) to (3) where the portfolio weights are obtained by respectively solving the MCVaR portfolio and MV portfolio problems. First note that the cost function and the number of exceedance for the MV portfolios are all smaller than those for the MCVaR portfolios. This is not because the MV portfolios perform better, but rather due to the fact that the CVaRs are not minimized in the MV problem, and thus larger CVaRs give smaller cost function and number of exceedance. We should only compare the cost function and the number of exceedance between the GC-GARCH, MA and DCC for the same MCVaR or MV portfolios with the same $\alpha$. We note that the cost functions of the portfolios with GC-GARCH and MA are all smaller than those of the DCC-GARCH model. The number of exceedance in the portfolios with GC-GARCH and MA are all smaller than those in DCC except for $\alpha=0.05$. Furthermore, the MA has fewer days of exceedance than the GC-GARCH for MCVaR portfolio at $\alpha=0.01$ and MV at $\alpha=0.005,0.01$. This suggests that the GC-GARCH model and MA work much better than the DCC-GARCH model in covering the losses using CVaR, and the model averaging using GC-GARCH can further improve the performance over the GC-GARCH using the MAP network.
	
\begin{figure}[H]
		\centering
		\includegraphics[width=14cm]{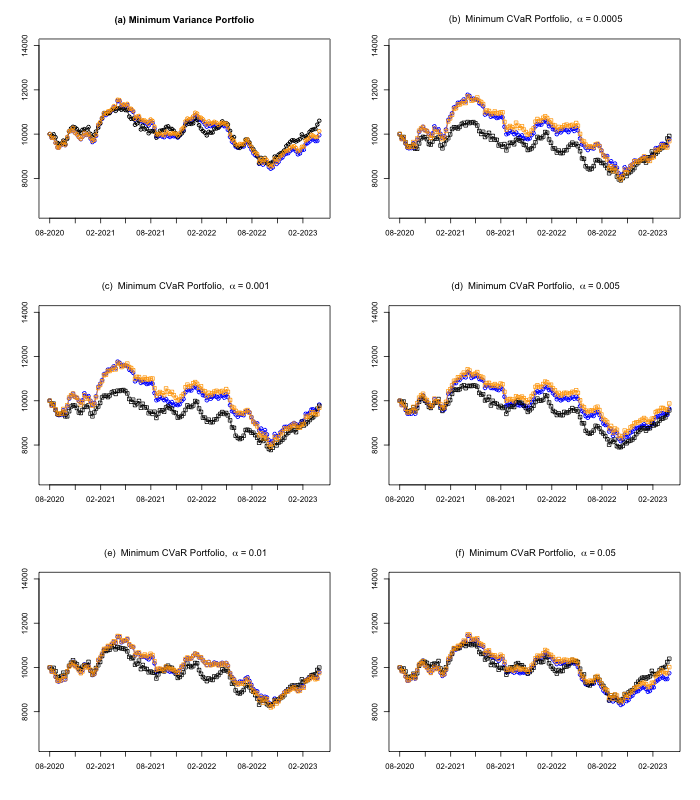}
		\caption{The time series of the cumulative values of the weekly rebalanced portfolios according to \eqref{eqt:proceeds}. The blue curves indicate the portfolios with the GC-GARCH model using the MAP network, the orange curves indicate the portfolios with the GC-GARCH model using model averaging, and the black curves indicate the portfolios obtained from the DCC-GARCH model.}
		\label{fig:strategy_1}
	\end{figure}

	We now compare the investment performance of strategy 1, where we rebalance the portfolio on the first day of every week. \autoref{fig:strategy_1} shows the time series of the cumulative values of the weekly rebalanced portfolios according to \eqref{eqt:proceeds}. The blue curves indicate the portfolios with the GC-GARCH model using the MAP network, the orange curves indicate the portfolios with the GC-GARCH model using model averaging, and the black curves indicate the portfolios obtained from the DCC-GARCH model. \autoref{fig:strategy_1} (a) shows the cumulative values of the minimum variance portfolios. The performances of GC-GARCH, MA and DCC are similar in this case. Figures \ref{fig:strategy_1} (b) to (f) show the minimum CVaR portfolios with $\alpha=0.0005,0.001,0.005,0.01,$ and $0.05$. The values of the portfolios using GC-GARCH and MA are on average higher than those using DCC. The portfolios generate higher profits for smaller $\alpha$. This supports that the GC-GARCH model outperforms the DCC-GARCH model when we take the tails into account. However, we observe that the cumulative values of the portfolios are falling from 2021 to 2022 in all portfolios in \autoref{fig:strategy_1}. Even though the CVaRs have been minimized in the experiment, some of the losses could still be significant, and we need to modify the strategy to avoid huge potential losses.
	
\begin{figure}[H]
		\centering
		\includegraphics[width=13cm]{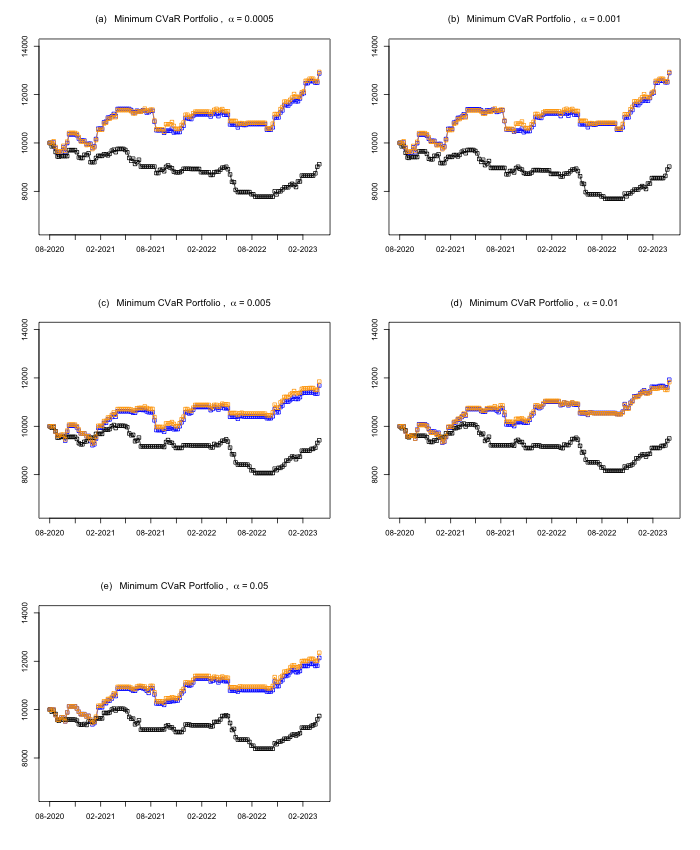}
		\caption{The time series of the cumulative values of the MCVaR portfolios using strategy 2 with $w_{S}=8$. The blue curves indicate the portfolios with the GC-GARCH model using the MAP network, the orange curves indicate the portfolios with the GC-GARCH model using model averaging, and the black curves indicate the portfolios obtained from the DCC-GARCH model.}
		\label{fig:investment_strategy_2_window_8_mvp}
	\end{figure}

\begin{figure}[H]
		\centering
		\includegraphics[width=13cm]{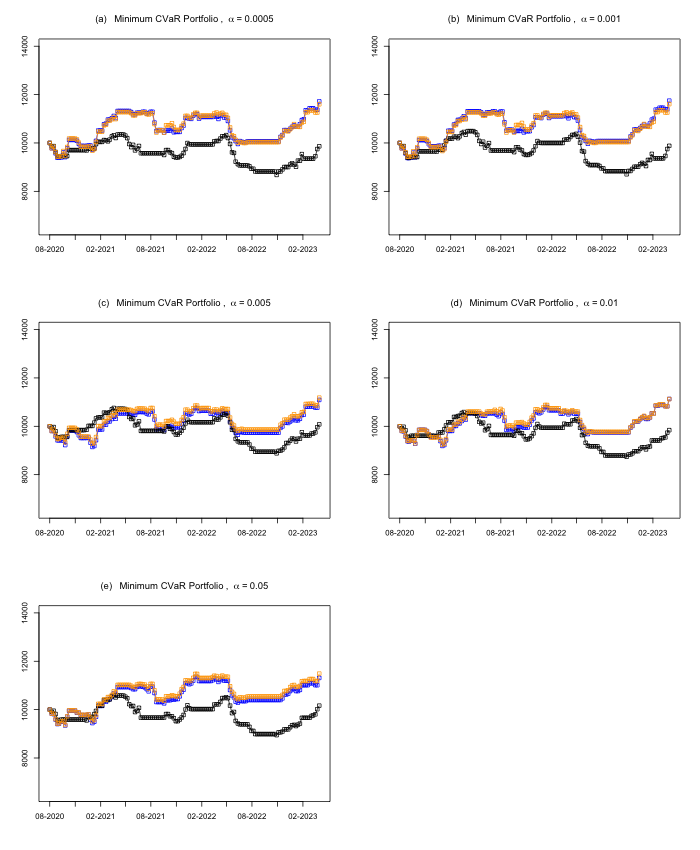}
		\caption{The time series of the cumulative values of the MCVaR portfolios using strategy 2 with $w_{S}=16$. The blue curves indicate the portfolios with the GC-GARCH model using the MAP network, the orange curves indicate the portfolios with the GC-GARCH model using model averaging, and the black curves indicate the portfolios obtained from the DCC-GARCH model.}
		\label{fig:investment_strategy_2_window_16_mvp}
	\end{figure}

\begin{figure}[H]
		\centering
		\includegraphics[width=15cm]{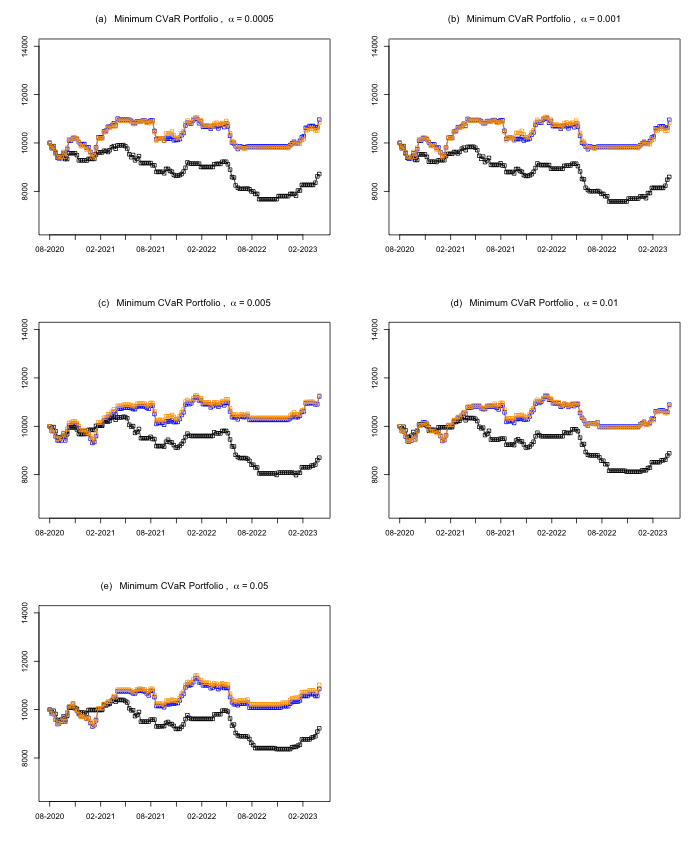}
		\caption{The time series of the cumulative values of the MCVaR portfolios using strategy 2 with $w_{S}=32$. The blue curves indicate the portfolios with the GC-GARCH model using the MAP network, the orange curves indicate the portfolios with the GC-GARCH model using model averaging, and the black curves indicate the portfolios obtained from the DCC-GARCH model.}
		\label{fig:investment_strategy_2_window_32_mvp}
	\end{figure}

	Strategy 2 tries to avoid investing in the stock market whenever the predicted 1-day ahead CVaR is higher than the previous $w_S$ weeks average CVaR in \eqref{eq:average_CVaR}. \autoref{fig:investment_strategy_2_window_8_mvp}, \autoref{fig:investment_strategy_2_window_16_mvp}, and \autoref{fig:investment_strategy_2_window_32_mvp}  show the cumulative values of the minimum CVaR portfolios with respectively $w_S=8,16,$ and $32$. The cumulative values of the portfolios with GC-GARCH with the MAP network (blue curves) and GC-GARCH with model averaging (orange curves) are on average growing from 2020 to 2023, whereas the cumulative values of the DCC-GARCH portfolios (black curves) tend to be falling. To investigate in more detail, \autoref{tab:avg_return_excluded} shows the numbers of weeks invested in the stock market and the average returns on the weeks excluded from investing in the stock market for the investment experiment using strategy 2. The number of weeks is similar across different $\alpha$ and $w_S$ using different models. The average returns on the weeks excluded are all negative for GC-GARCH and MA, and are much smaller than those in DCC-GARCH. This suggests that the strategy with the GC-GARCH model successfully avoids days with high losses, and the portfolios are able to grow when we take the tails into account. The GC-GARCH model has good ability to capture the extreme scenarios from the tails. Strategy 2 performs well for different $w_S=8,16,$ and $32$, especially when we take a small $w_S=8$, in which case the average returns on the weeks excluded from investing in the stock markets are ranged from $-0.21$\% to $-0.16$\%, as shown in \autoref{tab:avg_return_excluded}.

	\begin{table}[]
		\centering
		\begin{tabular}{llllllll}
			&  & \multicolumn{3}{c}{Cost function} & \multicolumn{3}{c}{Number of CVaR exceedance} \\ \hline
			& $\alpha$ & GC-GARCH & MA & DCC & GC-GARCH & MA & DCC \\ \hline
			MCVaR portfolio & 0.0005 & 0.0445 & 0.0454 & 0.0582 & 10 & 10 & 13 \\
			& 0.001 & 0.0464 & 0.0451 & 0.0589 & 9 & 9 & 13 \\
			& 0.005 & 0.0449 & 0.0502 & 0.0535 & 11 & 11 & 15 \\
			& 0.01 & 0.0532 & 0.0533 & 0.0545 & 15 & 12 & 15 \\
			& 0.05 & 0.0883 & 0.0765 & 0.0929 & 26 & 26 & 22 \\ \hline
			MV portfolio & 0.0005 & 0.0026 & 0.0094 & 0.0158 & 2 & 2 & 6 \\
			& 0.001 & 0.0109 & 0.0172 & 0.0273 & 3 & 3 & 10 \\
			& 0.005 & 0.0319 & 0.0412 & 0.0434 & 9 & 8 & 14 \\
			& 0.01 & 0.0450 & 0.0487 & 0.0493 & 13 & 11 & 14 \\
			& 0.05 & 0.0864 & 0.0740 & 0.0913 & 26 & 25 & 22
		\end{tabular}
		\caption{The cost function and the number of days that the losses exceed the CVaR for (1) GC-GARCH model using the MAP network, (2) GC-GARCH model with model averaging (marked MA in the table), and (3) DCC-GARCH model. The portfolio weights are obtained by respectively solving the MCVaR and MV problem.}
		\label{tab:back_testing}
	\end{table}
	
	\begin{table}[]
		\centering
		\scriptsize
		\begin{tabular}{llllllll}
			&  & \multicolumn{3}{c}{Number of weeks invested in the stock market} & \multicolumn{3}{c}{Average return excluded from investment (\%)} \\ \hline
			{} & $\alpha$ & {GC-GARCH} & {MA} & {DCC} & {GC-GARCH} & {MA} & {DCC} \\ \hline
			$w_S=8$ & 0.0005 & 79 & 79 & 78 & -0.21 & -0.16 & 0.05 \\
			& 0.001 & 79 & 79 & 78 & -0.21 & -0.15 & 0.05 \\
			& 0.005 & 78 & 78 & 74 & -0.16 & -0.12 & 0.00 \\
			& 0.01 & 79 & 79 & 74 & -0.16 & -0.12 & 0.02 \\
			& 0.05 & 80 & 80 & 75 & -0.17 & -0.14 & 0.03 \\ \hline
			$w_S=16$ & 0.0005 & 80 & 80 & 72 & -0.15 & -0.14 & 0.00 \\
			& 0.001 & 80 & 80 & 73 & -0.14 & -0.14 & -0.02 \\
			& 0.005 & 82 & 82 & 75 & -0.12 & -0.10 & -0.05 \\
			& 0.01 & 85 & 85 & 72 & -0.11 & -0.10 & -0.01 \\
			& 0.05 & 79 & 79 & 73 & -0.12 & -0.11 & 0.00 \\ \hline
			$w_S=32$ & 0.0005 & 89 & 89 & 76 & -0.10 & -0.09 & 0.08 \\
			& 0.001 & 89 & 89 & 77 & -0.09 & -0.09 & 0.08 \\
			& 0.005 & 87 & 87 & 78 & -0.13 & -0.08 & 0.05 \\
			& 0.01 & 90 & 90 & 75 & -0.10 & -0.08 & 0.06 \\
			& 0.05 & 87 & 87 & 74 & -0.09 & -0.07 & 0.07
		\end{tabular}
		\caption{The number of weeks invested in the stock market and the average returns on the weeks excluded from investing in the stock market for the investment experiment using strategy 2.}
		\label{tab:avg_return_excluded}
	\end{table}

	\section{Discussion and conclusion}
	\label{section:discussion_and_conclusion}
	We propose the GC-GARCH model to reduce the number of correlation parameters using the idea motivated from the CAPM theory, which uses the risk factors (market indexes) to explain the co-movements among a portfolio of stocks. The GC-GARCH model also captures the co-movements in the tails among risk factor and stock returns through the use of pair-copula construction. The risk factors are factorized using the Bayesian network, which allows us to ``rank'' the stocks from topological orders in BN. The edges in the BN indicate the flow of information. The time-varying specification of the parameters further flexibilizes the modeling. We adopt the three-stage estimation procedure, where we first estimate the parameters in the marginal distribution, then in the DAG copulas and finally the stock copulas. We have shown that the estimation procedures provide accurate estimates in Section \ref{section:simulation_study}. The moving-window investment experiment in Section \ref{section:empirical_study} show that the GC-GARCH models using the highest scoring network and model averaging give better performances than the traditional DCC-GARCH model. The GC-GARCH has better ability to cover extreme losses, to avoid risk and to make profits from investing in the market.
	
	There are several things we have not addressed in this paper. We assume that the underlying network in the GC-GARCH model is fixed. However, the underlying networks in financial markets are likely to be changing dynamically. We use the $t$-copula, which is an elliptical copula, in the GC-GARCH model. However, the tail dependence of the returns could be asymmetric. We could employ the GC-GARCH model using other families of copulas, such as the Gumbel copula and the Clayton copula. We can address these points in future work.
	
	\section*{Acknowledgement}
	\noindent
	This work was supported by the Hong Kong RGC General Research Fund (grant number 16507322).
	
\printbibliography

	\newpage
	\appendix
	\renewcommand{\thesubsection}{\Alph{section}.\arabic{subsection}}



	\section{An algorithm for subsetting the DAG space}
	\label{section:subset_DAG}
\begin{figure}[H]
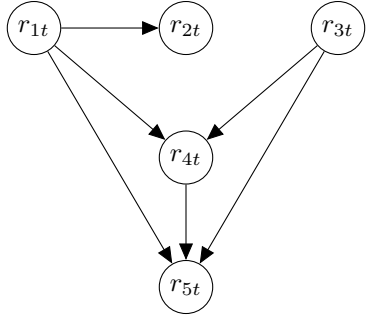

	\centering
	\tikz{
		\node[latent] (1) {$r_{1t}$};%
		\node[latent,xshift=2cm] (2) {$r_{2t}$}; %
		\node[latent,xshift=4cm] (3) {$r_{3t}$}; %
		\node[latent,below=of 2] (4) {$r_{4t}$};
		\node[latent,below=of 4] (5) {$r_{5t}$};
		\edge {1} {5} 
		\edge {1} {2}
		\edge {1} {4}
		\edge {3} {4,5}
		\edge {4} {5}
	}
	\caption{An example DAG that passes the test.}    \label{fig:easy_DAG}
	\end{figure}

	We provide an algorithm to subset the DAG space to reduce computation burden. Let $\pi_1=\{r_{i[1]t},\ldots,r_{i[k]t}\}$. We define $\pi_s^+$ ($s=1,2,\ldots$) to be the last element in $\pi_s$, and $\pi_s^-$ to be the set obtained by removing the last element in $\pi_s$. Then, we have $\pi_s=\pi_s^-\cup\pi_s^+$. We define a set of cumulative parents of a variable $r_{it}$ to be a set containing $\{r_{i[1],t},\ldots,r_{i[k],t}\}$ for some $k$. For example, consider the variable $r_{5t}$ in the network in \autoref{fig:easy_DAG}, when we use the topological order $(1,2,3,4,5)$ for pair-copula construction, the sets $\{r_{1,t}\}$, $\{r_{1,t},r_{3,t}\}$ and $\{r_{1,t},r_{3,t},r_{4,t}\}$ are considered as sets of cumulative parents, whereas the sets $\{r_{3,t}\}$, $\{r_{1,t},r_{4,t}\}$, $\{r_{3,t},r_{4,t}\}$ are examples that are not sets of cumulative parents. 
	
	To check whether the term $\pi_s^-$ contains a cumulative set of parents of $\pi_s^+$, we do the following test, and we include the DAG in the reduced space when all conditional distributions in the network pass the test.
	
	\begin{enumerate}[Step 1.]
		\item Take $\tilde{\pi}_s = \mathbf{r}_{\pa(\pi_s^+)} \cap \pi_s^-$, where $\mathbf{r}_{\pa(\pi_s^+)}$ is the set containing the variables that are parents of $\pi_s^+$.
		\item
		\begin{itemize}
			\item  If $\tilde{\pi}_s=\varnothing$, the test is passed.
			\item If $\tilde{\pi}_s\ne\varnothing$, and if $\tilde{\pi}_s$ contains a set of cumulative parents of $\pi_s^+$, then we take $\pi_{s+1}=\tilde\pi_{s}$ and update $s$ {to} $s+1$. Go back to Step 1.
			\item If $\tilde{\pi}_s\ne\varnothing$, and $\tilde{\pi}_s$ does not contain a set of cumulative parents of $\pi_s^+$, then we conclude that the calculation of $F^{[t]}(r_{i[k],t}|r_{i[1],t},\ldots,r_{i[k-1],t})$ requires numerical integration and the test is failed.
		\end{itemize}
	\end{enumerate}
	We reduce the DAG space by considering only those networks passing the above test. {Even though there exist some cases where we still need to calculate conditional distributions using numerical integration, the possibility of involving numerical integration in the likelihood calculation has been greatly reduced.}
	
	For example, consider the DAG in \autoref{fig:easy_DAG}, which involves the copulas 
	$$
	c^{[t]}_{2,1}(\cdot,\cdot),c^{[t]}_{4,1}(\cdot,\cdot),c^{[t]}_{4,3|1}(\cdot,\cdot),c^{[t]}_{5,1}(\cdot,\cdot),c^{[t]}_{5,3|1}(\cdot,\cdot),c^{[t]}_{5,4|3,1}(\cdot,\cdot),
	$$
	we first try to check, directly without using the above algorithm, if we can compute a copula density directly without using integration. We consider the second conditional distribution in\\ $c_{5,4|3,1}^{[t]}(F^{[t]}(r_{5,t}|r_{3,t},r_{1,t}),F^{[t]}(r_{4,t}|r_{3,t},r_{1,t}))$:
	$$
	F^{[t]}(r_{4t}|r_{3,t},r_{1,t})=h_{4,3|1}^{[t]}(F^{[t]}(r_{4,t}|r_{1,t}),F^{[t]}(r_{3,t}|r_{1,t})).
	$$ 
	Note that $F^{[t]}(r_{4,t}|r_{1,t})=h_{4,1}^{[t]}(F^{[t]}(r_{4,t}),F^{[t]}(r_{1,t}))$ can be directly computed using the $h$ function, and $F^{[t]}(r_{3,t}|r_{1,t})=F^{[t]}(r_{3,t})$ can be directly computed from the marginal distribution as $r_{3,t}$ and $r_{1,t}$ are independent.
	
	If we use the algorithm above, the copula $c_{5,4|3,1}^{[t]}(F^{[t]}(r_{5,t}|r_{3,t},r_{1,t}),F^{[t]}(r_{4,t}|r_{3,t},r_{1,t}))$ corresponds to $i=5$ and $k=3$. Then, $\pi_1=\{r_{5[1],t},r_{5[2],t},r_{5[3],t}\}=\{r_{1,t},r_{3,t},r_{4,t}\}$, $\pi_1^+=\{r_{4,t}\}$, and $\pi_1^-=\{r_{1,t},r_{3,t}\}$. The second conditional distribution in the copula can be written as $F^{[t]}(\pi_1^+|\pi_1^-)$. 
	\begin{enumerate}
	\item For $s=1$, we take $\tilde\pi_1=\mathbf{r}_{\pa(\pi_1^+)}\cap \pi_1^-=\mathbf{r}_{\pa(r_{4,t})}\cap \{r_{1,t},r_{3,t}\}=\{r_{1,t},r_{3,t}\}\cap\{r_{1,t},r_{3,t}\}=\{r_{1,t},r_{3,t}\}\ne \varnothing$. Since $\tilde\pi_1$ is cumulative parents of $\pi_1^+=r_{4,t}$, we take $\pi_2=\tilde \pi_1$ and update $s=2$. 
	\item  Now, $\pi_2=\{r_{1,t},r_{3,t}\}$, $\pi_2^+=\{r_{3,t}\}$ and $\pi_2^-=\{r_{1,t}\}$. Then, $\tilde\pi_2=\mathbf{r}_{\pa(r_{3t})}\cap \{r_{1,t}\}=\varnothing \cap \{r_{1,t}\}=\varnothing$. Then, $F^{[t]}(r_{4,t}|r_{3,t},r_{1,t})$ passes the test.
	\end{enumerate}
	
	We also give a counterexample. Consider the network in \autoref{fig:example_difficult_conditional_probability}. Consider the copula $c_{5,4|3,1}^{[t]}(F^{[t]}(r_{5t}|r_{3t},r_{1t}),$ $F^{[t]}(r_{4t}|r_{3t},r_{1t}))$. We focus on the second argument $F^{[t]}(r_{4t}|r_{3t},r_{1t}))$. Using the algorithm, the copula corresponds to $i=5$ and $k=3$. Then, $\pi_1=\{r_{5[1],t},r_{5[2],t},r_{5[3],t}\}=\{r_{1,t},r_{3,t},r_{4,t}\}$, $\pi_1^+=\{r_{4,t}\}$, and $\pi_1^-=\{r_{1t},r_{3t}\}$. For $s=1$, we take $\tilde\pi_1=\mathbf{r}_{\pa(\pi_1^+)}\cap \pi_1^-=\{r_{2,t},r_{3,t}\}\cap\{r_{1,t},r_{3,t}\}=\{r_{3,t}\}\ne \varnothing$. However, the parents of $r_{4t}$ are $\{r_{2,t},r_{3,t}\}$. Then, $\tilde\pi_1$ is not a set of cumulative parents of $\pi_1^+$, and thus we exclude the DAG in the reduced space. This part corresponds to \eqref{eqt:difficult_integration}, where we need to compute the integration to marginalize out $r_{2t}$. Without a set of cumulative parents, the computation must require the use of integration.
	

	\section{Parameter estimation results in simulation (S2)}
	\setcounter{figure}{0} 
	\setcounter{table}{0} 
	\subsection{Parameters in the DAG copulas}
	\label{section:parameters_in_the_DAG_copulas}
	The estimation results for simulation (S2) are presented in \autoref{tab:m_10_dag_median_a_b} for $a_{xy|z}$ and $b_{xy|z}$, and \autoref{tab:m_10_dag_median_p_v} for $\bar\varphi_{xy|z}$ and $v_{xy|z}$. Most of the true values are contained in the 95\% credible intervals. The estimates are also close to the true values for most of the parameters. The performances are similar to the case in simulation (S1). It can be concluded that the MCMC estimation method in Section \ref{section:DAG_copula_MCMC_estimation} is reliable.
	
	\begin{table}[H]
	\centering
	\scriptsize
	\begin{tabular}{llllllll}
		Parameter & True value & Mean of estimates & Mean MAE & Median of estimate & Median MAE & $q_{0.05}$ & $q_{0.95}$ \\ \hline
		$a_{10,3}$ & $\mathbf{0.05^*}$ & 0.0439 & 0.0221 & 0.043 & 0.018 & 0.0004 & 0.0797 \\
		$a_{10,9|3}$ & $\mathbf{0.03^*}$ & 0.0123 & 0.0244 & 0.0011 & 0.0276 & 0 & 0.0743 \\
		$a_{2,1}$ & $\mathbf{0.04^*}$ & 0.0417 & 0.0217 & 0.0406 & 0.017 & 0.0052 & 0.0791 \\
		$a_{3,1}$ & $\mathbf{0.06^*}$ & 0.0609 & 0.0202 & 0.0587 & 0.0154 & 0.0295 & 0.0933 \\
		$a_{3,2|1}$ & $\mathbf{0.02^*}$ & 0.0245 & 0.029 & 0.0161 & 0.0215 & 0 & 0.0805 \\
		$a_{4,1}$ & $\mathbf{0.03^*}$ & 0.0275 & 0.0283 & 0.0207 & 0.0227 & 0 & 0.0732 \\
		$a_{4,2|1}$ & $\mathbf{0.03^*}$ & 0.032 & 0.0343 & 0.0265 & 0.0229 & 0 & 0.1115 \\
		$a_{5,1}$ & $\mathbf{0.06^*}$ & 0.0573 & 0.0284 & 0.0567 & 0.0222 & 0.0123 & 0.1043 \\
		$a_{5,3|1}$ & $\mathbf{0.03^*}$ & 0.026 & 0.028 & 0.0185 & 0.0224 & 0.0001 & 0.0827 \\
		$a_{5,4|1,3}$ & $\mathbf{0.09^*}$ & 0.0817 & 0.0432 & 0.0826 & 0.0344 & 0.0138 & 0.1538 \\
		$a_{6,5}$ & $\mathbf{0.08^*}$ & 0.0596 & 0.0466 & 0.0541 & 0.0415 & 0.0002 & 0.1512 \\
		$a_{7,1}$ & $\mathbf{0.09^*}$ & 0.0845 & 0.0209 & 0.0831 & 0.0169 & 0.0538 & 0.1188 \\
		$a_{7,2|1}$ & $\mathbf{0.05^*}$ & 0.051 & 0.0382 & 0.0461 & 0.0282 & 0.0007 & 0.1096 \\
		$a_{7,5|1,2}$ & $\mathbf{0.09^*}$ & 0.0845 & 0.0376 & 0.0836 & 0.0278 & 0.0314 & 0.1393 \\
		$a_{8,1}$ & $\mathbf{0.07^*}$ & 0.072 & 0.0297 & 0.069 & 0.0223 & 0.0345 & 0.1177 \\
		$a_{8,3|1}$ & $\mathbf{0.06^*}$ & 0.0582 & 0.0338 & 0.0577 & 0.0253 & 0.0015 & 0.1178 \\
		$a_{9,1}$ & $\mathbf{0.07^*}$ & 0.0685 & 0.0225 & 0.0686 & 0.0173 & 0.0339 & 0.1062 \\
		$a_{9,4|1}$ & $\mathbf{0.02^*}$ & 0.0143 & 0.018 & 0.0086 & 0.0155 & 0 & 0.0465 \\
		$a_{9,7|1,4}$ & $\mathbf{0.02^*}$ & 0.0199 & 0.025 & 0.0128 & 0.0172 & 0 & 0.074 \\ \hline
		$b_{10,3}$ & $\mathbf{0.92^*}$ & 0.8629 & 0.1962 & 0.9196 & 0.0812 & 0.4181 & 0.9614 \\
		$b_{10,9|3}$ & $\mathbf{0.9^*}$ & 0.5729 & 0.4059 & 0.7276 & 0.3801 & 0.0001 & 0.9993 \\
		$b_{2,1}$ & $\mathbf{0.86^*}$ & 0.7981 & 0.185 & 0.842 & 0.1066 & 0.3114 & 0.9503 \\
		$b_{3,1}$ & $\mathbf{0.91^*}$ & 0.8974 & 0.0681 & 0.9123 & 0.0339 & 0.8237 & 0.9508 \\
		$b_{3,2|1}$ & $\mathbf{0.95^*}$ & 0.7064 & 0.3385 & 0.8519 & 0.2598 & 0.001 & 0.9946 \\
		$b_{4,1}$ & $\mathbf{0.92^*}$ & 0.7247 & 0.3232 & 0.8879 & 0.2292 & 0.0243 & 0.9953 \\
		$b_{4,2|1}$ & $\mathbf{0.95^*}$ & 0.7817 & 0.3202 & 0.9366 & 0.1857 & 0.0208 & 0.9933 \\
		$b_{5,1}$ & $\mathbf{0.86^*}$ & 0.7725 & 0.2147 & 0.8378 & 0.1238 & 0.1073 & 0.9369 \\
		$b_{5,3|1}$ & $\mathbf{0.94^*}$ & 0.7385 & 0.324 & 0.9026 & 0.224 & 0.0028 & 0.9947 \\
		$b_{5,4|1,3}$ & $\mathbf{0.81^*}$ & 0.7245 & 0.2275 & 0.7883 & 0.1429 & 0.0438 & 0.9443 \\
		$b_{6,5}$ & $\mathbf{0.84^*}$ & 0.6638 & 0.3074 & 0.7869 & 0.2269 & 0.004 & 0.9621 \\
		$b_{7,1}$ & $\mathbf{0.86^*}$ & 0.8457 & 0.0742 & 0.8549 & 0.0376 & 0.7567 & 0.9115 \\
		$b_{7,2|1}$ & $\mathbf{0.86^*}$ & 0.7247 & 0.2877 & 0.8401 & 0.1923 & 0.0195 & 0.9834 \\
		$b_{7,5|1,2}$ & $\mathbf{0.84^*}$ & 0.786 & 0.1899 & 0.8399 & 0.1043 & 0.3452 & 0.9407 \\
		$b_{8,1}$ & $\mathbf{0.86^*}$ & 0.8085 & 0.175 & 0.857 & 0.0863 & 0.4541 & 0.9301 \\
		$b_{8,3|1}$ & $\mathbf{0.86^*}$ & 0.7497 & 0.2502 & 0.8415 & 0.1544 & 0.0419 & 0.9521 \\
		$b_{9,1}$ & $\mathbf{0.84^*}$ & 0.8152 & 0.1225 & 0.8359 & 0.0649 & 0.679 & 0.924 \\
		$b_{9,4|1}$ & $\mathbf{0.93^*}$ & 0.7019 & 0.3498 & 0.8918 & 0.2601 & 0.0124 & 0.9952 \\
		$b_{9,7|1,4}$ & $\mathbf{0.96^*}$ & 0.7551 & 0.3261 & 0.9283 & 0.2201 & 0.0124 & 0.996
	\end{tabular}
	\caption{The true values, the mean, mean absolute error (MAE) of the mean, the median, the MAE of the median, the 5th quantile ($q_{0.05}$), and the 95th quantile ($q_{0.95}$) of the posterior median estimates of $a_{xy|z}$ and $b_{xy|z}$ out of the 200 replications for simulation (S2).}
	\label{tab:m_10_dag_median_a_b}
	\end{table}
	\begin{table}[H]
	\centering
	\scriptsize
	\begin{tabular}{llllllll}
		Parameter & True value & Mean of estimates & Mean MAE & Median of estimate & Median MAE & $q_{0.05}$ & $q_{0.95}$ \\ \hline
		$\bar\varphi_{10,3}$ & $\mathbf{0.24^*}$ & 0.2831 & 0.1249 & 0.2781 & 0.0965 & 0.0987 & 0.4595 \\
		$\bar\varphi_{10,9|3}$ & $-0.41$ & $-$0.2534 & 0.0469 & $-$0.256 & 0.1573 & $-$0.327 & $-$0.186 \\
		$\bar\varphi_{2,1}$ & $\mathbf{-0.78^*}$ & $-$0.7702 & 0.0283 & $-$0.7711 & 0.0232 & $-$0.8198 & $-$0.7228 \\
		$\bar\varphi_{3,1}$ & $\mathbf{-0.36^*}$ & $-$0.355 & 0.1126 & $-$0.3551 & 0.0847 & $-$0.5275 & $-$0.1755 \\
		$\bar\varphi_{3,2|1}$ & $\mathbf{-0.02^*}$ & $-$0.0202 & 0.0807 & $-$0.0187 & 0.06 & $-$0.1422 & 0.0986 \\
		$\bar\varphi_{4,1}$ & $\mathbf{0.26^*}$ & 0.2593 & 0.0645 & 0.2602 & 0.0495 & 0.1531 & 0.3621 \\
		$\bar\varphi_{4,2|1}$ & $\mathbf{0.02^*}$ & 0.0185 & 0.1042 & 0.0203 & 0.074 & $-$0.1257 & 0.1474 \\
		$\bar\varphi_{5,1}$ & $\mathbf{0.54^*}$ & 0.5194 & 0.0548 & 0.5168 & 0.0473 & 0.4325 & 0.6079 \\
		$\bar\varphi_{5,3|1}$ & $\mathbf{0.12^*}$ & 0.1144 & 0.0722 & 0.1122 & 0.0559 & 0.0094 & 0.2255 \\
		$\bar\varphi_{5,4|1,3}$ & $\mathbf{-0.26^*}$ & $-$0.2428 & 0.0687 & $-$0.2419 & 0.0558 & $-$0.3642 & $-$0.1406 \\
		$\bar\varphi_{6,5}$ & $\mathbf{0.29^*}$ & 0.195 & 0.0811 & 0.1994 & 0.1055 & 0.0771 & 0.3028 \\
		$\bar\varphi_{7,1}$ & $\mathbf{-0.47^*}$ & $-$0.4649 & 0.0742 & $-$0.468 & 0.0598 & $-$0.5728 & $-$0.3435 \\
		$\bar\varphi_{7,2|1}$ & $\mathbf{-0.42^*}$ & $-$0.4121 & 0.0632 & $-$0.4126 & 0.0489 & $-$0.5067 & $-$0.3083 \\
		$\bar\varphi_{7,5|1,2}$ & $\mathbf{0.03^*}$ & 0.0127 & 0.1103 & 0.0185 & 0.0777 & $-$0.1545 & 0.1786 \\
		$\bar\varphi_{8,1}$ & $\mathbf{0.43^*}$ & 0.4018 & 0.0776 & 0.4074 & 0.0638 & 0.2633 & 0.532 \\
		$\bar\varphi_{8,3|1}$ & $\mathbf{0.5^*}$ & 0.489 & 0.0651 & 0.4921 & 0.0527 & 0.3823 & 0.6006 \\
		$\bar\varphi_{9,1}$ & $\mathbf{-0.69^*}$ & $-$0.6862 & 0.0413 & $-$0.6894 & 0.0327 & $-$0.7534 & $-$0.6112 \\
		$\bar\varphi_{9,4|1}$ & $\mathbf{0.2^*}$ & 0.1871 & 0.0506 & 0.1869 & 0.04 & 0.1082 & 0.2713 \\
		$\bar\varphi_{9,7|1,4}$ & $0.74$ & 0.5693 & 0.0704 & 0.5564 & 0.1725 & 0.4759 & 0.7001 \\ \hline
		$v_{10,3}$ & $\mathbf{7.9^*}$ & 9.9883 & 3.1891 & 9.6558 & 2.8627 & 5.7493 & 15.934 \\
		$v_{10,9|3}$ & $6.67$ & 14.649 & 3.764 & 14.3274 & 8.0097 & 8.9711 & 21.157 \\
		$v_{2,1}$ & $\mathbf{7.97^*}$ & 7.9942 & 2.2888 & 7.7657 & 1.8153 & 4.9135 & 12.4083 \\
		$v_{3,1}$ & $\mathbf{7.91^*}$ & 8.6289 & 2.7028 & 8.0335 & 1.9919 & 5.6216 & 14.5503 \\
		$v_{3,2|1}$ & $\mathbf{6.19^*}$ & 6.6826 & 2.0876 & 6.2403 & 1.5549 & 4.1974 & 10.8303 \\
		$v_{4,1}$ & $\mathbf{7.27^*}$ & 8.0482 & 2.4939 & 7.5606 & 1.8723 & 5.0499 & 12.1542 \\
		$v_{4,2|1}$ & $\mathbf{9.98^*}$ & 9.9089 & 3.3581 & 9.082 & 2.6316 & 5.7849 & 16.4214 \\
		$v_{5,1}$ & $\mathbf{5.41^*}$ & 5.8013 & 1.5465 & 5.4467 & 1.1149 & 3.8427 & 8.8646 \\
		$v_{5,3|1}$ & $\mathbf{5.26^*}$ & 6.0324 & 1.6404 & 5.6626 & 1.3035 & 4.0945 & 8.8652 \\
		$v_{5,4|1,3}$ & $\mathbf{7.47^*}$ & 8.14 & 2.2743 & 7.8704 & 1.7779 & 5.0633 & 13.045 \\
		$v_{6,5}$ & $\mathbf{5.27^*}$ & 8.34 & 2.4175 & 7.6211 & 3.1237 & 5.2655 & 12.8108 \\
		$v_{7,1}$ & $\mathbf{9.75^*}$ & 10.0117 & 3.058 & 9.1846 & 2.4324 & 6.2343 & 15.6522 \\
		$v_{7,2|1}$ & $\mathbf{8.17^*}$ & 8.7747 & 2.8075 & 8.2079 & 2.0742 & 5.1824 & 13.6875 \\
		$v_{7,5|1,2}$ & $\mathbf{6.35^*}$ & 8.0564 & 2.6115 & 7.6028 & 2.271 & 4.7868 & 12.7598 \\
		$v_{8,1}$ & $\mathbf{9.98^*}$ & 10.0132 & 3.3093 & 9.4096 & 2.5236 & 6.014 & 16.2632 \\
		$v_{8,3|1}$ & $\mathbf{6.79^*}$ & 7.6349 & 2.6396 & 7.0186 & 1.9867 & 4.4484 & 12.7467 \\
		$v_{9,1}$ & $\mathbf{6.86^*}$ & 7.1971 & 1.9426 & 6.7725 & 1.5081 & 4.8446 & 10.8707 \\
		$v_{9,4|1}$ & $\mathbf{6.65^*}$ & 7.7157 & 2.4767 & 7.2372 & 1.9109 & 4.5809 & 12.0884 \\
		$v_{9,7|1,4}$ & $\mathbf{5.52^*}$ & 8.6611 & 2.6835 & 8.2426 & 3.2541 & 4.9682 & 13.7083
	\end{tabular}
	\caption{The true values, the mean, mean absolute error (MAE) of the mean, the median, the MAE of the median, the 5th quantile ($q_{0.05}$), and the 95th quantile ($q_{0.95}$) of the posterior median estimates of $v_{xy|z}$ and $\bar\varphi_{xy|z}$ out of the 200 replications for simulation (S2).}
	\label{tab:m_10_dag_median_p_v}
	\end{table}
	
	\subsection{Parameters in the stock copula}
	\label{section:appendix_simulation_S2_stock_parameters}
\begin{figure}[H]
	\centering
	\includegraphics[width=13cm]{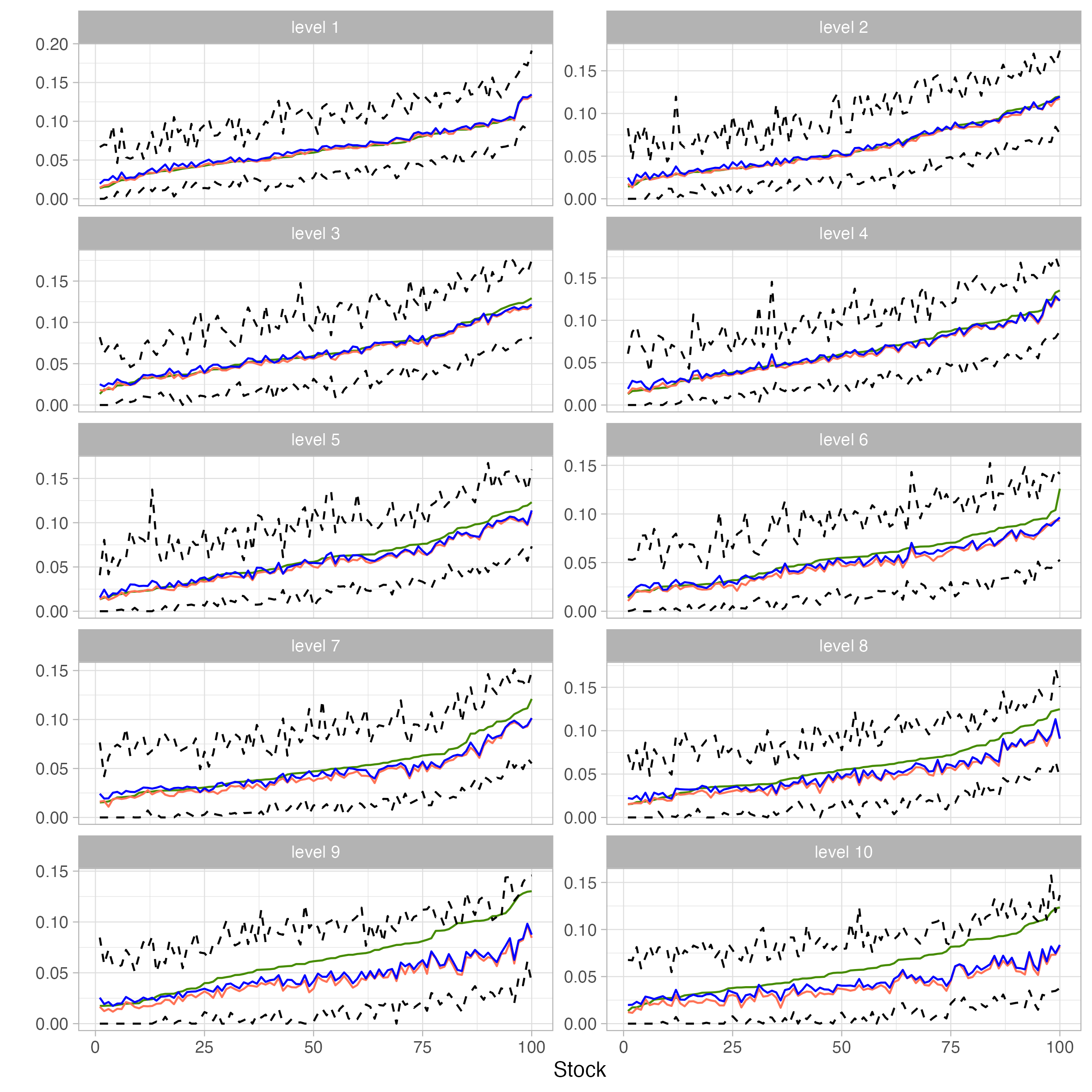}
	\caption{A graph summarizing the parameter estimates for $a_{xy|z}$ by level in simulation (S2). The copulas are in ascending order based on the true values of $a_{xy|z}$ in each level for a better visualization. The green curves indicate true values, the blue curves indicate mean estimates, the orange curves indicate the median estimates, and the pair of dashed curves represent the 90\% credible intervals of the sequential estimates out of the 200 replications.}
	\label{fig:axy_stock_m_10}
	\end{figure}

\begin{figure}[H]
	\centering
	\includegraphics[width=13cm]{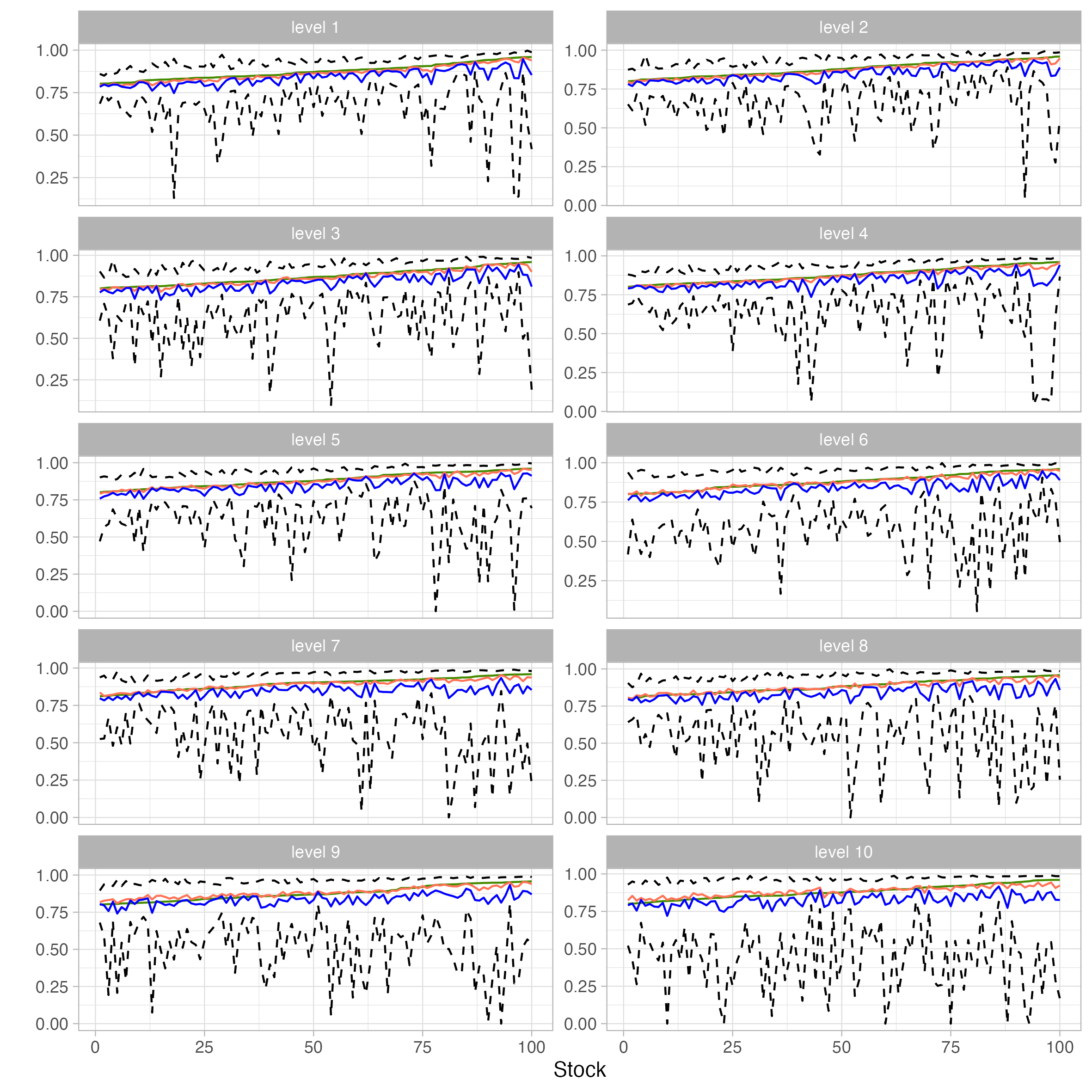}
	\caption{A graph summarizing the parameter estimates for $b_{xy|z}$ by level in simulation (S2). The copulas are in ascending order based on the true values of $b_{xy|z}$ in each level for a better visualization. The green curves indicate true values, the blue curves indicate mean estimates, the orange curves indicate the median estimates, and the pair of dashed curves represent the 90\% credible intervals of the sequential estimates out of the 200 replications.}
	\label{fig:bxy_stock_m_10}
	\end{figure}

\begin{figure}[H]
	\centering
	\includegraphics[width=13cm]{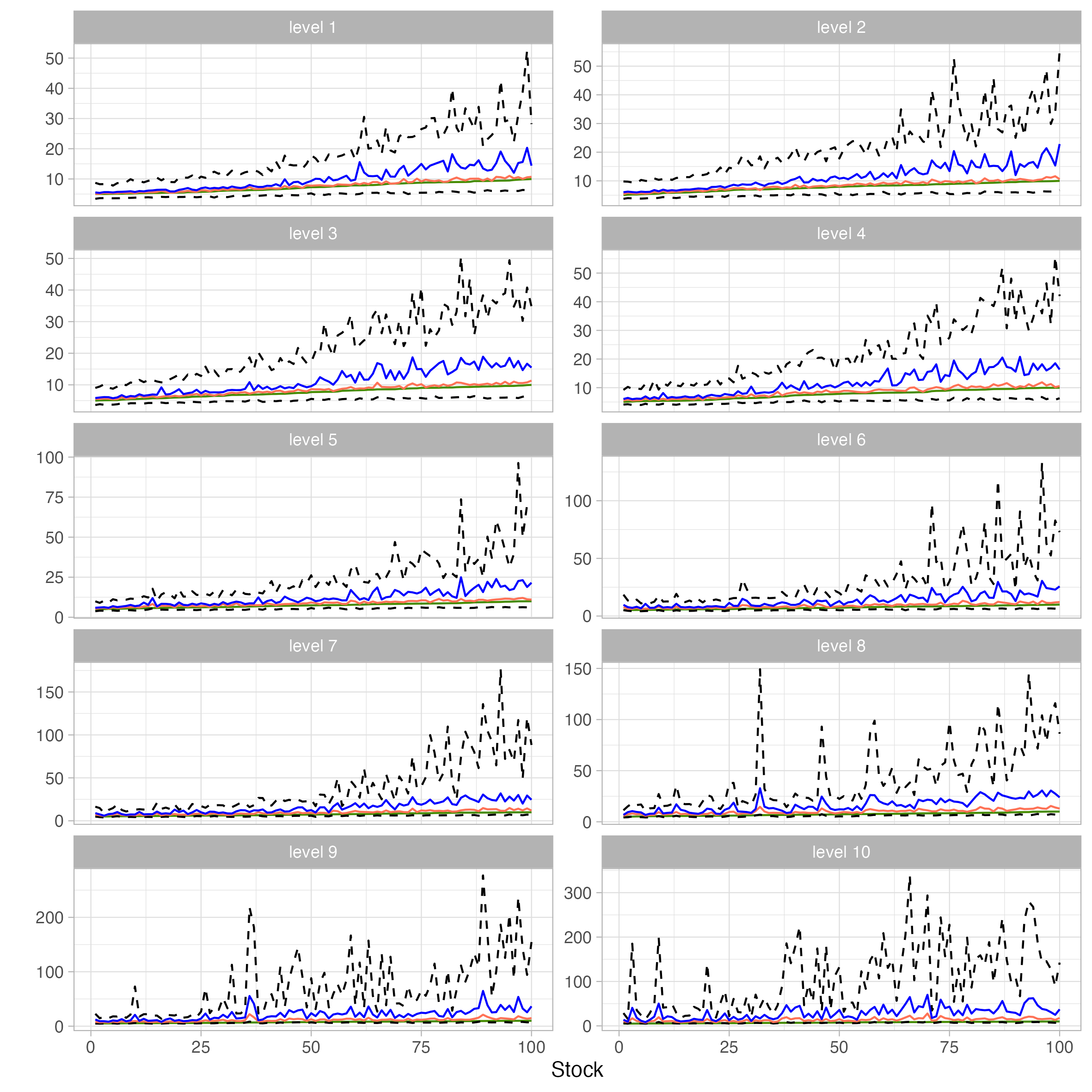}
	\caption{A graph summarizing the parameter estimates for $v_{xy|z}$ by level in simulation (S2). The copulas are in ascending order based on the true values of $v_{xy|z}$ in each level for a better visualization. The green curves indicate true values, the blue curves indicate mean estimates, the orange curves indicate the median estimates, and the pair of dashed curves represent the 90\% credible intervals of the sequential estimates out of the 200 replications.}
	\label{fig:vxy_stock_m_10}
	\end{figure}

\begin{figure}[H]
	\centering
	\includegraphics[width=13cm]{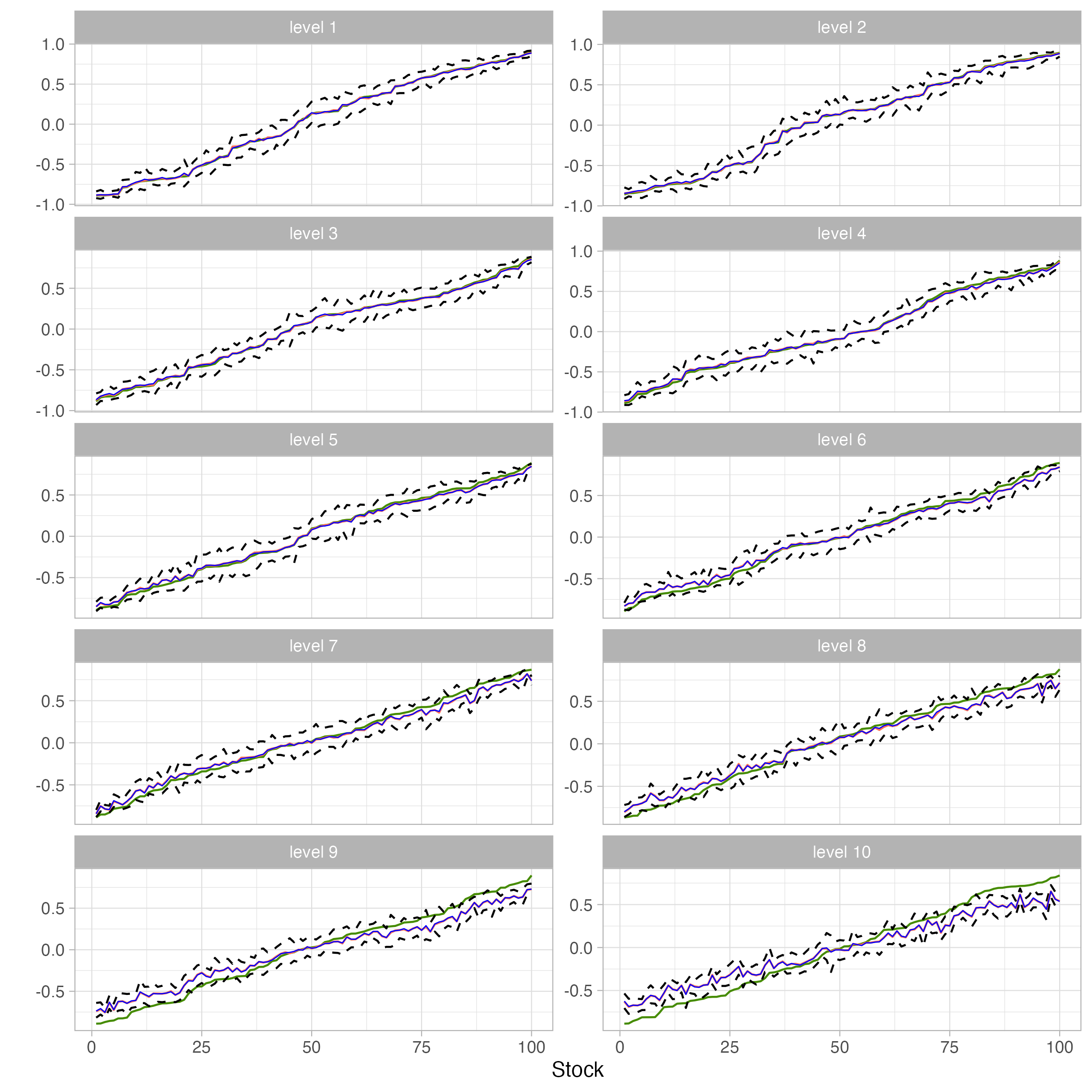}
	\caption{A graph summarizing the parameter estimates for $\bar\varphi_{xy|z}$ by level in simulation (S2). The copulas are in ascending order based on the true values of $\bar\varphi_{xy|z}$ in each level for a better visualization. The green curves indicate true values, the blue curves indicate mean estimates, the orange curves indicate the median estimates, and the pair of dashed curves represent the 90\% credible intervals of the sequential estimates out of the 200 replications.}
	\label{fig:phibar_stock_m_10}
	\end{figure}

	This section gives the sequential estimation results for simulation (S2). \autoref{fig:axy_stock_m_10}, \autoref{fig:bxy_stock_m_10}, \autoref{fig:vxy_stock_m_10} and \autoref{fig:phibar_stock_m_10} are four graphs summarizing the parameter estimation for $a_{xy|z}$, $b_{xy|z}$, $v_{xy|}z$, and $\bar\varphi_{xy|z}$. These graphs are similar to the graphs in \autoref{fig:axy_stock_m_8}, \autoref{fig:bxy_stock_m_8}, \autoref{fig:vxy_stock_m_8} and \autoref{fig:phibar_stock_m_8} presented in Section \ref{section:parameters_in_the_stock_copulas}, and the observations regarding the estimation results are similar to those in Section \ref{section:parameters_in_the_stock_copulas}.
	
	\newpage
	\section{Comparing the sequential estimator and the posterior mean/median estimators}
	\label{section:comparing_sequential_estimates}
	\setcounter{figure}{0} 
	\setcounter{table}{0} 
	In our empirical study, we conduct a moving-window study with a total of 173 windows, in each of which we include 92 stocks. Estimating stock copulas using MCMC sampling is infeasible in this case. For $m=10$, $p=92$ and $T=750$, for each stock, it took 6 hours to run 20,000 iterations in the MCMC sampling. The sequential estimation, however, only takes 30 seconds.
	
\begin{figure}[H]
	\centering
	\begin{subfigure}{0.7\textwidth}
		\centering
		\includegraphics[width=7.5cm]{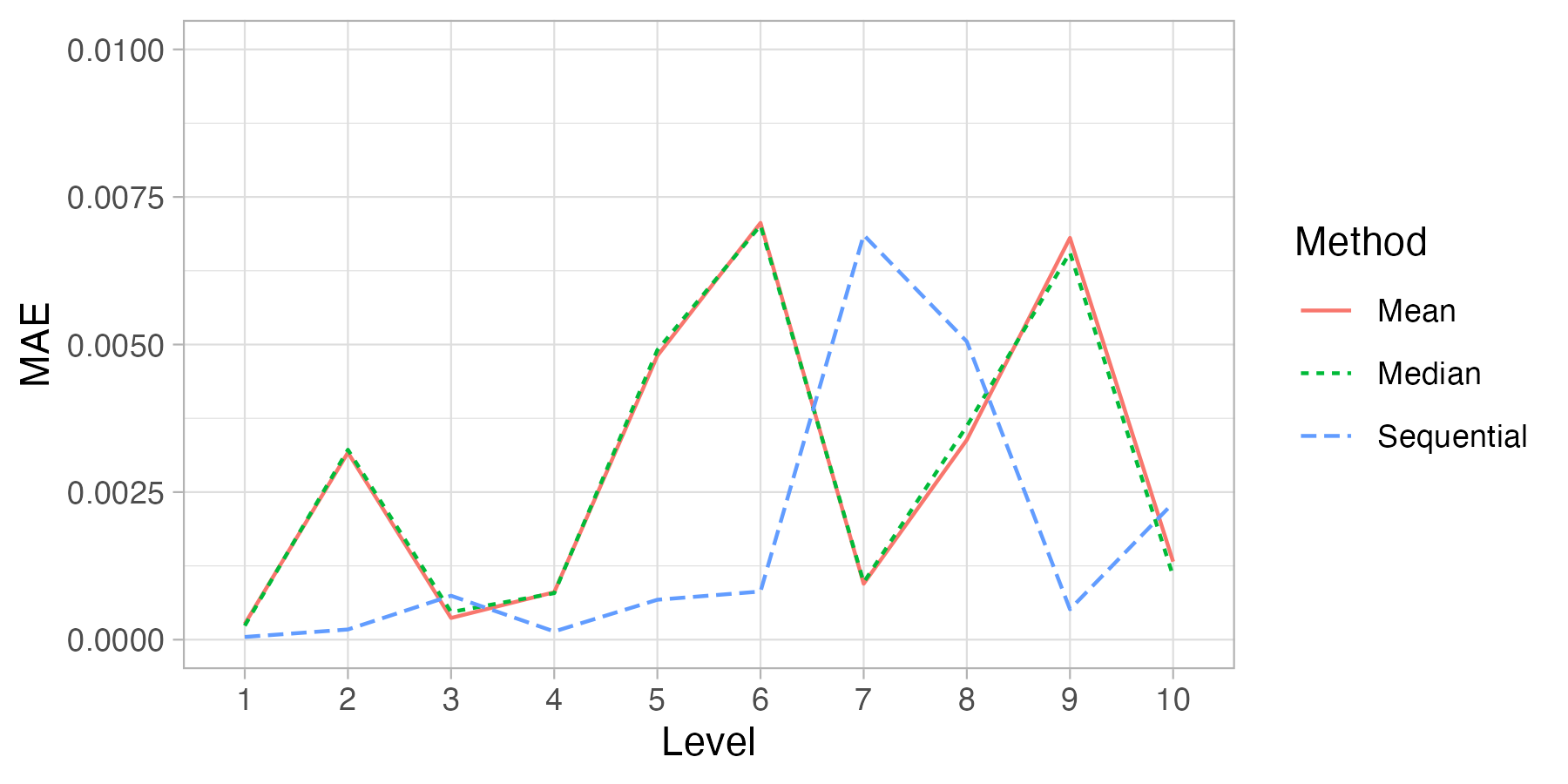}
		\caption{The average MAEs for $\bar\varphi_{xy|z}$.}
		\label{fig:phibar_mae}
	\end{subfigure}
	
	\begin{subfigure}{0.7\textwidth}
		\centering
		\includegraphics[width=7.5cm]{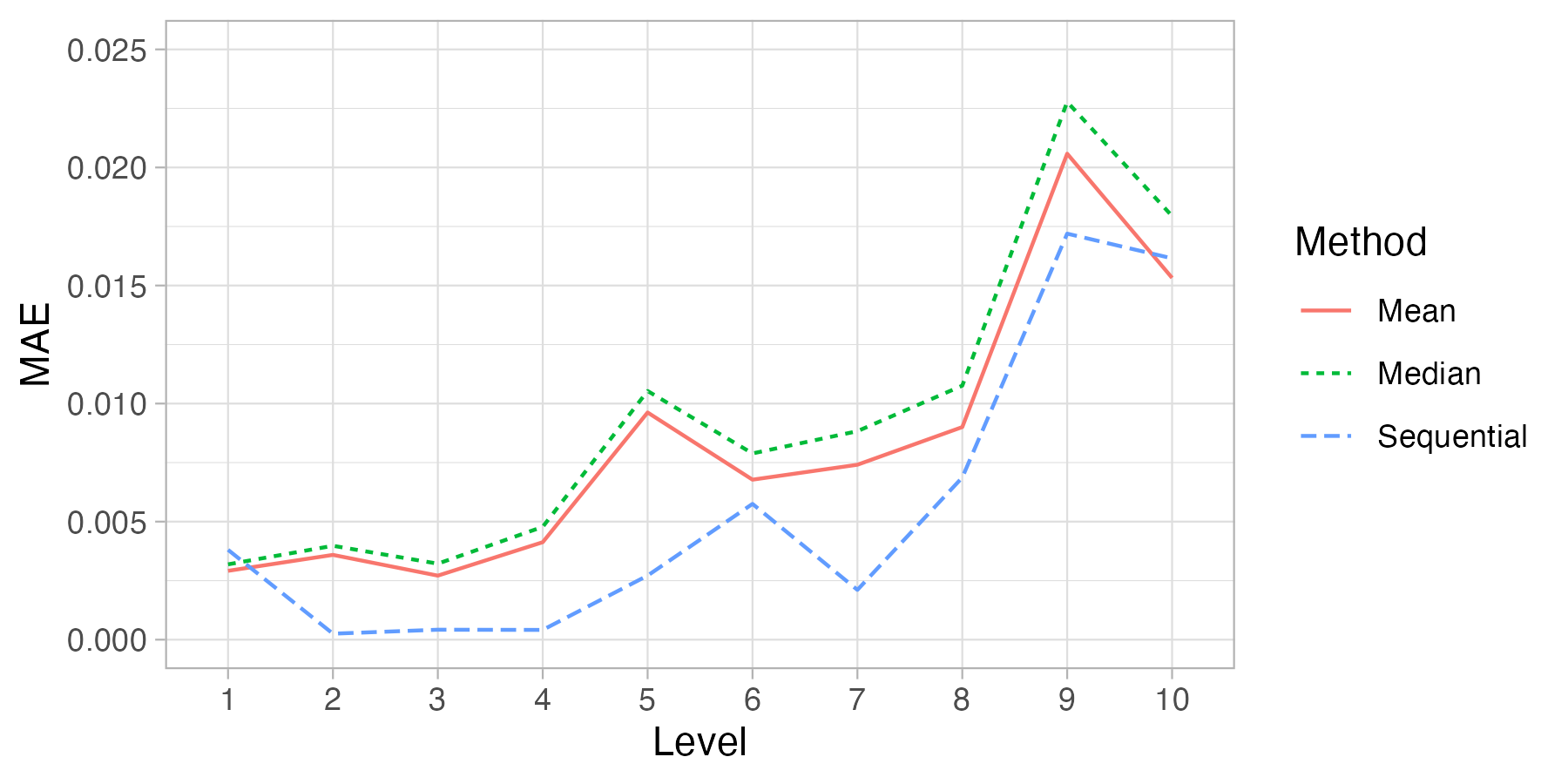}
		\caption{The average MAEs for $a_{xy|z}$.}
		\label{fig:axy_mae}
	\end{subfigure}
	
	\begin{subfigure}{0.7\textwidth}
		\centering
		\includegraphics[width=7.5cm]{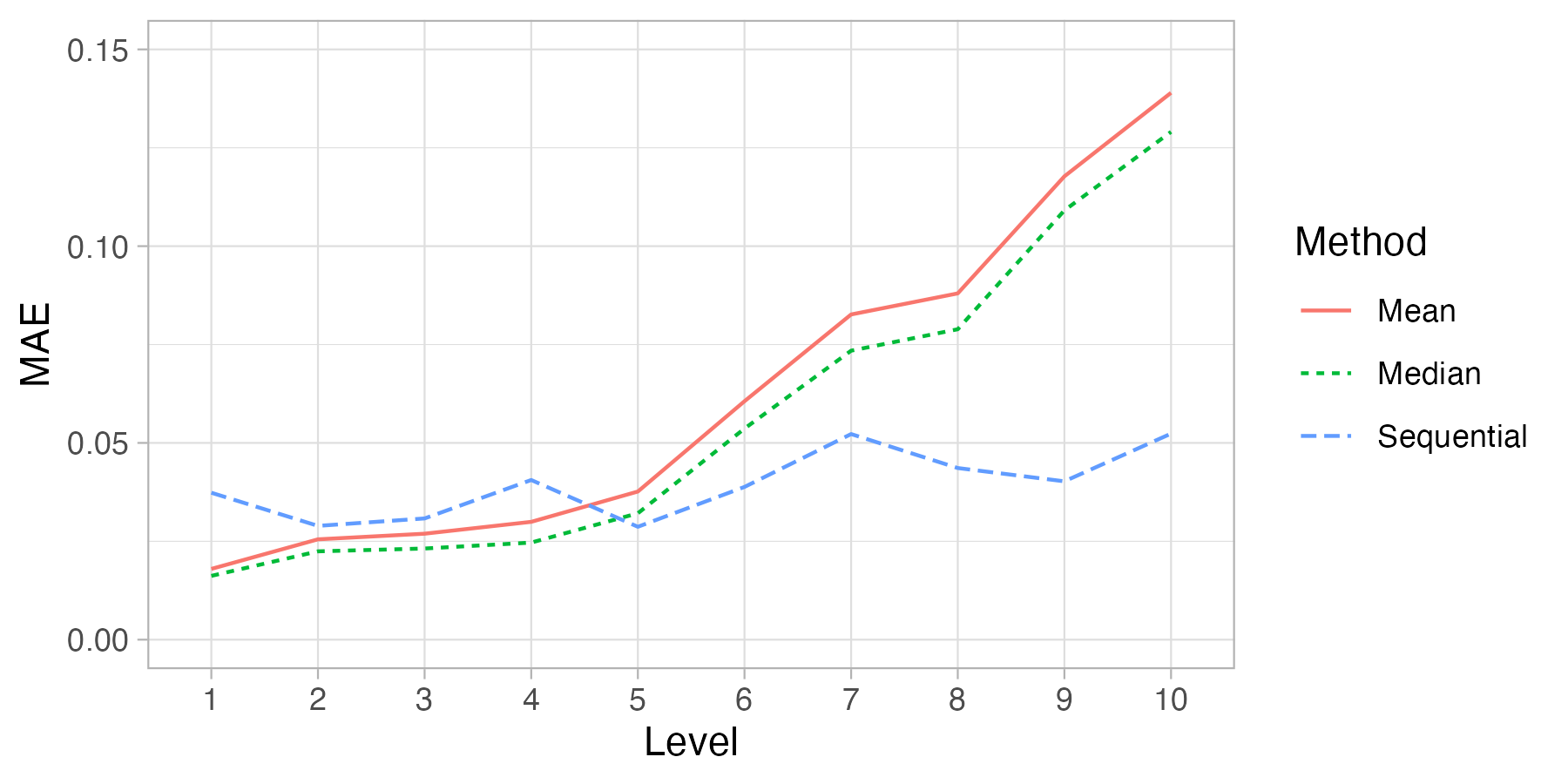}
		\caption{The average MAEs for $b_{xy|z}$.}
		\label{fig:bxy_mae}
	\end{subfigure}
	
	\begin{subfigure}{0.7\textwidth}
		\centering
		\includegraphics[width=7.5cm]{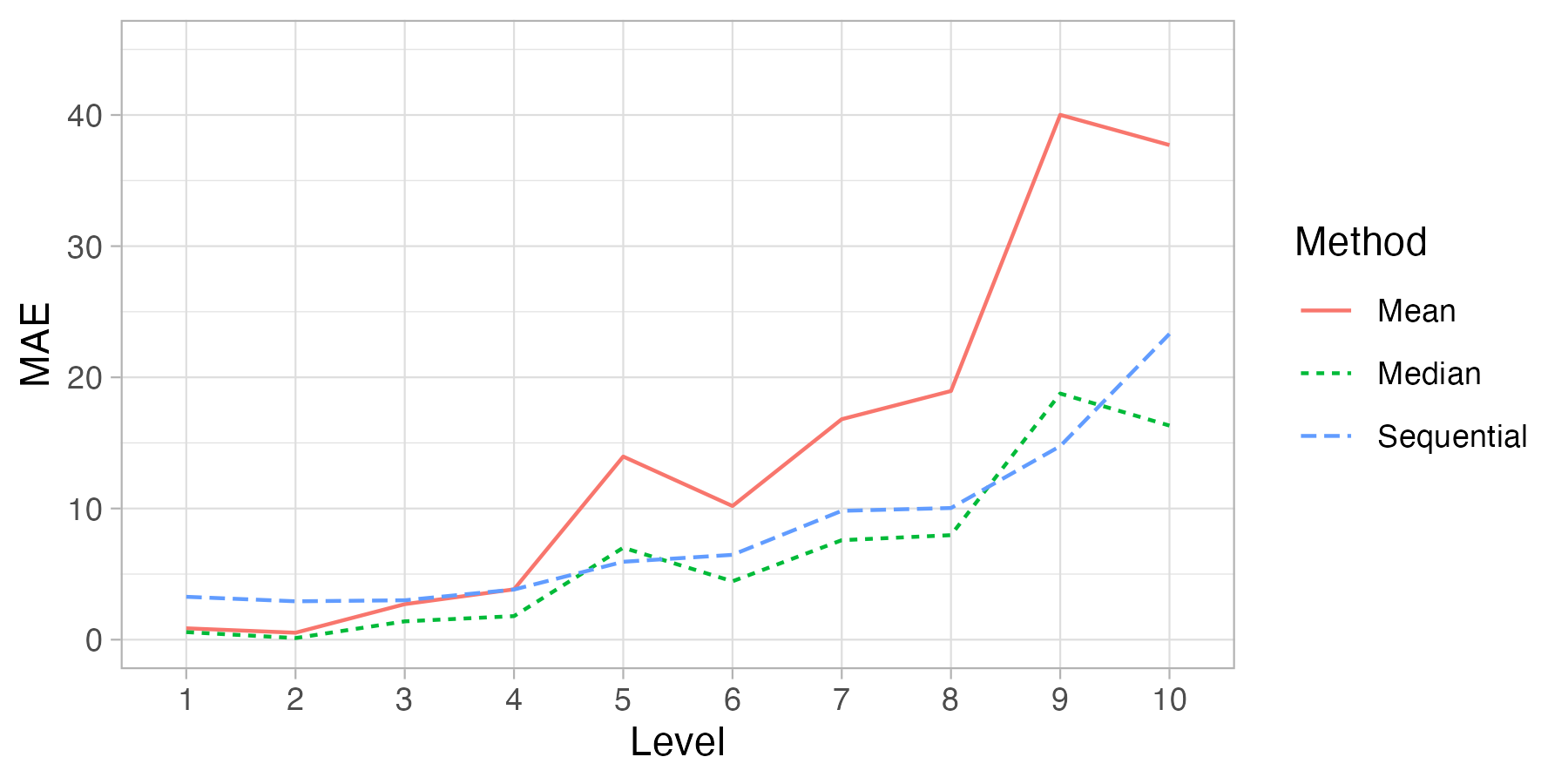}
		\caption{The average MAEs for $v_{xy|z}$.}
		\label{fig:vxy_mae}
	\end{subfigure}
	\caption{The time series plots of the average MAEs of the sequential estimates, the posterior mean and posterior median estimates of (a) $\overline\varphi_{xy|z}$, (b) $a_{xy|z}$, (c) $b_{xy|z}$, and (d) $v_{xy|z}$, plotted by level.}
	\label{fig:all_mae_seq}
	\end{figure}

	In this section, we compare the sequential estimates to the posterior mean/median estimates. We conduct 20 replications with different data sets generated from the GC-GARCH model with the same underlying network in \autoref{fig:true_graph_S2} as in simulation (S2). The true parameters are also fixed across replications. We calculate the MAEs of the estimates of all 100 stocks, by level. \autoref{fig:all_mae_seq} shows the time series plots of the average MAE, within the same level, of the sequential estimates and the posterior mean/mode/median estimates of (a) $\bar\varphi_{xy|z}$, (b) $a_{xy|z}$, (c) $b_{xy|z}$, and (d) $v_{xy|z}$. The x-axes are the levels.
	
	\begin{itemize}
	\item In (a), we observe that the MAEs of the sequential estimates of $\bar\varphi_{xy|z}$ (dashed blue line) are similar to the MAEs of the mean and median MCMC estimates (respectively the solid red and dotted green lines).
	\item In (b), the MAEs for $a_{xy|z}$ using sequential estimation are smaller than the MAEs of the MCMC estimates in most of the levels.
	\item In (c), we observe that the sequential estimates of $b_{xy|z}$ have slightly larger MAEs for the first four levels, and the sequential estimates actually work better for the subsequent levels.
	\item In (d), the MAEs of the sequential estimates and the MCMC median estimates of $v_{xy|z}$ are similar.
	\end{itemize}
	Summarizing the above four points, we observe that sequential estimates work as well as the MCMC estimates.


	\section{List of stocks}
	
	\label{section:List_of_stocks}
	\begin{table}[H]
	\scriptsize
	\centering
	\begin{tabular}{llll}
		Name            & Stock symbol & Name            & Stock symbol \\ \hline
		CKH HOLDINGS    & 0001.HK      & CSPC PHARMA     & 1093.HK      \\
		CLP HOLDINGS    & 0002.HK      & CHINA RES LAND  & 1109.HK      \\
		HK \& CHINA GAS & 0003.HK      & CK ASSET        & 1113.HK      \\
		HSBC HOLDINGS   & 0005.HK      & YANKUANG ENERGY & 1171.HK      \\
		POWER ASSETS    & 0006.HK      & SINO BIOPHARM   & 1177.HK      \\
		HANG SENG BANK  & 0011.HK      & CHINA RAIL CONS & 1186.HK      \\
		HENDERSON LAND  & 0012.HK      & BYD COMPANY     & 1211.HK      \\
		SHK PPT         & 0016.HK      & ABC             & 1288.HK      \\
		NEW WORLD DEV   & 0017.HK      & AIA             & 1299.HK      \\
		GALAXY ENT      & 0027.HK      & PICC GROUP      & 1339.HK      \\
		MTR CORPORATION & 0066.HK      & ICBC            & 1398.HK      \\
		HANG LUNG PPT   & 0101.HK      & PSBC            & 1658.HK      \\
		TSINGTAO BREW   & 0168.HK      & CRRC            & 1766.HK      \\
		GEELY AUTO      & 0175.HK      & GF SEC          & 1776.HK      \\
		ALI HEALTH      & 0241.HK      & CHINA COMM CONS & 1800.HK      \\
		CITIC           & 0267.HK      & PRADA           & 1913.HK      \\
		WH GROUP        & 0288.HK      & COSCO SHIP HOLD & 1919.HK      \\
		CHINA RES BEER  & 0291.HK      & SANDS CHINA LTD & 1928.HK      \\
		TINGYI          & 0322.HK      & CHOW TAI FOOK   & 1929.HK      \\
		SINOPEC CORP    & 0386.HK      & MINSHENG BANK   & 1988.HK      \\
		HKEX            & 0388.HK      & COUNTRY GARDEN  & 2007.HK      \\
		CHINA RAILWAY   & 0390.HK      & ANTA SPORTS     & 2020.HK      \\
		TECHTRONIC IND  & 0669.HK      & CHINA VANKE     & 2202.HK      \\
		CHINA OVERSEAS  & 0688.HK      & SHENZHOU INTL   & 2313.HK      \\
		TENCENT         & 0700.HK      & PING AN         & 2318.HK      \\
		CHINA TELECOM   & 0728.HK      & MENGNIU DAIRY   & 2319.HK      \\
		AIR CHINA       & 0753.HK      & PICC P\&C       & 2328.HK      \\
		CHINA UNICOM    & 0762.HK      & LI NING         & 2331.HK      \\
		LINK REIT       & 0823.HK      & GREATWALL MOTOR & 2333.HK      \\
		PETROCHINA      & 0857.HK      & PRU             & 2378.HK      \\
		XINYI GLASS     & 0868.HK      & SUNNY OPTICAL   & 2382.HK      \\
		ZHONGSHENG HLDG & 0881.HK      & BOC HONG KONG   & 2388.HK      \\
		CNOOC           & 0883.HK      & CPIC            & 2601.HK      \\
		HUANENG POWER   & 0902.HK      & CHINA LIFE      & 2628.HK      \\
		CONCH CEMENT    & 0914.HK      & ENN ENERGY      & 2688.HK      \\
		CHINA LONGYUAN  & 0916.HK      & STANCHART       & 2888.HK      \\
		CCB             & 0939.HK      & ZIJIN MINING    & 2899.HK      \\
		CHINA MOBILE    & 0941.HK      & BANKCOMM        & 3328.HK      \\
		LONGFOR GROUP   & 0960.HK      & CICC            & 3908.HK      \\
		XINYI SOLAR     & 0968.HK      & CM BANK         & 3968.HK      \\
		SMIC            & 0981.HK      & BANK OF CHINA   & 3988.HK      \\
		CITIC BANK      & 0998.HK      & CMOC            & 3993.HK      \\
		CKI HOLDINGS    & 1038.HK      & CITIC SEC       & 6030.HK      \\
		HENGAN INT'L    & 1044.HK      & CMSC            & 6099.HK      \\
		CHINA SOUTH AIR & 1055.HK      & CEB BANK        & 6818.HK      \\
		CHINA SHENHUA   & 1088.HK      & HTSC            & 6886.HK     
	\end{tabular}
	\caption{The $p=92$ stocks included in the portfolio.}
	\label{tab:list_of_stocks}
	\end{table}

		\end{document}